\documentclass{aa}

\usepackage{graphicx}
\usepackage{txfonts}
\usepackage{float}
\usepackage{hyperref}
\usepackage{url}
\usepackage{subfig}
\usepackage{longtable}
\usepackage{tabularx}
\usepackage{multirow}

\usepackage{xcolor}
\usepackage{natbib}
\def\pasp{PASP}
\def\apj{ApJ}
\def\aap{A\&A}
\def\mnras{MNRAS}
\def\aj{AJ}

\def\apjs{ApJS}
\def\aaps{Astron. Astrophys. Suppl. Ser.}
\def\memsai{Mem. Societa Astronomica Italiana}
\def\actaa{Acta Astronomica}

\def\apjl{Astrophysical Journal, Letters}

\begin{document}

   \title{Digging deeper into the dense Galactic globular cluster Terzan 5 with Electron-Multiplying CCDs\thanks{Based on data collected with the Danish 1.54 m telescope at ESO’s La Silla Observatory.}}

   \subtitle{Variable star detection and new discoveries}

   \author{R. Figuera Jaimes\inst{\ref{inst:mas}, \ref{inst:puc}, \ref{inst:standrews}}
          \and M. Catelan\inst{\ref{inst:puc}, \ref{inst:mas}}
          \and K. Horne\inst{\ref{inst:standrews}}
          \and J. Skottfelt\inst{\ref{inst:open}}
          \and C. Snodgrass\inst{\ref{inst:royal}}
          \and M. Dominik\inst{\ref{inst:standrews}}
          \and U. G. J{\o}rgensen\inst{\ref{inst:niels}}
          \and J. Southworth\inst{\ref{inst:keele}}
          \and M. Hundertmark\inst{\ref{inst:rechen}}
          \and P. Longa-Pe{\~n}a\inst{\ref{inst:antof}}
          \and S. Sajadian\inst{\ref{inst:isfahan}}
          \and J. Tregolan-Reed\inst{\ref{inst:ata}}
          \and T. C. Hinse\inst{\ref{inst:sdu}}
          \and M. I. Andersen\inst{\ref{inst:niels2}}
          \and M. Bonavita\inst{\ref{inst:royal}}
          \and V. Bozza\inst{\ref{inst:salerno}, \ref{inst:sezione}}
          \and M. J. Burgdorf\inst{\ref{inst:hamburg}}
          \and L. Haikala\inst{\ref{inst:ata}}
          \and E. Khalouei\inst{\ref{inst:seoul}}
          \and H. Korhonen\inst{\ref{inst:eso}}
          \and N. Peixinho\inst{\ref{inst:coimbra}}
          \and M. Rabus\inst{\ref{inst:concep}}
          \and S. Rahvar\inst{\ref{inst:sharif}}
          }
   \institute{Millennium Institute of Astrophysics MAS, Nuncio Monsenor Sotero Sanz 100, Of. 104, Providencia, Santiago, Chile\\ \label{inst:mas}
              \email{rfiguera@astro.puc.cl}
        \and Instituto de Astrof\'isica, Facultad de F\'isica, Pontificia Universidad Cat\'olica de Chile, Av. Vicu\~na Mackenna 4860, 7820436 Macul, Santiago, Chile \label{inst:puc}
        \and Centre for Exoplanet Science, SUPA, School of Physics \& Astronomy, University of St Andrews, North Haugh, St Andrews KY16 9SS, UK \label{inst:standrews}
        \and Centre for Electronic Imaging, Department of Physical Sciences, The Open University, Milton Keynes, MK7 6AA, UK \label{inst:open}
        \and Institute for Astronomy, University of Edinburgh, Royal Observatory, Edinburgh EH9 3HJ, UK \label{inst:royal}
        \and Centre for ExoLife Sciences, Niels Bohr Institute, University of Copenhagen, {\O}ster Voldgade 5, 1350 Copenhagen, Denmark \label{inst:niels}
        \and Astrophysics Group, Keele University, Staffordshire, ST5 5BG, UK \label{inst:keele}
        \and Astronomisches Rechen-Institut, Zentrum f{\"u}r Astronomie der Universit{\"a}t Heidelberg (ZAH), 69120 Heidelberg, Germany \label{inst:rechen}
        \and Centro de Astronom\'ia, Universidad de Antofagasta, Avenida Angamos 601, Antofagasta 1270300, Chile.\label{inst:antof}
        \and Department of Physics, Isfahan University of Technology, Isfahan 84156-83111, Iran \label{inst:isfahan}
        \and Instituto de Astronomia y Ciencias Planetarias, Universidad de Atacama, Copayapu 485, Copiapo, Chile \label{inst:ata}
        \and University of Southern Denmark, Department of Physics, Chemistry and Pharmacy, SDU-Galaxy, Campusvej 55, 5230 Odense M, Denmark \label{inst:sdu}
        \and Niels Bohr Institute, University of Copenhagen, Jagtvej 128, 2200 Copenhagen, Denmark \label{inst:niels2}
        \and Dipartimento di Fisica "E.R. Caianiello", Universit{\`a} di Salerno, Via Giovanni Paolo II 132, 84084, Fisciano, Italy \label{inst:salerno}
        \and Istituto Nazionale di Fisica Nucleare, Sezione di Napoli, Napoli, Italy \label{inst:sezione}
        \and Universit{\"a}t Hamburg, Faculty of Mathematics, Informatics and Natural Sciences, Department of Earth Sciences, Meteorological Institute, Bundesstra\ss{}e 55, 20146 Hamburg, Germany\label{inst:hamburg}
        \and Astronomy Research Center, Research Institute of Basic Sciences, Seoul National University, 1 Gwanak-ro, Gwanak-gu, Seoul 08826, Korea\label{inst:seoul}
        \and European Southern Observatory (ESO), Alonso de Córdova 3107, Vitacura, Santiago, Chile\label{inst:eso}
        \and Instituto de Astrof\'{i}sica e Ci\^{e}ncias do Espa\c{c}o, Departamento de F\'{i}sica, Universidade de Coimbra, 3040-004 Coimbra, Portugal\label{inst:coimbra}
        \and Departamento de Matem\'atica y F\'isica Aplicadas, Facultad de Ingenier\'ia, Universidad Cat\'olica de la Sant\'isima Concepci\'on, Alonso de Rivera 2850, Concepci\'on, Chile\label{inst:concep}
        \and Department of Physics, Sharif University of Technology, PO Box 11155-9161 Tehran, Iran\label{inst:sharif}
}

\date{Accepted June 26, 2024}

\abstract
   %context heading (optional) leave it empty if necessary  
   {High frame-rate imaging  was employed to mitigate the effects of atmospheric turbulence (seeing) in observations of globular cluster Terzan 5.}
   %aims heading (mandatory)
   {High-precision time-series photometry has been obtained with the highest angular resolution so far taken in the crowded central region of Terzan 5, with ground-based telescopes, and ways to avoid saturation of the brightest stars in the field observed.}
   %methods heading (mandatory)
   {The Electron-Multiplying Charge Coupled Device (EMCCD) camera installed at the Danish 1.54-m telescope at the ESO La Silla Observatory was employed to produce thousands of short-exposure time images (ten images per second) that were stacked to produce the normal-exposure-time images (minutes).  We employed difference image analysis in the stacked images to produce high-precision photometry using the DanDIA pipeline.}
   %results heading (mandatory)
   {Light curves of 1670 stars with 242 epochs were analyzed in the crowded central region of Terzan 5 to statistically detect variable stars in the field observed. We present a possible visual counterpart outburst at the position of the pulsar J1748-2446N, and the visual counterpart light curve of the low-mass X-ray binary CX 3. Additionally, we present the discovery of 4 semiregular variables. We also present updated ephemerides and properties of the only RR Lyrae star previously known in the field covered by our observations in Terzan 5. Finally, we report a significant displacement of two sources by $\sim0.62$ and $0.59$~arcseconds with respect to their positions in previous images available in the literature.}
   %conclusions heading (optional), leave it empty if necessary
   {}

\maketitle

\section{Introduction}

Globular clusters are very intriguing stellar systems formed by hundreds of thousands of gravitationally-bound stars. They have a high spherical symmetry and, in general, a high concentration of stars towards their centre. They also host the oldest stars in our galaxy. Therefore, they are similar to fossils and can help us to understand the early formation and evolution of our galaxy, the Milky Way.

In this paper, we focused our attention on the globular cluster Terzan 5. According to the Catalog of Parameters for Milky Way Globular Clusters \citep[][2010 version]{harriscatalog}, Terzan 5 is located at RA(J2000) $=$ 17:48:04.8 and Dec(J2000) $=$ $-$24:46:45. It has a metallicity of $[\mathrm{Fe}/\mathrm{H}]=-0.23$, a reddening $E(B-V)=2.28$~mag, a $V$ magnitude level of the horizontal branch (HB) $V_{\mathrm{HB}}=22.04$~mag, distance modulus $(m-M)_\mathrm{V}=21.27$~mag, absolute visual magnitude $M_\mathrm{V}=-7.42$~mag, and distance from the Sun $R_{\odot}=6.9$~kpc.

\cite{ortolani96+02} produced a $VI$ colour-magnitude diagram (CMD) with images taken with the ESO New Technology Telescope (NTT) at ESO La Silla, Chile, with seeing around $0.34^{\arcsec}$- $0.50^{\arcsec}$, and derived a reddening of $E(B-V)=2.49$ and a distance from the Sun of $R_{\odot}=5.6$~kpc.

\cite{origlia11+08} carried out a spectroscopic study where 33 red-giant members of Terzan 5 were analyzed. The spectra were taken using the NIRSPEC instrument at the Keck II telescope at W.\ M.\ Keck Observatory, Hawaii. The resulting spectra have a resolution of $R=25000$. Inspection of the iron abundance distribution showed that the cluster has two stellar populations, the first with an average metallicity of about $[\mathrm{Fe}/\mathrm{H}]=-0.25 \pm 0.07$~rms and the second with about $[\mathrm{Fe}/\mathrm{H}]=+0.27 \pm 0.04$~rms. 
In the Catalog of Variable Stars in Galactic Globular Clusters \citep[][August 2019 update]{Clement01+09}, and in the literature available so far, there are 13 variable stars known in the field of Terzan 5.

The first variables reported were two Mira variables (V1-V2) that were discovered by \cite{spinrad74+02} using observations taken with photographic plates ($\sim$20 epochs) with telescopes at Lick, Cerro Tololo, and Palomar observatories.

Later, \cite{edmons01+03} discovered two more variables using data taken with the {\em Hubble Space Telescope (HST)}. One (V3) corresponds to the first RR Lyrae RR0 type discovered in Terzan 5 with a period of 0.60 days, and the second (V4) is a blue faint sinusoidal variable with a period of about 7 hr. The authors consider that this star is a possible eclipsing blue straggler or (less likely) the infrared counterpart to the low-mass X-ray binary discovered by \cite{makishima81+19} and also analyzed by \cite{johnston95+02} and \cite{verbunt95+03}.

\cite{sloan10+09} discovered five Mira variables (V5-V9) using the SIRIUS near-infrared camera at the 1.4 m Infrared Survey Facility telescope at the South African Astronomical Observatory, which produced simultaneous photometry in the $J$, $H$, and $K_s$ bands.

Finally, \cite{origlia19+09} discovered four more variables, one Mira type (V10) and three RR Lyrae RR0 type (V11, V12, and V13). In this work, they used 24 images taken with the {\em HST} Wide Field Planetary Camera 2, 12 images in F606W and 12 in the F814W passbands.

There are also, as of the time of this study, 42 millisecond pulsars\footnote{\url{http://www.naic.edu/~pfreire/GCpsr.html}} (MSPs) known in the field of Terzan 5 \citep[see e.g.,][and references therein]{ransom05+06, ransom18+09, cadelano18+07, bahramian22+13}.

It is well established that MSPs are neutron stars rotating at incredibly fast and precise periods, typically in the range of milliseconds. The strength of their magnetic fields is believed to significantly influence their rotation rates. In the context of globular clusters, MSPs are known to form binary systems where they possibly accrete matter from their companions, leading to changes in their magnetic fields and rotational speeds \citep[see e.g.,][and references therein]{bhattacharyya22+02}. Globular clusters host a large population of MSPs, with approximately 300 known in 38 clusters. Notably, Terzan 5 stands out with over 20 more detections than the well-studied cluster 47 Tucanae$^{1}$, making it an ideal target for investigating these phenomena further. Additionally, there have been observational indications of electromagnetic signals from MSP companions in the visual range, which offers the potential to quantify these variations and gain insights into their physical origins, such as interactions with companions or the ejection of accreted material. Visual detections not only provide independent confirmation of variations in these systems but also facilitate precise positional measurements for the observed targets \citep[see e.g.,][]{breton13+08, pallanca16+cosmiclab}.

Furthermore, globular clusters encompass a wide array of stars in various stages of evolution. Among them, numerous types of variable stars have been identified, including Miras, type II Cepheids, RR Lyrae, SX Phoenicis, and eclipsing binaries, among others \citep[e.g.,][and references therein]{Clement01+09, clement17, belloni21+01, lugger23+05}. Some clusters exhibit a higher abundance of these variables than others. A comprehensive census, detection, and characterization of the variable stars within these stellar systems present an excellent opportunity to deepen our understanding of stellar evolution theories and the formation and evolution of globular clusters. Pulsating variable stars, such as RR Lyrae, have proven invaluable as independent estimators of physical properties within globular clusters, including metallicities, distances, and the Oosterhoof dichotomy. Moreover, given the evidence of multiple stellar populations within globular clusters, the study of variable stars can contribute to their detection and characterization. Worthy of note is the fact that globular clusters harbour the oldest stars in our galaxy, making an accurate census and characterization of these stars crucial for advancing our knowledge of early galaxy formation and evolution \citep[see e.g.,][and references therein]{catelan2015pulsating, figuera18phdt}.

Terzan 5, located in the Galactic bulge, poses unique challenges for observations due to its complex nature. The presence of background stars greatly affects the observations of this cluster. Furthermore, Terzan 5 is known for its strong differential reddening, adding another layer of complexity. Additionally, the target itself is faint due to strong obscuration from heavy extinction, with a $V$ magnitude level of approximately 22.04 mag on the HB. With a central luminosity density of 5.14, it stands as one of the densest known globular clusters \citep[][2010 version]{harriscatalog}.

Given these factors, studying the densely populated central region of Terzan 5 poses significant difficulties, including issues related to crowding, blending, and saturation of stars. Blending and saturation can result in the loss of variability information for some of the variables. To address these challenges and achieve a comprehensive census of the variable stars in the central region, the utilization of an EMCCD camera, high frame-rate imaging, and difference image analysis (DIA) becomes crucial. This study marks the first time-series observational program specifically designed to accurately detect and capture the complete phase variation of variable stars within the densely crowded central region of Terzan 5.

In Sect.~\ref{sec:observations}, we present the instruments used, the observations taken, and the pipelines and techniques employed to produce the photometry. In Sect.~\ref{sec:var_det}, the techniques employed to search and detect variable stars are explained. The CMD of Terzan 5 is given in Sect.~\ref{sec:cmd}. Known variables in the field observed are presented in Sect.~\ref{sec:known_var}. In Sect.~\ref{sec:new_var}, new variables and candidates discovered in this study are shown. Stars flagged as variables in the {\em Gaia} survey are discussed in Sec.~\ref{sec:gaia}. In Sec.~\ref{sec:motion}, we present the case of two stars that showed a significant displacement. Finally, conclusions are given in Sect.~\ref{sec:concl}.

\section{Instruments and Observations}\label{sec:observations}

\subsection{Telescope}

To carry out observations, the Danish 1.54 m telescope\footnote{\url{https://www.eso.org/public/teles-instr/lasilla/danish154/}} was employed. This telescope is located at an altitude of 2375 m in ESO's La Silla Observatory, Chile at 70$^{\circ}$44$^{\arcmin}$07$^{\arcsec}$.662W 29$^{\circ}$15$^{\arcmin}$14$^{\arcsec}$.235S.

\subsection{Detector}

The telescope is equipped with an Andor Technology iXon+897 EMCCD camera \citep{lesser15, howell06}, which has a 512 $\times$ 512 array of 16 $\mu$m pixels, a pixel scale of 0$\farcs$09 per pixel and a total field of view of $\sim45\times45$ arcsec$^{2}$.

For the purpose of this research, the camera was configured to work at a frame rate of 10 Hz (this is 10 images per second) and an EM gain of $300\,\mathrm{e}^-/$photon. The camera is placed behind a dichroic mirror, which works as a long-pass filter.
Considering the mirror and the sensitivity of the camera, it is possible to cover a wavelength range between 650 nm to 1050 nm \citep{skottfelt15+14}, corresponding roughly to a combination of SDSS $\mathrm{i}^{\prime}+\mathrm{z}^{\prime}$ filters \citep{bessell05}.

\subsection{Observations}

\begin{figure}
  \centering
  \includegraphics[width=0.98\hsize]{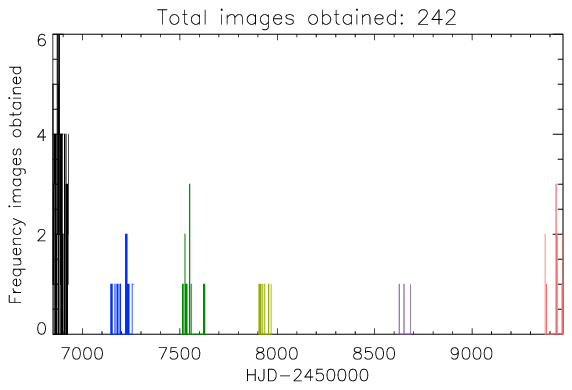}
  \caption{Histogram of observations obtained after the stacking in the field of the globular cluster Terzan 5. The colours represent the images obtained for each year: black (group 1) is 2014, blue (group 2) is 2015, green (group 3) is 2016, olive (group 4) is 2017, purple (group 5) is 2019, and rose (group 6) is 2021.}
         \label{fig:histo_obs}
\end{figure}

The MINDSTEp consortium program aimed to explore the Milky Way bulge for exoplanet detection and characterization, with observations occurring annually from April to September. However, in cases where observations of the Galactic bulge were not possible, globular clusters were utilized as alternatives. Terzan 5 was one of the chosen globular clusters for these observations. A histogram in Fig.~\ref{fig:histo_obs} displays the distribution of science images obtained between 2014 and 2021, showing that a total of 242 science images were obtained within the field of Terzan 5.

\subsection{Exposure times}

The exposure time aimed for each of the mentioned images was 10 minutes (600 seconds). To achieve this, the camera was continuously taking images (at the rate of 10 images per second) for the total exposure time of 10 minutes and then all the images obtained during the 10 minutes were stacked to produce a single observation \citep[see][for a detailed explanation of the technique employed to perform the stacking]{Figuera16+mindstep,Skottfelt15+mindstep}. That is, a 10-minute exposure time observation is the result of stacking 6000 images that were continuously taken. We produced 161 images with a total exposure time of 600 s, 71 images in the range of 550 - 559 s, 7 images in the range of 500 - 549 s, 2 images in the range of 450 to 499 s, and 1 image in the interval of 400 to 449 s. 

Science images with 600 s mean that all frames were used, while science images with different exposure times mean that some frames were not included. This might be due to different reasons, such as poor seeing, telescope tracking issues, or limitations. The reference frame with a 0$\farcs$41 FWHM (see Sec. \ref{subsec:phot}) was intentionally obtained by discarding images with poor seeing and selecting only those with the best possible seeing to create the best reference frame possible for use in subsequent steps.

\subsection{Photometry}\label{subsec:phot}

Once the stacked images were produced, the next step was to analyze the images for scientific purposes. It means that the next step is to extract the photometry in each of the images obtained. Nowadays, there are several methods and techniques to achieve this goal \citep[see e.g.,][and references therein]{catelan23}. In our case, we employed the DanDIA\footnote{DanDIA is built from the DanIDL library of IDL routines available at \url{http://www.danidl.co.uk}.} pipeline \citep{bramich08, bramich13+10}, which is based on difference image analysis \citep[DIA;][]{alard98+01, alard00}.

The reference frame used to perform the difference image analysis was built by stacking a total of 3000 images, which is equivalent to a science image with a 300s exposure time. The resulting mean FWHM measured in the stars in the reference frame was about 0$\farcs$41. 

%and it is shown in Figure~\ref{fig:finding_chart_terzan5}.

The instrumental magnitudes $m_{\mathrm{ins}}$ for each star in the light curves obtained are represented by
\begin{equation}\label{eq:inst_mag}
m_{\mathrm{ins}}(t)=17.5-2.5\log(f_{\mathrm{tot}}(t)),
\end{equation}
where $f_{\mathrm{tot}}(t)$ is the total flux in ADU/s defined as 
\begin{equation}\label{eq:total_flux}
f_{\mathrm{tot}}(t)=f_{\mathrm{ref}}+\frac{f_{\mathrm{diff}}(t)}{p(t)},
\end{equation}
where $f_{\mathrm{ref}}$ (ADU/s) corresponds to the reference fluxes of each star detected in the reference frame, $f_{\mathrm{diff}}(t)$ (ADU/s) is the difference fluxes measured in the difference images at the position of each star detected in the reference frame, and $p(t)$ is the photometric scale factor used to scale the reference frame to each image \citep{bramich11+04}.

For a comprehensive explanation of how the pipeline works, we point the reader to \cite{Figuera16+mindstep} and \cite{bramich08}.

The light curve information, photometric measurements, and fluxes of all variable stars studied in this work are available through the CDS\footnote{\url{http://cds.u-strasbg.fr/}} database with the format given in Table \ref{tab:electronic_data}.

\begin{table*}
\caption{Time-series photometry for all known and new variables in the field of view covered in Terzan 5.
The standard $M_{\mathrm{std}}$ and instrumental $M_{\mathrm{ins}}$ magnitudes are listed in columns 5 and 6, respectively, corresponding to the cluster, variable star, filter, and epoch (UTC) of mid-exposure
listed in columns 1-4, respectively. The uncertainty in $M_{\mathrm{ins}}$ is listed in column 7, which also corresponds to the uncertainty in $M_{\mathrm{std}}$. For completeness, we also list the quantities $f_{\mathrm{ref}}$, $f_{\mathrm{diff}}$ and $p$ from Eq. \ref{eq:total_flux} in columns 8, 10, and 12, along with
the uncertainties $\sigma_{\mathrm{ref}}$ and $\sigma_{\mathrm{diff}}$ in columns 9 and 11, respectively. This is an extract from the full table, which is available with the electronic version of the article at the CDS.}
\label{tab:electronic_data}
\centering
\tabcolsep=0.08cm
\begin{tabular}{cccccccccccccc}
\hline\hline
Cluster & var & Filter & HJD & $M_{\mathrm{std}}$ & $M_{\mathrm{ins}}$ & $\sigma_m$ & $f_{\mathrm{ref}}$ & $\sigma_{\mathrm{ref}}$ & $f_{\mathrm{diff}}$ & $\sigma_{\mathrm{diff}}$ & $p$ \\
        & ID  &        & (d) &       (mag)        &       (mag)        &   (mag)    & (ADU s$^{-1}$)    &     (ADU s$^{-1}$)     &   (ADU s$^{-1}$)   & (ADU s$^{-1}$) &     \\
\hline
Terzan5 & V3  & I & 2456847.71238 & 18.274 & 9.569 & 0.017 & 1321.348 & 329.766 & +001166.327 & 161.418 & 7.0252\\
Terzan5 & V3  & I & 2456848.49854 & 18.421 & 9.716 & 0.032 & 1321.348 & 329.766 & -000190.850 & 331.537 & 8.6296\\
\vdots   & \vdots & \vdots &      \vdots   & \vdots &\vdots &\vdots &\vdots     &  \vdots     &   \vdots     &  \vdots  & \vdots\\
Terzan5 & V14 & I & 2456847.71238 & 14.633 & 5.928 & 0.001 & 40810.543 & 333.083 & +012128.637 & 379.316 & 7.0252\\
Terzan5 & V14 & I & 2456848.49854 & 14.639 & 5.933 & 0.002 & 40810.543 & 333.083 & +013131.433 & 548.535 & 8.6296\\
\vdots   & \vdots & \vdots &      \vdots   & \vdots &\vdots &\vdots &\vdots     &  \vdots     &   \vdots     &  \vdots  & \vdots\\
Terzan5 & V15 & I & 2456847.71238 & 14.916 & 6.211 & 0.002 & 33569.503 & 339.078 & -005486.698 & 360.373 & 7.0252\\
Terzan5 & V15 & I & 2456848.49854 & 14.914 & 6.208 & 0.002 & 33569.503 & 339.078 & -006093.916 & 532.514 & 8.6296\\
\vdots   & \vdots & \vdots &      \vdots   & \vdots &\vdots &\vdots &\vdots     &  \vdots     &   \vdots     &  \vdots  & \vdots\\
\hline

\end{tabular}
\end{table*}

\subsection{Photometric calibration}

\begin{figure}[htp!]
\centering
\includegraphics[width=\hsize]{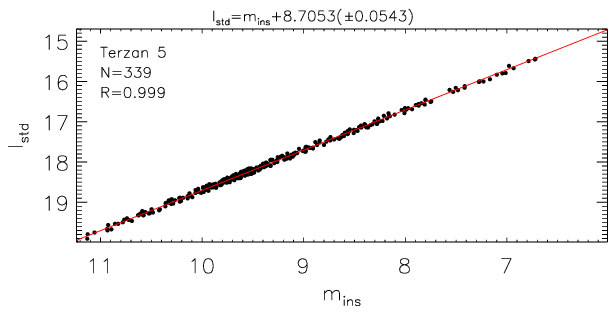}
\caption{Standard $I$ magnitude available in \citet{ortolani96+02} as a function of the instrumental magnitude. The red line is the fit that best matches the data and is given in the title of this plot and Equation~\ref{eq:std_mag}. The correlation coefficient is 0.999.}
\label{fig:std_calibration}
\end{figure}

The resulting instrumental magnitudes were transformed to the standard Johnson-Kron-Cousins photometric system \citep{landolt92} by taking advantage of the $I$ magnitudes electronically available in \citet{ortolani96+02}. The linear transformation found between both magnitudes is 
\begin{equation}\label{eq:std_mag}
  I=m_{inst}+(8.7053\pm0.0543), 
\end{equation}
where $m_{inst}$ corresponds to the instrumental magnitudes and $I$ is the resulting standard magnitude. Fig.~\ref{fig:std_calibration} shows the data used (black points) and the best fit obtained (red line). A total of 339 stars were used in the fit and the obtained correlation coefficient is 0.999.

\subsection{Astrometry}\label{subsec:astrometry}

\begin{figure*}
  \centering
  \includegraphics[width=\hsize]{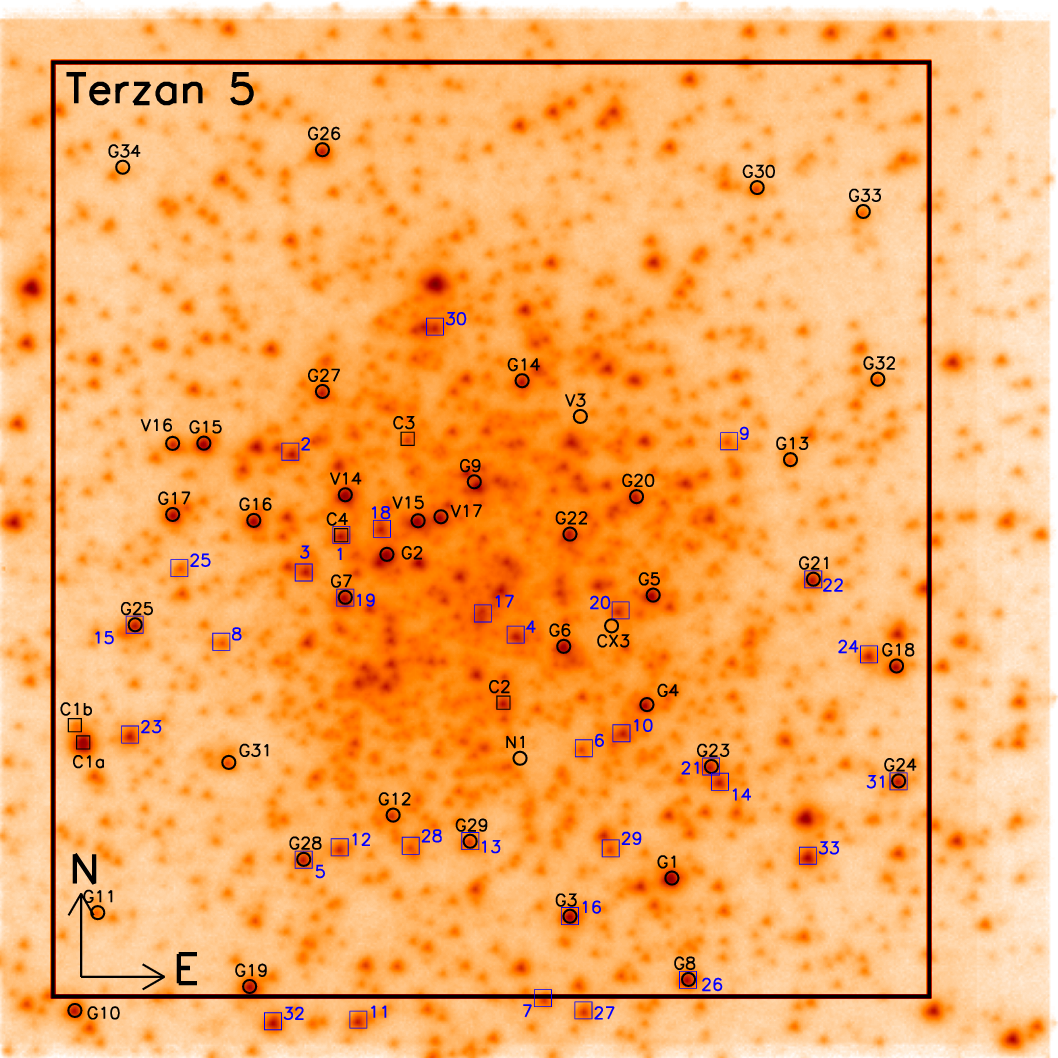}
  \caption{Finding chart for globular cluster Terzan 5. Black labels are variable stars studied in this work. Circled labels correspond to new discoveries and previously known variable stars while squared labels are the candidate variables. Blue labels are the 33 red giant stars studied in \cite{origlia11+08}. The average FWHM obtained for the stars in the reference frame was about 0$\farcs$41.
 The big black square is the region explored in our study for new variable star detection and characterization which corresponds to a field of view of about $38^{\arcsec}\times 40^{\arcsec}$. Labeled stars lying outside the black box correspond to stars previously studied in the literature.}
         \label{fig:finding_chart_terzan5}
\end{figure*}

To do the astrometry correction to our reference frame, an {\em HST} image available in the {\em HST} archive was employed (file name: hst$\_$12933$\_$01$\_$acs$\_$wfc$\_$f814w$\_$jc3801iu$\_$drc.fits). We matched stars in the field of our reference frame with those in the field of the {\em HST} image by using the GAIA software \citep[Graphical Astronomy and Image Analysis Tool;][]{Draper2000}. A total of 260 stars were matched over the entire field of view. The root-mean-square (RMS) scatter obtained was approximately $0\farcs03$ ($\sim$0.3 pixels). The {\em HST} astrometry was also corroborated with Gaia astrometry and with the STScI\footnote{\url{https://www.stsci.edu/hst/instrumentation/wfc3/news}} documentation. The final reference frame with the resulting astrometric solution was employed to produce the finding chart shown in Fig.~\ref{fig:finding_chart_terzan5}, which contains the position and identification of all variable stars studied in this work. Table~\ref{table:ephemerides_var} provides the equatorial J2000 celestial coordinates for all variables in this study.

\section{Variable star detection}\label{sec:var_det}

To search for variable stars in the field covered by our reference image, we explored several techniques aiming to have a semi-automatic detection of variable candidates. To search for variable stars in the field covered by our reference image, we explored several techniques, aiming to have a semi-automatic detection of variable candidates. Our final selection, which is displayed in Fig.~\ref{fig:rms_sb_terzan5}, is based on a judicious combination of different criteria, as well as careful inspection of the difference images, as discussed in the following subsections.

\subsection{Root-mean-square diagram}\label{sec:rms}

We started using the RMS calculated for all stars detected versus their mean $I$ magnitude obtained with the final photometry. However, on this occasion, we fitted a polynomial model to the data to define where most of the stars are located. Thereafter, we defined a threshold that was shifted by a factor of 3 over the model. Any star placed over the threshold was selected as a candidate to be variable subject to further scrutiny of its light curve and examination in the difference images.

\subsection{$\cal S_B$ statistic}\label{sec:sb}

Another approach used was the $\cal S_B$ statistic defined in \cite{figuera13+04} for semi-automatic detection of variable star candidates. The $\cal S_B$ statistic is defined as 
\begin{equation}\label{eq:sb_index}
{\cal S_B}=\left(\frac{1}{NM}\right)\sum_{i=1}^M\left(\frac{r_{i,1}}{\sigma_{i,1}}+\frac{r_{i,2}}{\sigma_{i,2}}+...+\frac{r_{i,k_i}}{\sigma_{i,k_i}}\right)^2,
\end{equation}
where $N$ is the number of data points for a given light curve, and $M$ is the number of groups formed of time-consecutive residuals of the same sign from a constant-brightness light curve model (e.g., mean or median). The residuals $r_{i,1}$ to $r_{i,k_i}$ correspond to the $i$th group of $k_i$ time-consecutive residuals of the same sign with corresponding uncertainties $\sigma_{i,1}$ to $\sigma_{i,k_i}$. The $\cal S_B$ statistic is larger in value for light curves with long runs of consecutive data points above or below the mean, which is the case for variable stars with periods longer than the typical photometric cadence.

\begin{figure}
  \centering
  \includegraphics[width=\hsize]{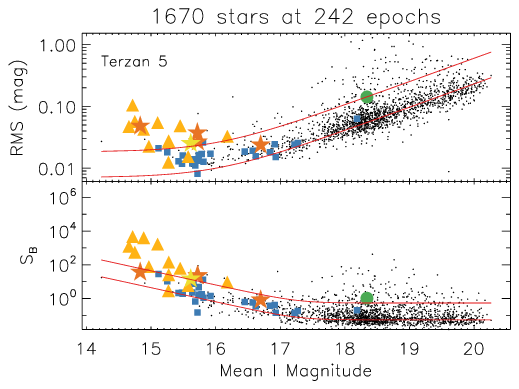}
  \caption{RMS and $\cal S_B$ index versus the mean $I$ magnitude in the globular cluster Terzan 5. The green-circled point is the previously known RR Lyrae V3, dark-yellow-triangle points are the semiregular variables, orange five-pointed points are the stars selected as candidate variables, blue squares are stars flagged as long-period variables (LPVs) in {\em Gaia} Data Release 3 (DR3), and light-yellow five-pointed star is an eclipsing binary candidate. The lower red lines correspond to the fits applied to the data, while the upper red lines are the shifts applied to define the thresholds for variable candidate detection.}
         \label{fig:rms_sb_terzan5}
\end{figure}

In Fig.~\ref{fig:rms_sb_terzan5}, the $\cal S_B$ is plotted against the mean $I$ magnitude for all stars detected in our analysis. Candidates for variability were selected based on their position above the threshold, set at ten times the model $\cal S_B$ values. These candidates, along with those obtained in Sect.~\ref{sec:rms} were further scrutinized in the difference images, as described in Sec. \ref{sec:sum_diff}. Fig.~\ref{fig:rms_sb_terzan5} presents the final selection of variable stars determined through the combined analysis discussed in Sec. \ref{sec:rms} and Sec. \ref{sec:sum_diff}.

A comparison between the RMS and the $\cal S_B$ is shown in Fig.~\ref{fig:rms_sb_terzan5}. All newly discovered variables exhibited high RMS and $\cal S_B$ values well exceeding the thresholds. {\em Gaia} stars (G1-G9) flagged as variables, {\em for} which we found a period and classified as semi-regular, also exceeded the thresholds, except for G7, G8, and G9. In the case of G9, this is due to the fact that it was not automatically detected in the reference frame. The remaining fraction of stars with RMS and $\cal S_B$ values that resembled nonvariable stars corresponded to {\em Gaia}-identified LPVs, mainly stars with low amplitudes listed in Table~\ref{table:ephemerides_var} (G11-G34). These are depicted as blue squares in Fig.~\ref{fig:rms_sb_terzan5}, potentially indicating that they are at the threshold of uncertainties. Further exploration of variables at RMS and $\cal S_B$ values below thresholds may require methods to detrend our data.

\subsection{Analysis of difference images}\label{sec:sum_diff}

Difference image analysis is a powerful technique used for photometry in crowded fields \citep{alard98+01}. The technique involves subtracting a sequence of aligned images, taken of the same field, from a reference frame. The result of this subtraction is a set of difference images that capture the flux variation due to the changes in brightness of variable stars. When these difference images are stacked, the flux variation of these stars can be accumulated into a single image. Consequently, this procedure can be used as a method to potentially detect variable stars in observed fields. Stacked images can be represented as:

\begin{equation}\label{eq:sum_diff}
S_{ij}=\sum_k\frac{|D_{k,ij}|}{\sigma_{k,ij}},
\end{equation}

\noindent where $S_{ij}$ is the sum of all the difference images, $D_{k,ij}$ is the $k$th difference image, $\sigma_{k,ij}$ is the pixel uncertainty for each image $k$, and $i$ and $j$ denote pixel positions.

All the  variable star candidates obtained by using the RMS diagrams or the $\cal S_B$ statistic explained in Sects. \ref{sec:rms} and \ref{sec:sb} were inspected visually in the stacked images to confirm or refute their variability. Additionally, any other candidate detected in the stacked images that might have been initially overlooked (e.g., if it fell below the defined thresholds in the SB and RMS diagrams) was included in the inspection process. Difference images were also blinked to finally corroborate the variation of the stars selected. Difference fluxes for all variable stars studied in this work were significantly higher than the background level measured in the stacked image defined in Eq.~\ref{eq:sum_diff}, by at least 6$\sigma$. Only two of the LPVs previously reported in {\em Gaia} DR3 were slightly below the 6$\sigma$ level.

\section{Period search}\label{sec:per}

To corroborate or improve the period of previously known variable stars and to find possible periods for newly discovered variables, several very well-known techniques were implemented such as the string length, fast $\chi^2$, and least squares methods \citep[see, e.g.][respectively]{lafler65+01, palmer09, lenz14+01}.

From our previous experience we noticed that the string length is very useful in searching for short periods; however, because the technique has to test a significant number of periods and measure the dispersion in the resulting phased light curves for each period, we only searched for periods with this method in the range of 0.01 to 0.9 days.

To search for periods in other intervals, for example, from 0 to Nyquist frequency, we used the fast $\chi^2$ method, which is optimized for the detection of periods in irregularly sampled data with nonuniform errors. Finally, we also implemented the least squares technique to detect the strongest frequency that dominates the frequency spectrum in each light curve. In the cases where a discrepancy was evident between the frequencies found, we compared both frequencies plotting the phased light curves and chose the one that visually better phased the data. In the cases where differences between the frequencies were very small, and the phased light curves did not show any significant improvement between one and the other, we chose the strongest frequency found in the periodogram by the least squares method. The adopted periods in our analysis are given in Table~\ref{table:ephemerides_var}.

\section{Colour-magnitude diagram}\label{sec:cmd}

To build the CMD of Terzan 5, we employed the data already available in Table~2 in \cite{ortolani96+02}. The idea was to highlight in the CMD all variable stars detected in the field of our reference image. However, we noticed that the mentioned table did not have celestial coordinates available to compare with our reference frame, but instead the x and y physical coordinates of the image they used in their Fig.~1. For this reason, we obtained the mentioned frame from the ESO archive\footnote{\url{http://archive.eso.org/cms.html}} (file name: ONTT.1994-05-16T06:36:38.000.fits), for which we calculated the equatorial coordinates similar to that done in Sect.~\ref{subsec:astrometry}.

\begin{figure}
  \centering
  \includegraphics[width=0.98\hsize]{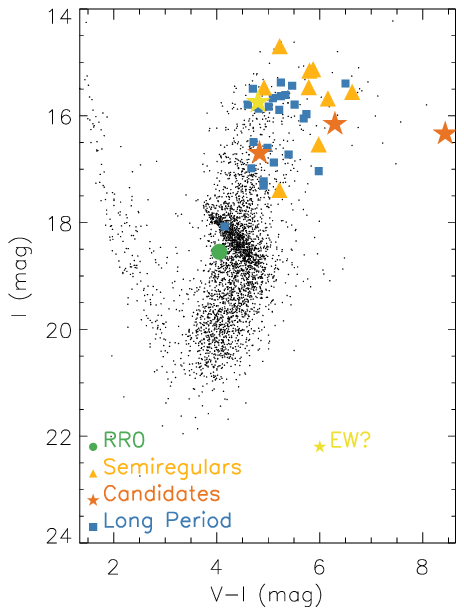}
  \caption{CMD obtained for the globular cluster Terzan 5. The vertical axis corresponds to the $I$ magnitude and the horizontal axis is the $V-I$ colour. The green-circled point is the previously known RR Lyrae V3, dark-yellow-triangle points are the semiregular variables, orange five-pointed points are the stars selected as candidate variables, blue squares are stars flagged as LPVs in {\em Gaia} DR3, and the light-yellow five-pointed star is an eclipsing binary candidate.}
         \label{fig:cmd_terzan5}
\end{figure}

In Table \ref{tab:table2_ortolani}, we include the magnitudes and colours obtained by \citet{ortolani96+02}, but in this case, the celestial coordinates have been included and in Fig.~\ref{fig:cmd_terzan5} the CMD is shown. 

In Fig.~\ref{fig:cmd_terzan5}, it is possible to see that most of the variables and candidates are located at the top of the red giant branch (RGB), and some are more toward the middle level of the RGB. The dispersion noticed is due to the known intense differential reddening associated with this cluster \citep[see e.g.,][and references therein]{origlia19+09, ortolani96+02}. The reddening effect can also be seen in the diagonally-extended HB which is located at $I\sim18.24$~mag \citep{ortolani96+02}. It is also possible to notice that the RR Lyrae V3 is located below the level of the HB by $\sim 0.31$~mag.

\begin{table*}
\caption{Information used to build the CMD for Terzan 5 with data listed in \cite{ortolani96+02}. Celestial coordinates have been added to the table. Column 1 is a star ID. Columns 2-3, are the x and y coordinates previously listed in the table. The $V$, $I$ magnitudes, and $(V-I)$ colours are given in columns 4-6, respectively. Columns 7-8, correspond to the right ascension and declination. This is an extract from the full table, which is available with the electronic version of the article at the CDS.}
\label{tab:table2_ortolani}
\centering
\begin{tabular}{cccccccc}
\hline\hline
ID & x & y & $V$ & $I$ & $V-I$ & RA & Dec \\
   & (pixels) & (pixels) & (mag) & (mag) & (mag) & (J2000) & (J2000) \\
\hline
1 &959.00 & 4.15 &24.345 &20.534 &3.811 &17:47:59.73 & -24:47:53.9\\
2 &385.37 & 4.63 &23.847 &19.903 &3.944 &17:48:05.16 & -24:47:50.5\\
\vdots & \vdots & \vdots & \vdots & \vdots & \vdots & \vdots & \vdots \\
5 &224.59 & 9.80 &24.793 &20.336 &4.457 &17:48:06.68 & -24:47:48.9\\
6 &986.43 & 11.48& 23.770& 19.611& 4.159& 17:47:59.47& -24:47:53.1\\
\vdots & \vdots & \vdots & \vdots & \vdots & \vdots & \vdots & \vdots \\
9 &222.02 & 22.28& 23.881& 19.128& 4.753& 17:48:06.70& -24:47:47.3\\
10& 679.69& 22.81& 24.514& 20.114& 4.400& 17:48:02.37& -24:47:49.9\\
\vdots & \vdots & \vdots & \vdots & \vdots & \vdots & \vdots & \vdots \\
\hline

\end{tabular}
\end{table*}

\section{Previously known variables}\label{sec:known_var}
 
As mentioned earlier, so far there are 13 known variable stars in the field of Terzan 5. The authors used their own nomenclature to label their discoveries and for practical reasons \citet{Clement01+09} adopted a new nomenclature in her catalogue as shown in table~\ref{table:known_var}, nomenclature which we will adopt in this article.

\begin{table*}
\caption{Known variable stars listed in the Catalog of Variable Stars in Galactic Globular Clusters \citep[][August 2019 update]{Clement01+09}. Column 1 is the star ID assigned in the catalog. Column 2 corresponds to the original star ID assigned by the authors in the published paper. Right ascension and declination are given in columns 3 and 4. The period reported for the variables is given in column 5 while columns 6-8 are the mean magnitude, amplitude, and colour for the mean magnitude and amplitude. Finally, column 9 is the type of variable reported.}            % title of Table
\label{table:known_var}
\centering             
\begin{tabular}{c c c c c c c c c}        
\hline\hline                 
ID   & ID &   RA     &     Dec     &  $P$ &  Mag & Amp & Colour & Type   \\    % table heading
Clement & Author & J2000   & J2000       &   d     &      &     &       &  \\
\hline                        
V1  & V     & 17:48:03.80 & -24:47:54.6 & 245.0   & 14.3 & 1.5  & I & M    \\
V2  & V$_s$ & 17:47:59.46 & -24:47:17.6 & 217.0   & 7.65 & 0.67 & K & M    \\
V3  & V1    & 17:48:05.14 & -24:46:38.2 & 0.6     & --   & 0.25 & R & RR0  \\
V4  & V2    & 17:48:05.06 & -24:46:52.5 & 0.291   & --   & 0.33 & R & E?   \\
V5  & V5    & 17:48:03.40 & -24:46:42.0 & 464.0   & 6.83 & 0.86 & K & M    \\
V6  & V6    & 17:48:09.25 & -24:47:06.3 & 269.0   & 7.50 & 0.61 & K & M    \\
V7  & V7    & 17:47:54.33 & -24:49:54.6 & 377.0   & 7.03 & 0.76 & K & M    \\
V8  & V8    & 17:48:07.18 & -24:46:26.6 & 261.0   & 7.44 & 0.80 & K & M    \\
V9  & V9    & 17:48:11.86 & -24:50:17.1 & 464.0   & 8.41 & 0.83 & K & M    \\
V10 & M6    & 17:47:53.2  & -24:44:34.0 & 455.0   &  --  &  --  &   & M    \\
V11 & RR1   & 17:48:02.8  & -24:47:47.5 & 0.72    &  --  &  --  &   & RR0  \\
V12 & RR2   & 17:48:08.2  & -24:45:42.1 & 0.64    &  --  &  --  &   & RR0  \\
V13 & RR3   & 17:48:04.3  & -24:47:37.7 & 0.89    &  --  &  --  &   & RR0  \\
\hline

\end{tabular}
\end{table*}

\subsection{RR Lyrae V3}

\begin{figure}[h]
  \centering
  \begin{tabular}{c}
    \includegraphics[width=0.97\hsize]{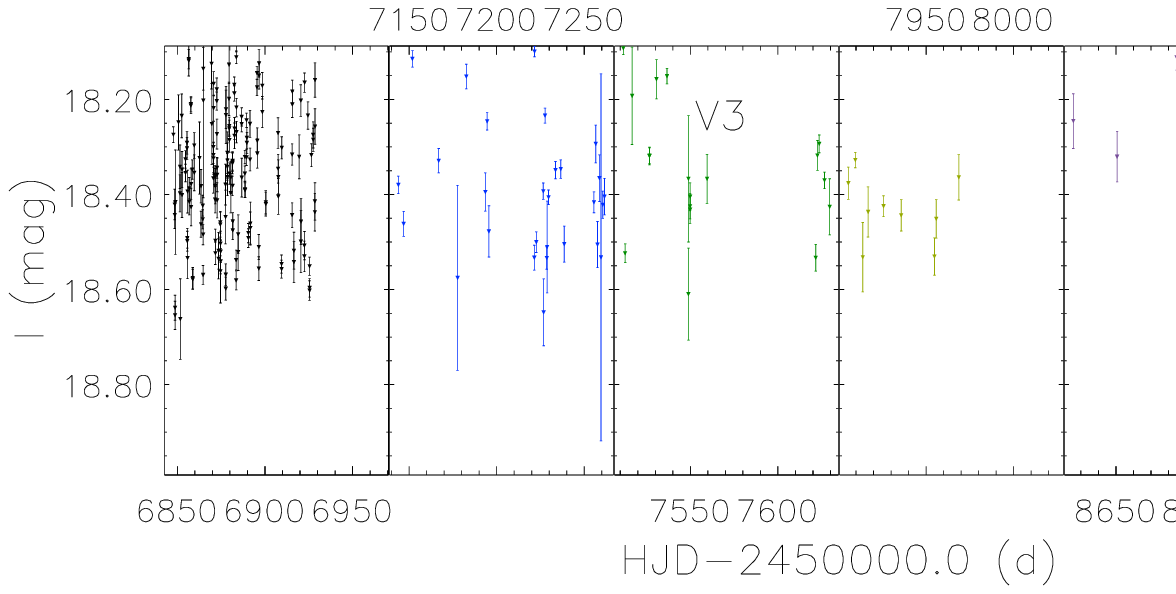} \\
    \\
    \includegraphics[width=0.97\hsize]{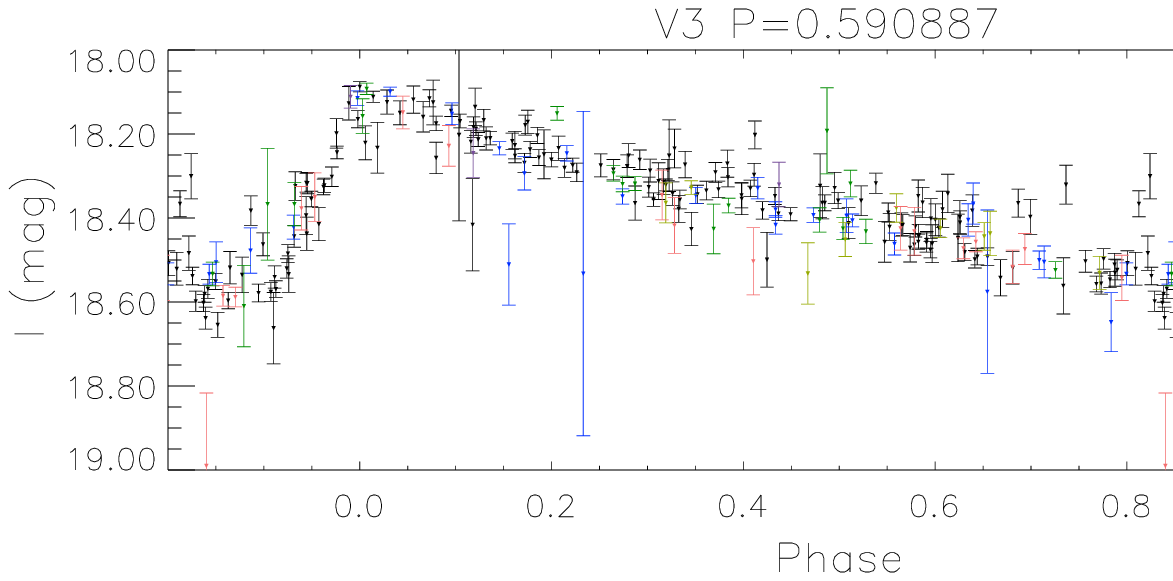} \\
  \end{tabular}
  \caption{Variable star V3. The top plot shows the light curve in $I$ magnitude versus HJD while at the bottom the $I$ magnitude is plotted versus the phase obtained with a period of 0.590887 d. The colours of the plots match those in Fig.~\ref{fig:histo_obs}.}
  \label{fig:V3}
\end{figure}

V3 is one of the two known variables whose coordinates fall inside the field of view covered by our observations. The light curve for this star is presented in Fig.~\ref{fig:V3}. The top plot is the light curve in magnitude against the Heliocentric Julian Date (HJD) version, while the bottom plot shows the phased version. To produce the very nice phased light curve shown, we detected that this star is pulsating in the fundamental mode typical for RR Lyrae RR0 type with a period of about 0.590887 $\pm$ 0.000013 days. The peak-to-peak amplitude of our light curve is 0.57 mag.

This star is highlighted as a green filled circle in the RMS and $\cal S_B$ index given in Fig.~\ref{fig:rms_sb_terzan5}. In the CMD shown in Fig.~\ref{fig:cmd_terzan5}, the star appears at an $I$-band magnitude of about 18.55 mag and a colour of 4.05 mag. The equivalent $V$ magnitude given in \citet{ortolani96+02} is about 22.60 mag.

\begin{figure}[h]
  \centering
  \includegraphics[width=\hsize]{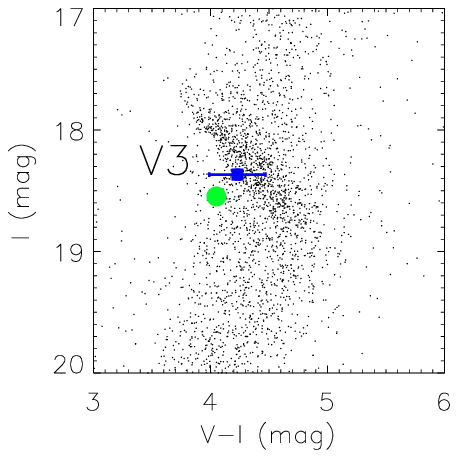}
  \caption{Zoom into the HB given in Fig.~\ref{fig:cmd_terzan5}. The green circle corresponds to the $I$ value given in \citet{ortolani96+02}. The blue square is the median $I$ magnitude obtained in our light curve.}
         \label{fig:zoom_hb}
\end{figure}

The position of this star (green point) is shown more in detail in Fig.~\ref{fig:zoom_hb}. However, the blue square is the position of the star using the mean $I$ magnitude obtained in this work. As the light curve obtained in this analysis covers the star's variation very well, the median $I$ magnitude represents an improvement in its CMD position, which brings it closer to the HB at 18.37 $\pm$ 0.04 mag, a difference of about 0.18 mag compared to the measurement in \citet{ortolani96+02}. The size of the blue square represents the vertical uncertainty.

To obtain an estimate of the uncertainty in the colour, we proceeded as follows. Ortolani's $V$ and $I$ data were taken almost simultaneously, hence their $V$ and $I$ magnitudes were captured roughly at the same phase. Since there is no reliable mean $V$ measurement, we adopted the $V$ difference of 0.18 mag mentioned above as a lower limit to the uncertainty in the colour, adding to this twice the uncertainty associated with our mean $I$ value. Taking into account the dispersion due to differential reddening, the inclination of the HB, and the new position of V3 in the CMD, it is likely that this star is a cluster member.

\begin{figure}[h]
  \centering
  \includegraphics[width=\hsize]{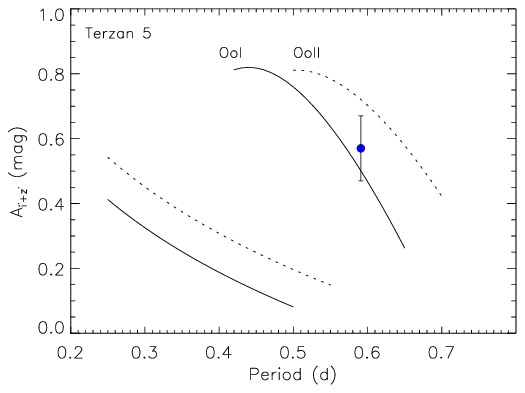}
  \caption{Period-amplitude diagram for RR Lyrae stars. The blue circle is star V3. The continuous line corresponds to Oosterhoff type I (OoI) and the dotted line is Oosterhoff type II (OoII) according to the models in \citet{kunder13+04}.}
         \label{fig:pad}
\end{figure}

If V3 is a cluster member, the period and amplitude can
be used to infer the Oosterhoff type of Terzan 5. In Fig.~\ref{fig:pad}, the
period-amplitude diagram for RR Lyrae stars is shown. Solid lines represent Oosterhoff type I (OoI) locus, as obtained with Equations 1 and 2 from \citet{kunder13+04}. The dashed lines represent the Oosterhoff type II (OoII) ridgeline, as defined by Equations 3-4, which consider $I$ amplitudes and periods. Previous research \citep{Figuera16+mindstep} has shown that $A_{\mathrm{i}^{\prime}+\mathrm{z}^{\prime}}$ amplitudes obtained in this project align well with the analysis obtained with equations in \citet{kunder13+04}, facilitating the proper classification of clusters with known Oosterhoff types. In this figure, the position of RR0 V3 is indicated by a blue circle. We measured the distance from the point to each of the reference loci, noticing that V3 is 0.08 mag away from the OoI locus, but it is approximately 0.16 mag away from the OoII reference line. However, since this analysis was conducted with only one RR Lyrae star, we consider this tentative classification as our initial attempt at determining its Oosterhoff type.

\subsection{Variable V4}

This is the second star previously reported that falls inside the field of view of our observations. It was reported as a possible eclipsing blue straggler with faint sinusoidal variability with a period of about 7 hr. \citet{edmons01+03} consider, in a much less likely scenario, that the star is the infrared counterpart to the low-mass X-ray binary. There is no optical information in this work. We did not find any variable source at the given coordinates and surrounding areas. We checked the position in the finding chart supplied by \cite{edmons01+03} in their Fig.~3, and still no detection was found. Perhaps this star is so faint that it was outside the limits of detection of our detector and the methodology employed in our observations. V4 was not detected either in \citet{ortolani96+02}.

\subsection{EXO 1745-248 / CXOGlb J174805.2-244647 / CX 3}

This is another object detected during the inspection of the difference images, located at the celestial coordinates RA(J2000) $=$17:48:05.213 and Dec(J2000) $=$-24:46:47.57. This target was initially identified as an X-ray burst in a study by \citet{makishima81+19} using data from the \textit{Hakucho} satellite, which observed 14 outbursts between August 5-21, 1980. \citet[][and references therein]{heinke03+05} also examined this target using \textit{Chandra} X-ray observations and noted irregular activity since its discovery. Designated as CX 3, it is classified as a low-mass X-ray binary (LMXB). 

In 2000, \citet{heinke03+05} conducted observations using the \textit{Chandra} X-Ray Observatory and detected an outburst at its position. They attempted to identify an infrared counterpart for this LMXB using \textit{HST} images, identifying two possible star candidates. However, they did not observe any variability, leading to uncertainty in their identification. They suggested that CX 3 may be an ultracompact LMXB, based on comparison with other sources in globular clusters.

Subsequent observations were carried out by \citet{wijnands05+07} in 2003 using the \textit{Chandra} Observatory, which found CX 3 in a quiescent state, as well as a hard spectral component. These findings were consistent with a study by \citet{heinke06+06}.

Lastly, \citet{ferraro15+06} conducted an observational program with \textit{HST} during a CX 3 bursting and made a visual detection at its position in data acquired on April 20, 2015. In their observations, CX 3 increased in brightness by approximately 3 mag. They were able to detect the target in their CMD and concluded it is a sub-giant-branch (SGB) star in a quiescent state, transitioning to the RGB. They proposed that CX 3 could be a peculiar object in the process of forming a radio MSP.

\begin{figure}[h]
  \centering
  \includegraphics[width=\hsize]{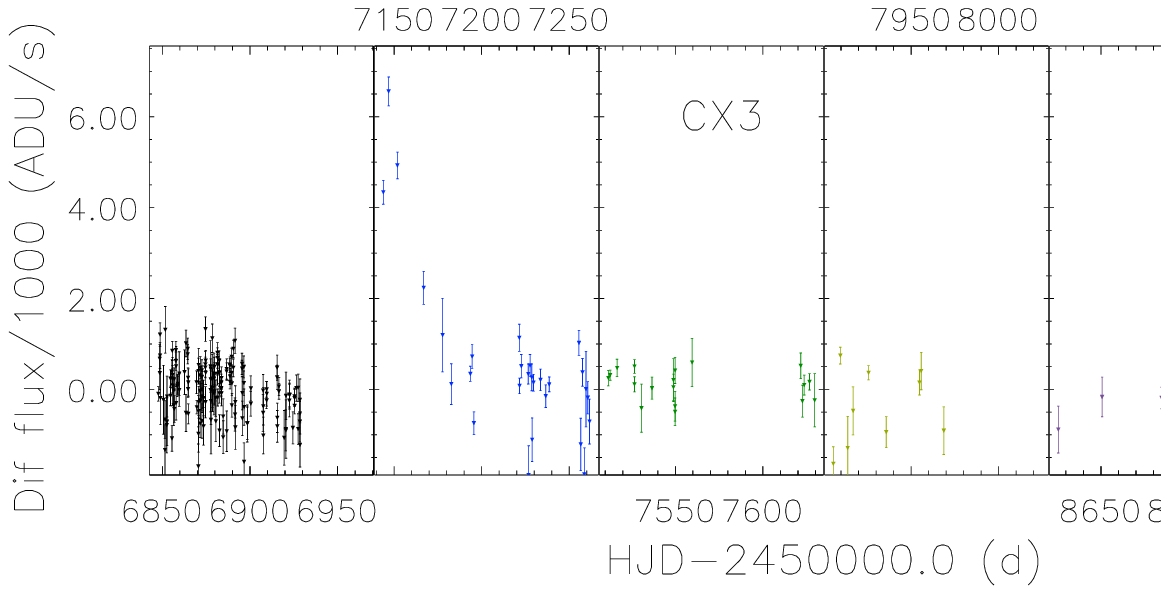}
  \caption{Variability detection at the position of CX 3. The colours of the plots match those in Fig.~\ref{fig:histo_obs}.}
         \label{fig:cx3}
\end{figure}

These observations align with our own visual counterpart detection shown in Fig.~\ref{fig:cx3} during the year 2015. We recorded a signal increase on May 01 (HJD 2457143.871736619), reaching its maximum on May 04 (HJD 2457146.899496661), and then decreasing until returning to its baseline around June 09 (HJD 2457182.670915530).

\section{New Discoveries}\label{sec:new_var}

In this section, the new variable sources detected and discovered in the field of Terzan 5 covered by our observations and the methodology employed are presented.

\subsection{The Unexpected Outburst (N1)}\label{sub:N1}

Inspection of the difference images (see Fig.~\ref{fig:2446Nzoom}) revealed an outburst (Fig.~\ref{fig:2446N}) that took place on HJD 2457177.682793049, located at position RA(J2000) $=$ 17:48:04.914 and Dec(J2000) $=$ $-$24:46:53.15. However, upon exploration of the reference frame at the given coordinates, no stars were found.

In an effort to better understand the origin of this signal, a review of the literature revealed that the closest reported object (separated by $0\farcs65$) corresponds to target J1748-2446N, also known as Ter5 N and Ter5-VLA27.

This nearby source, J1748-2446N, which could be associated with our visual detection, is an MSP that has been detected and analyzed in several radio observations. \citet{urquhart20+14} reported this target at position RA(J2000) $=$ 17:48:04.914 ($\pm 0\farcs05$) and Dec(J2000) $=$ -24:46:53.75 ($\pm0\farcs10$), using the Karl G. Jansky Very Large Array (resolution 0$\farcs$2 - 0$\farcs$04\footnote{\url{https://public.nrao.edu/telescopes/vla/}}) in the USA, with observations conducted in the frequency interval 2--8 GHz. \citet{prager17+06} reported coordinates of RA(J2000) $=$ 17:48:04.919 and Dec(J2000) $=$ -24:46:53.78 using the Green Bank Telescope. \cite{martsen+08} also detected this target using the same facility.

This MSP also has an X-ray counterpart detection \citep{bogdanov21+06, zhao22+01} at reported coordinates RA(J2000) $=$ 17:48:04.906 and Dec(J2000) $=$ -24:46:54.00 with an astrometric offset between the radio MSP position and the nearest X-ray source of $\Delta\alpha=0\farcs20$ and $\Delta\delta=0\farcs18$, respectively. The discovery paper \citep{ransom05+06} of this pulsar, was conducted using observations with the Green Bank Telescope, reporting properties for this target, such as a period of 8.66690 ms, an orbital period of 0.3855 d, and an eccentricity of 0.000045.

It is worth noting a slight difference between the coordinates reported in radio and the X-ray position. One might attribute this discrepancy to the pointing accuracy of the instruments. However, as noted in \cite{bogdanov21+06}, the uncertainties are often related to the X-ray sources while the radio MSP positions are determined with very high accuracy (i.e., to better than $0\farcs1$).

Similar offsets to that seen in our study have also been documented in other targets in the literature. \cite{bogdanov21+06} reported a position offset ($0\farcs55$) between the timing position and the X-ray position in Ter5 A, suggesting that Ter5 A is an eclipsing redback with timing difficulties. In their analysis, they also found similar issues with the position of other redback targets associated with timing, such as NGC 6397 A and M28 I, with offsets by $0\farcs811$ and $2^{\arcsec}$, respectively. Given that Ter 5 N has not exhibited eclipses, it does not appear to be a redback. Additionally, considering the precise astrometry results obtained in our analysis, the visual detection shown in Fig. \ref{fig:2446N} is very unlikely to be associated with the MPS.

To the best of our knowledge, the detection of a visual counterpart to this object has not been previously reported in the literature.

\begin{figure}[h]
  \centering
  \includegraphics[width=\hsize]{images/2446N_charts.eps}
  \caption{Zoom in to the signal found at the position of the visual detection N1. The top boxes show the last image taken before the detection (left), the one with the detection (middle), and the image taken after the registered detection (right) while the bottom boxes correspond to the difference images obtained from subtracting the reference frame from each of the science images shown at the top. Each box is about $6{\arcsec}\times6{\arcsec}$ in size, and the colour scale is logarithmic.}
         \label{fig:2446Nzoom}
\end{figure}

\begin{figure}[h]
  \centering
  \includegraphics[width=\hsize]{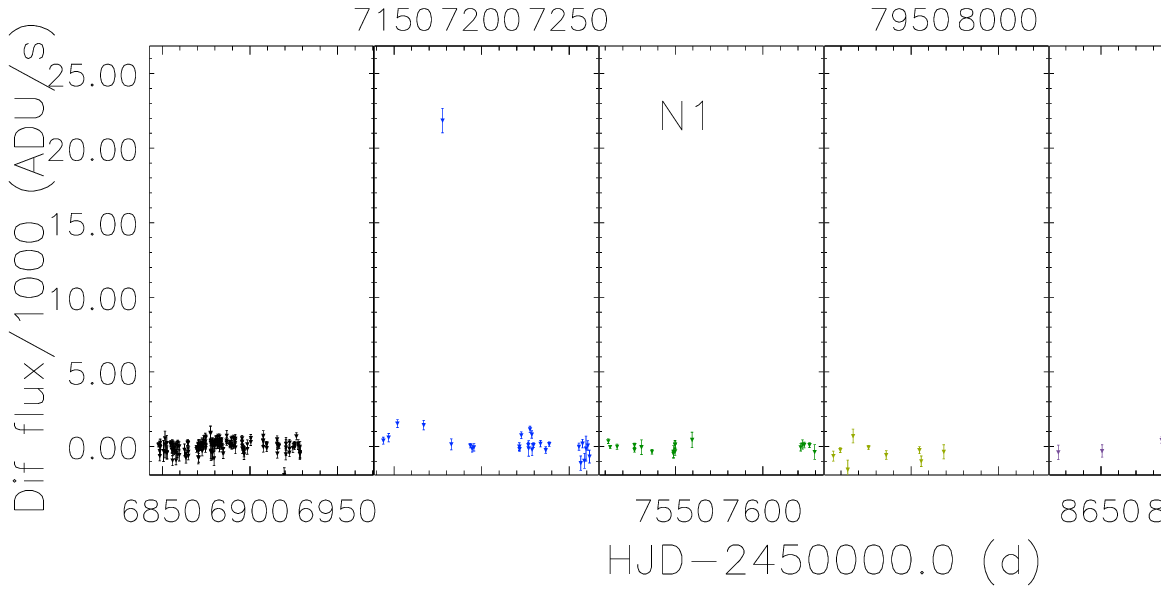}
  \caption{Light curve of source N1, located in the vicinity of PSR J1748-2446N and possibly associated with it (see Sect.~\ref{sub:N1}). The colours of the plots match those in Fig.~\ref{fig:histo_obs}.}
         \label{fig:2446N}
\end{figure}

\subsection{Semiregular variables}\label{subsec:SR_var}

(V14-V17): Several of the brightest stars had high RMS values and were also highlighted by the $\cal S_B$ index. This, combined with the difference images, helped us to confirm that these four stars indeed varied. The resulting light curve for each of the stars was analyzed to explore periodicity in their variation. The periods found range from about 46 to 193 days, with amplitudes from about 0.14 to 0.28 mag. These stars are located along the red giant branch in Fig.~\ref{fig:cmd_terzan5}, and have thus been classified as long-period semiregular variables. Their light curves are presented in Fig.~\ref{fig:phased_SR} and the period estimates are listed in Table~\ref{table:ephemerides_var}. A comparison between the frequencies found by least squares and fast $\chi^2$ methods is provided in Table~\ref{tab:sr_freq}, which are also highlighted in Fig.~\ref{fig:periodogram}.

\begin{figure*}
\centering
\subfloat[][]{
  \begin{tabular}{cc}
    \includegraphics[width=0.474\hsize]{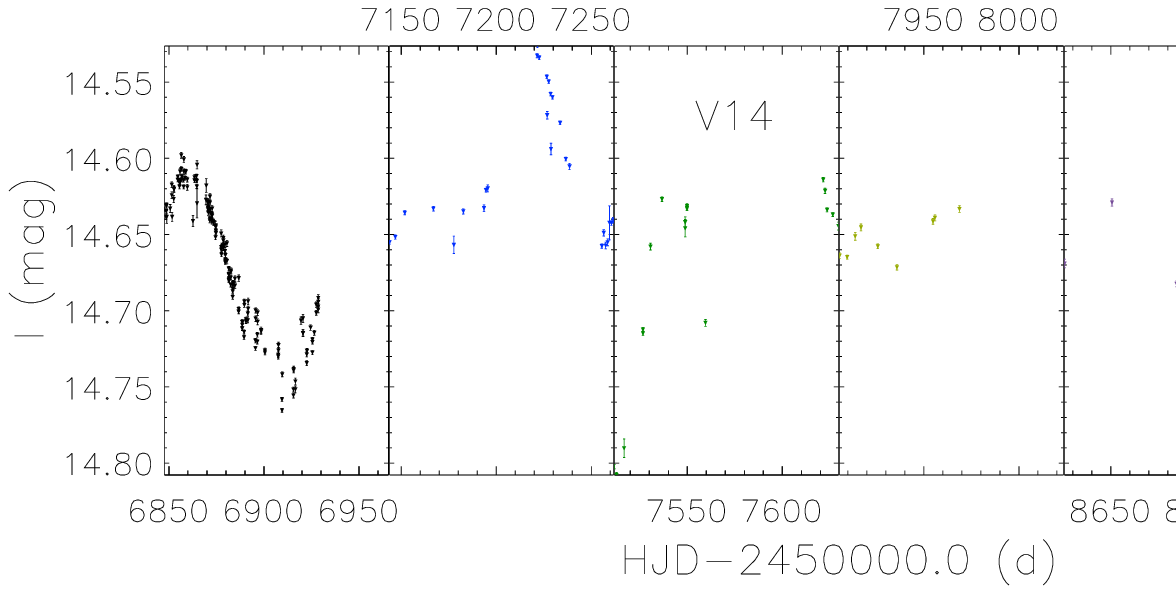} &
    \includegraphics[width=0.474\hsize]{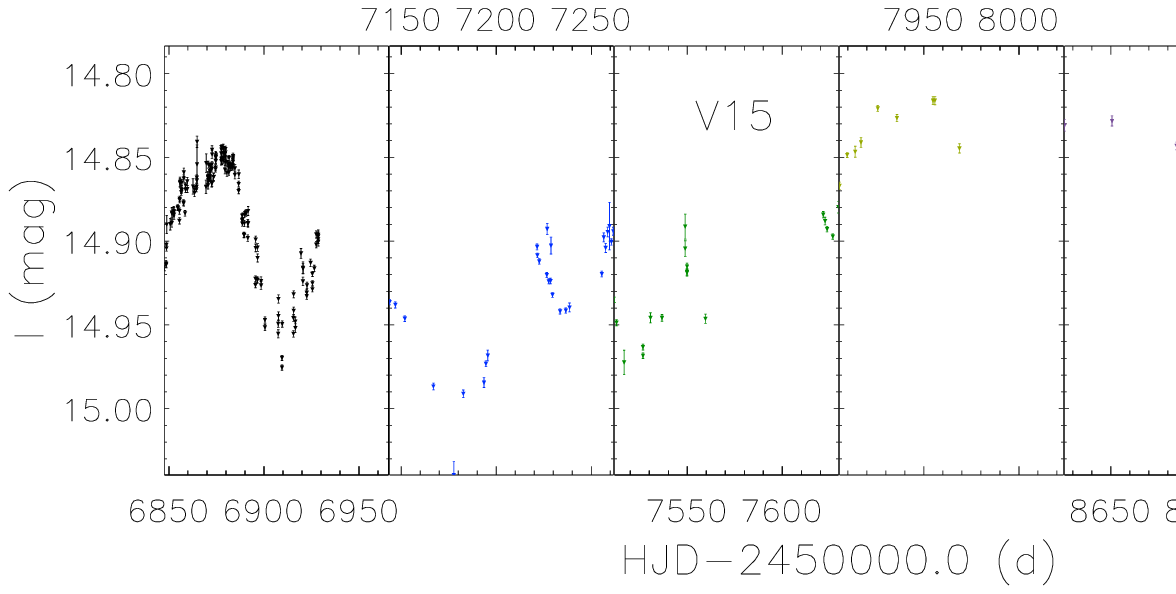} \\
    \includegraphics[width=0.474\hsize]{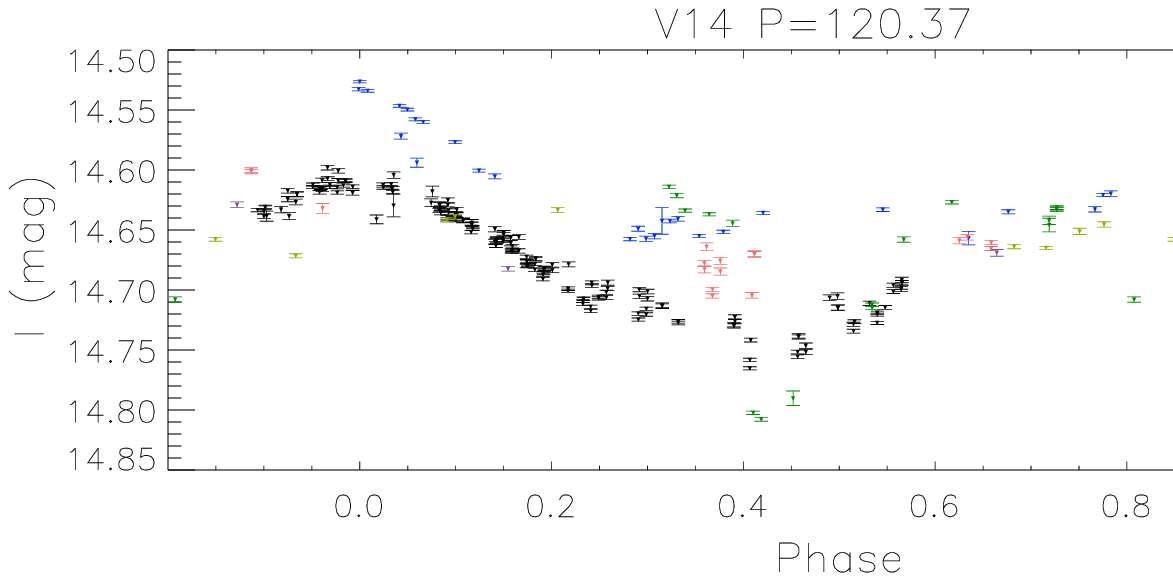} &
    \includegraphics[width=0.474\hsize]{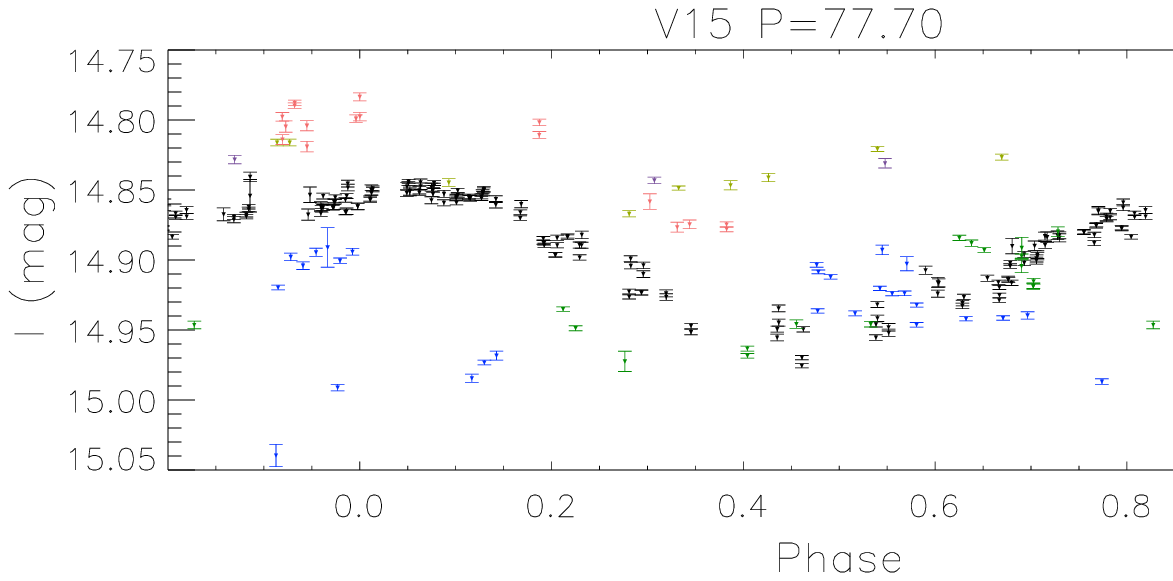} \\
     & \\
     & \\
    \includegraphics[width=0.474\hsize]{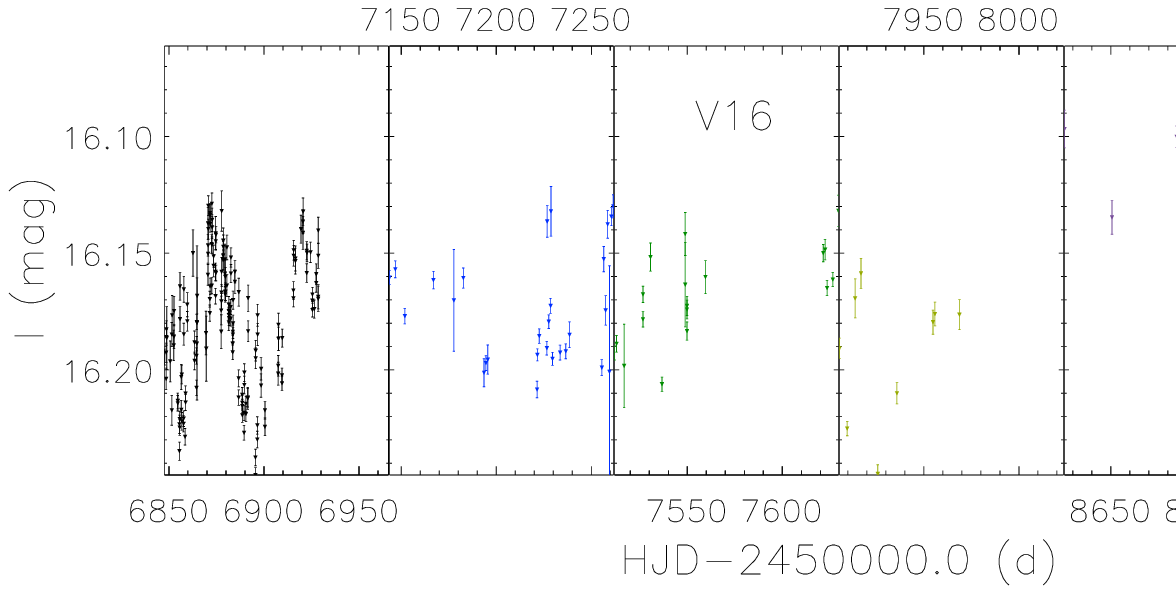} &
    \includegraphics[width=0.474\hsize]{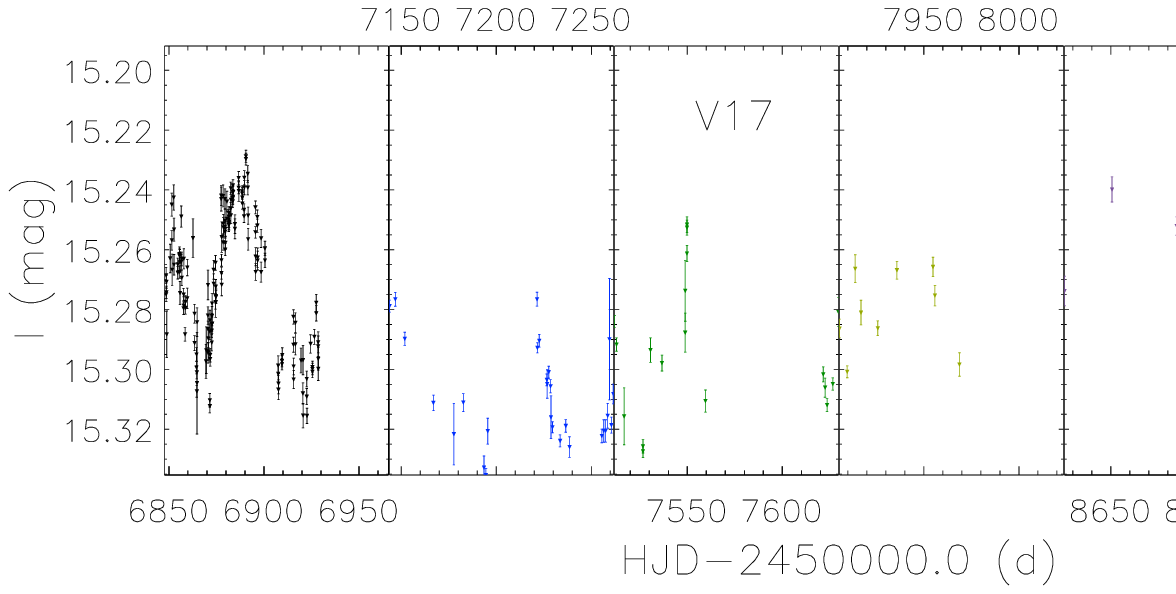} \\
    \includegraphics[width=0.474\hsize]{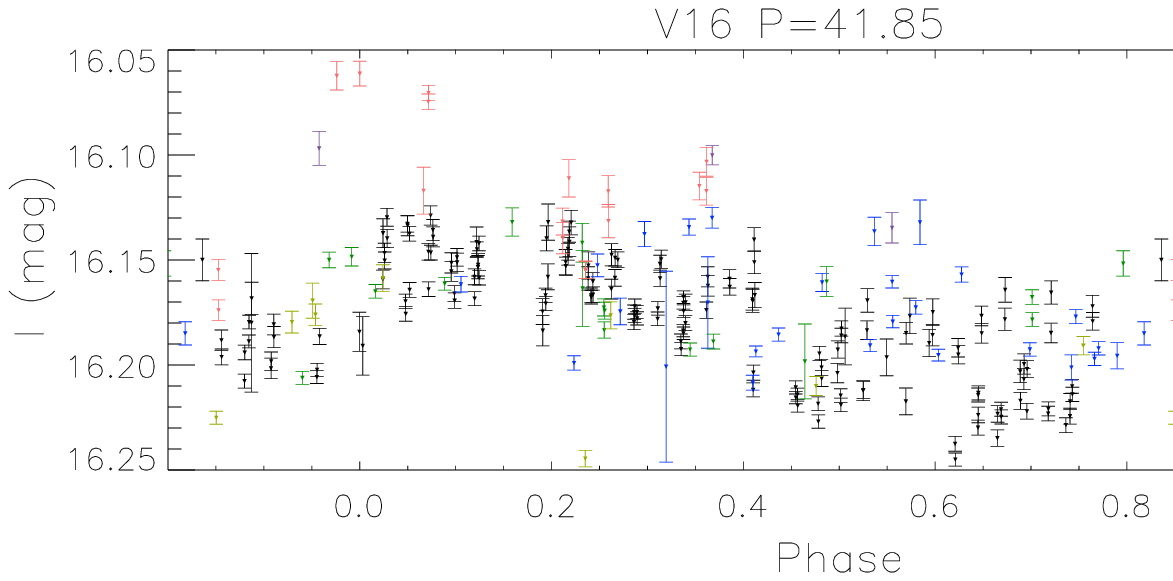} &
    \includegraphics[width=0.474\hsize]{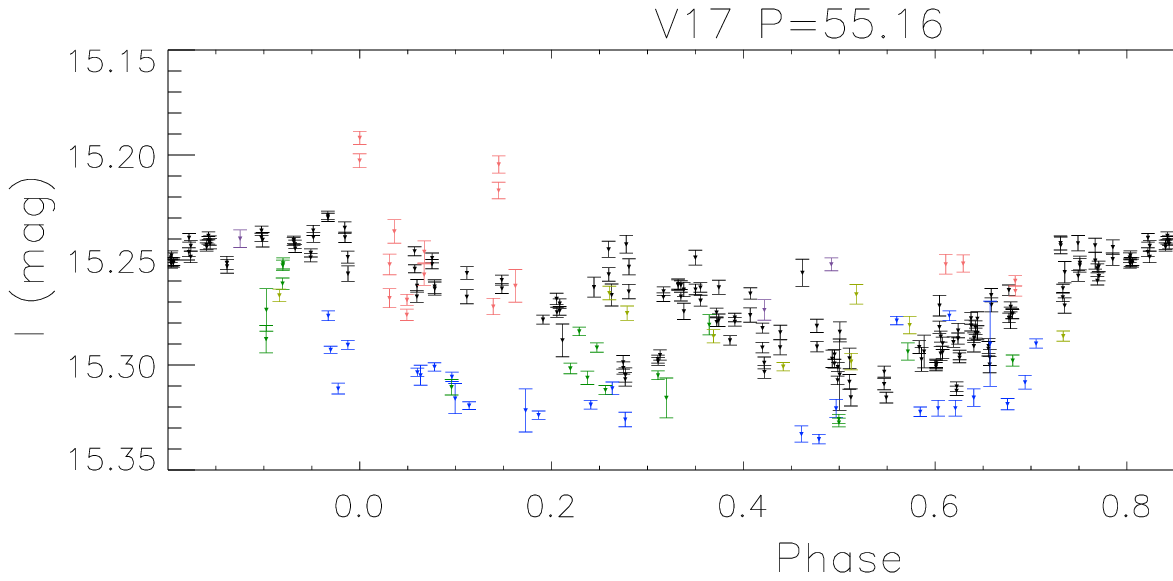} \\
    & \\
    & \\
    \includegraphics[width=0.474\hsize]{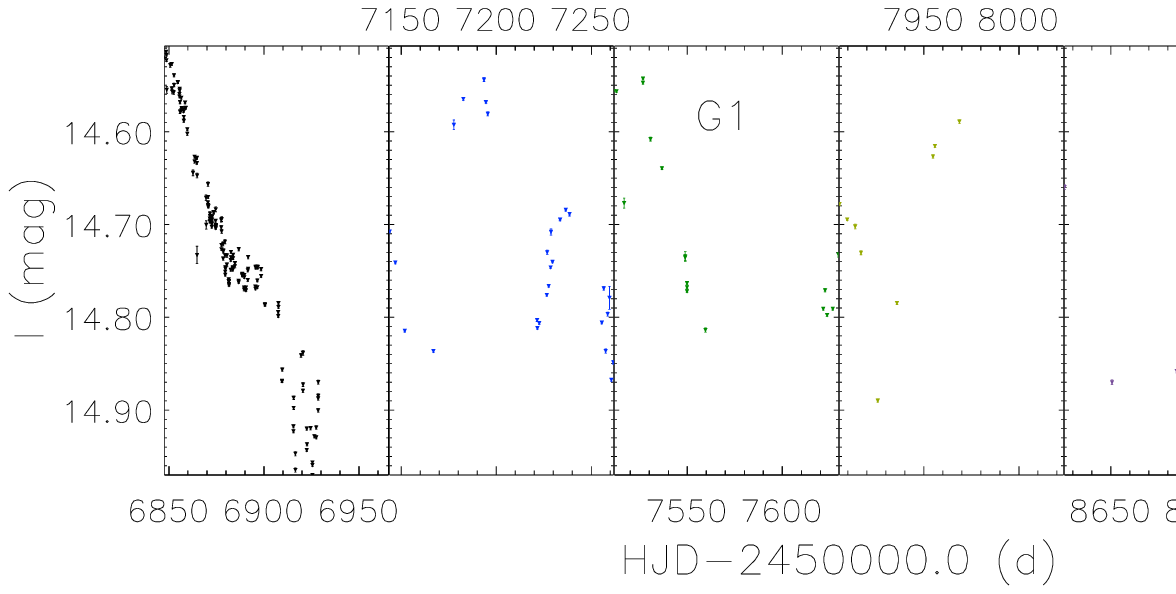} &
    \includegraphics[width=0.474\hsize]{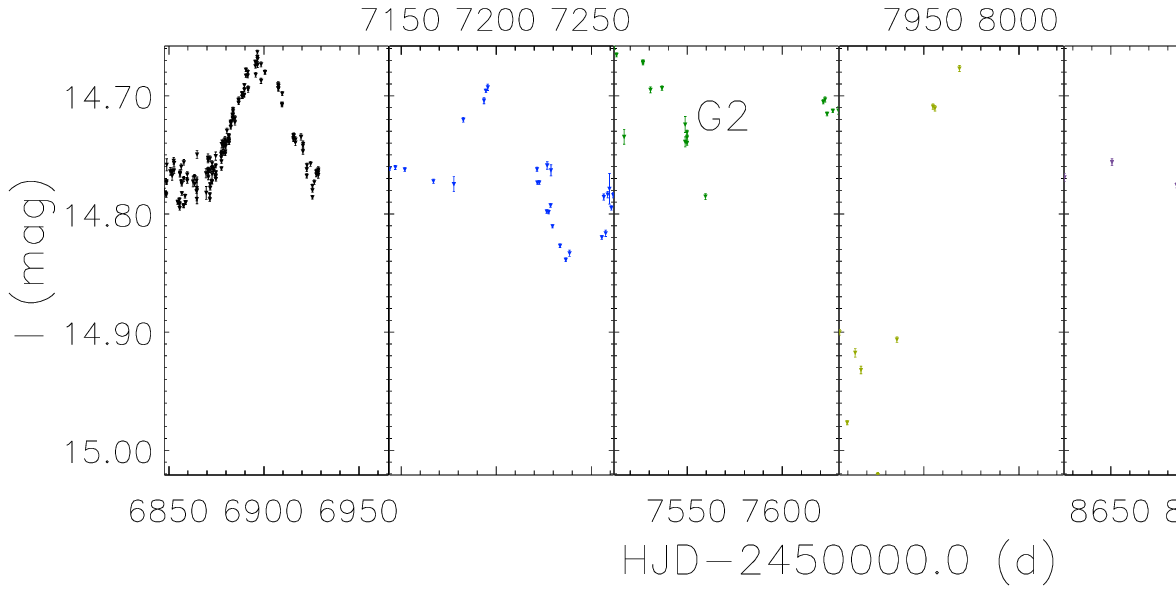} \\
    \includegraphics[width=0.474\hsize]{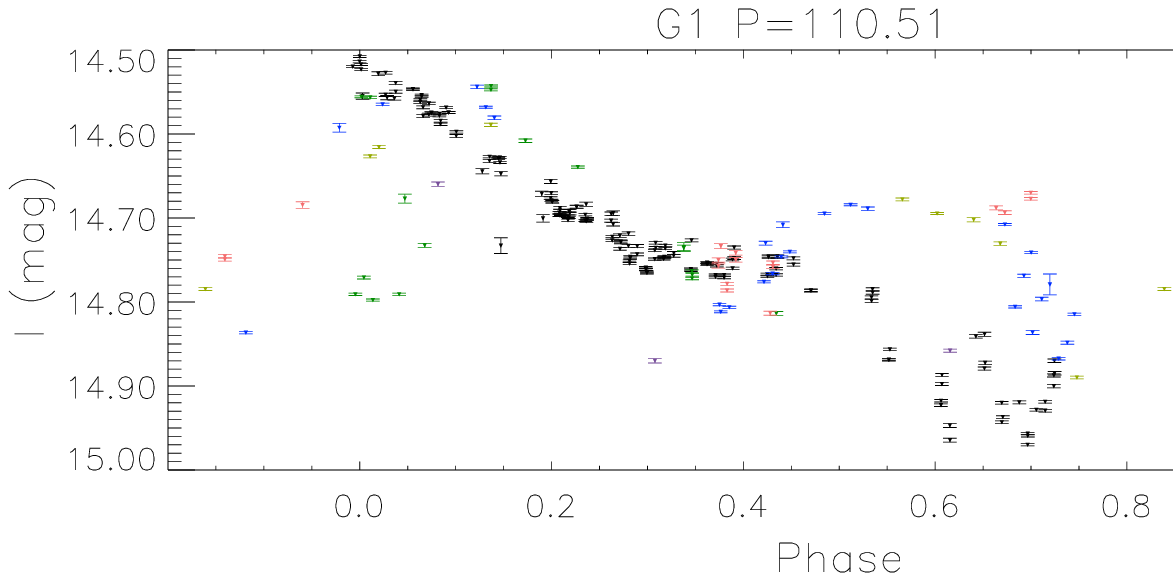} &
    \includegraphics[width=0.474\hsize]{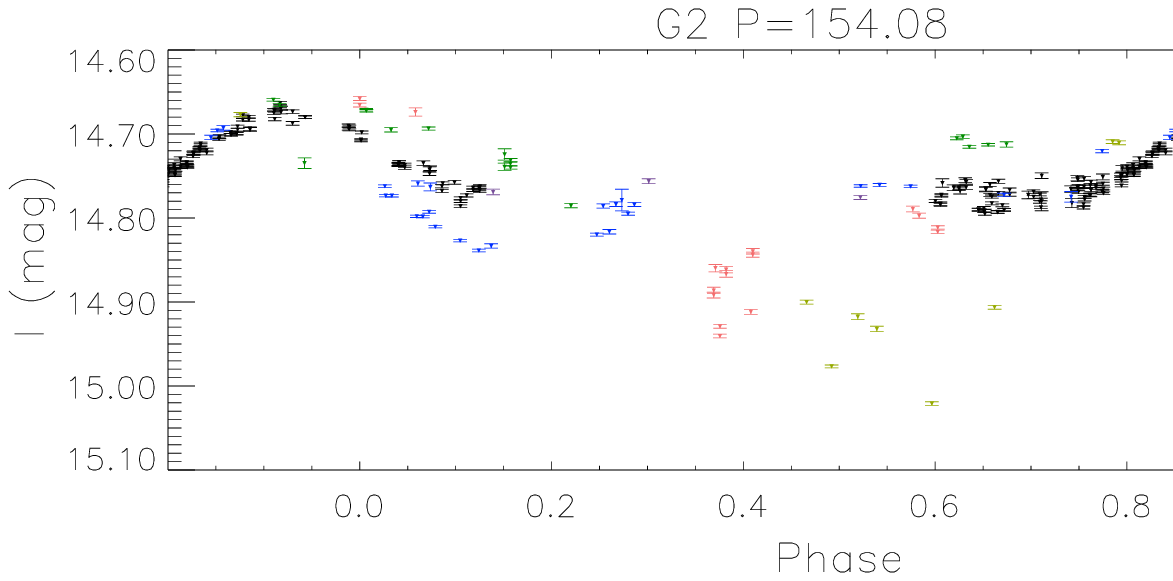} \\
  \end{tabular}
}
\caption{Light curves of newly discovered variables V14-V17 in the globular cluster Terzan 5, as well as variables classified as semiregular. The top plot shows the $I$-band light curves with time in units of HJD, while the bottom plot displays the phased light curves with periods and epochs listed in Table~\ref{table:ephemerides_var}. The colours of the plots match those in Fig.~\ref{fig:histo_obs}.}
\label{fig:phased_SR}
\end{figure*}

\begin{figure*}
\ContinuedFloat
\centering
\subfloat[][]{
  \begin{tabular}{cc}
    \includegraphics[width=0.474\hsize]{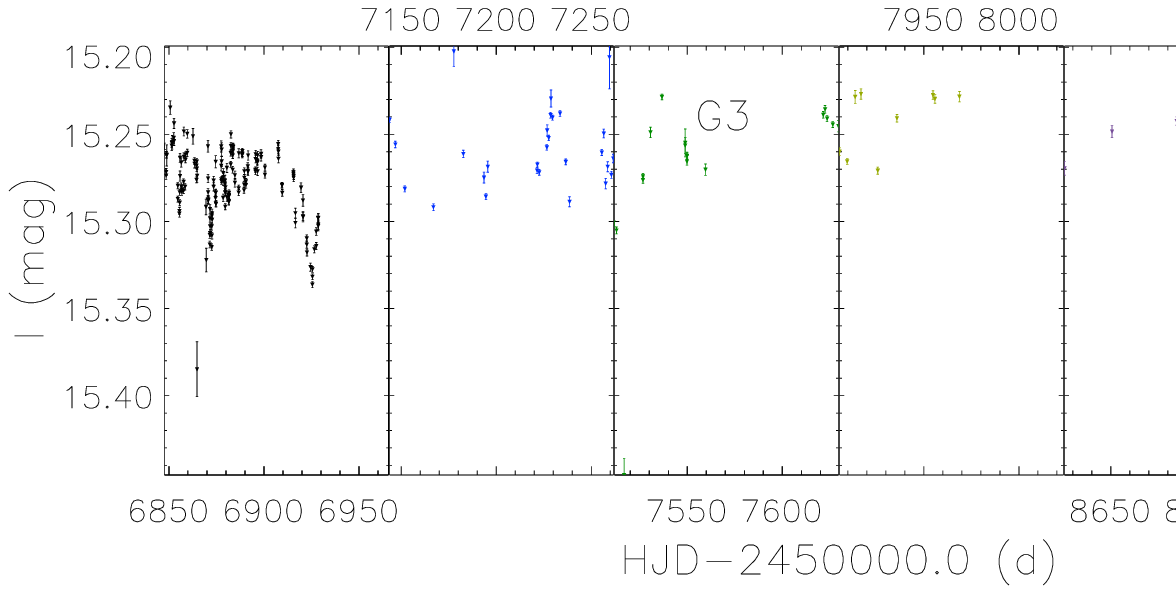} &
    \includegraphics[width=0.474\hsize]{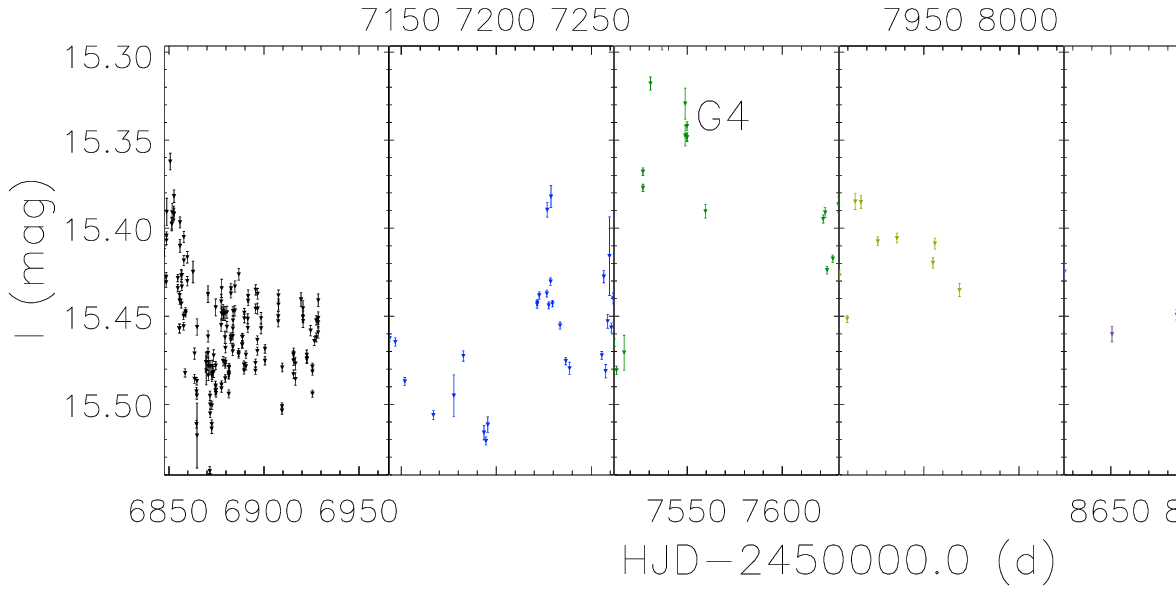} \\
    \includegraphics[width=0.474\hsize]{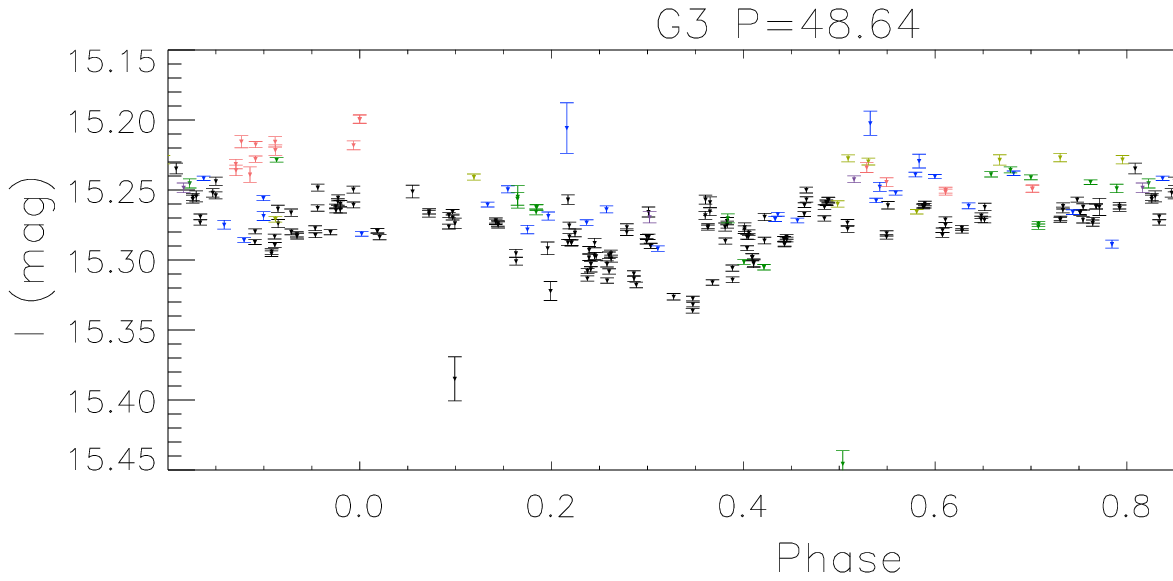} &
    \includegraphics[width=0.474\hsize]{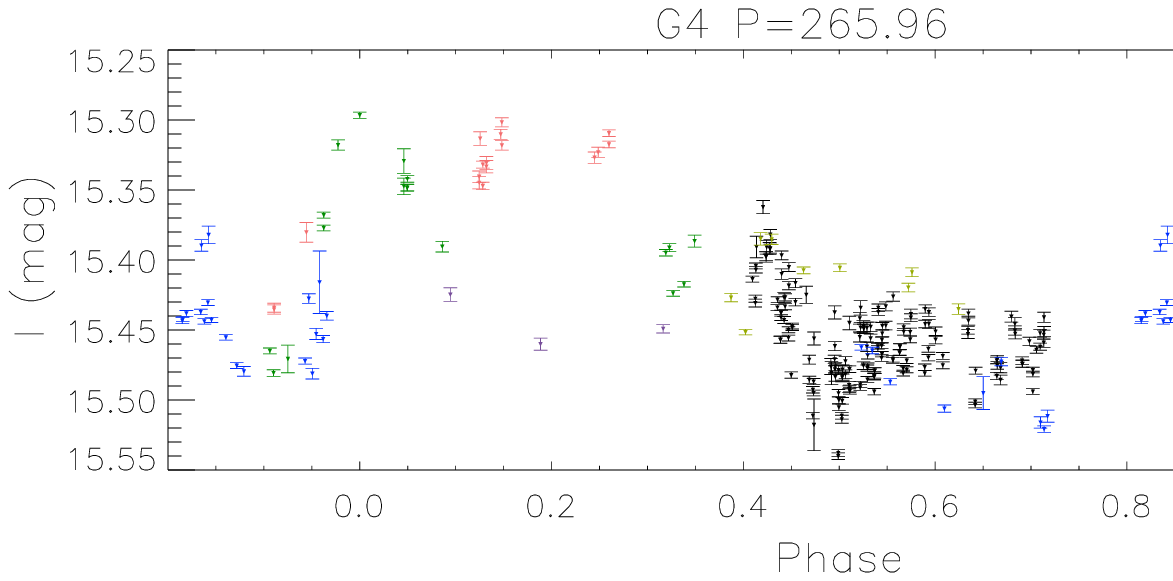} \\
    & \\
    & \\  
    \includegraphics[width=0.474\hsize]{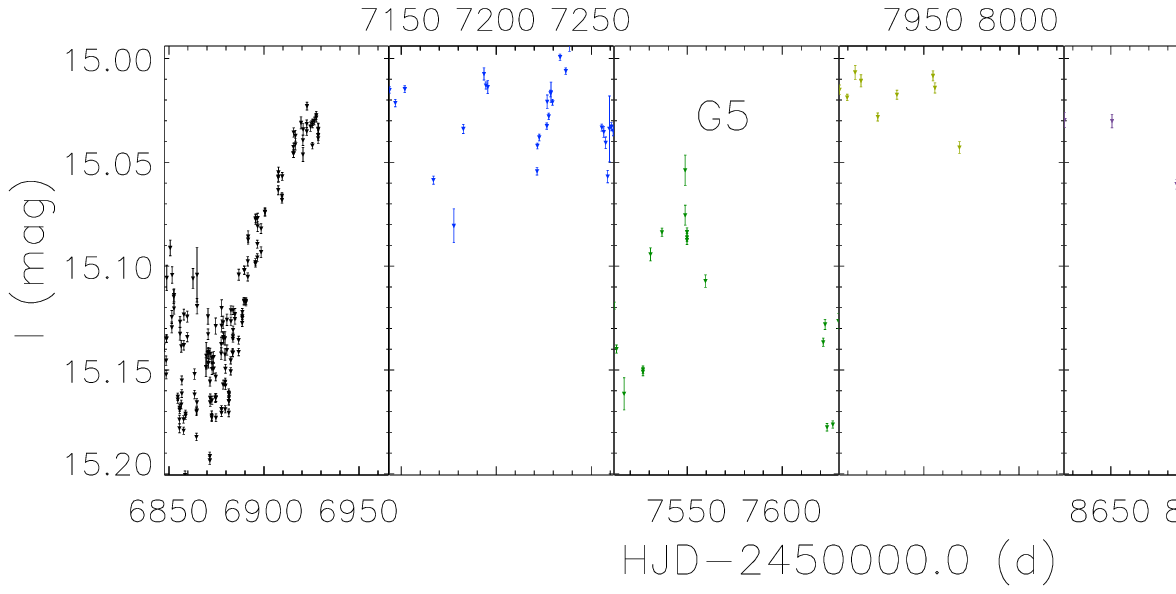} &
    \includegraphics[width=0.474\hsize]{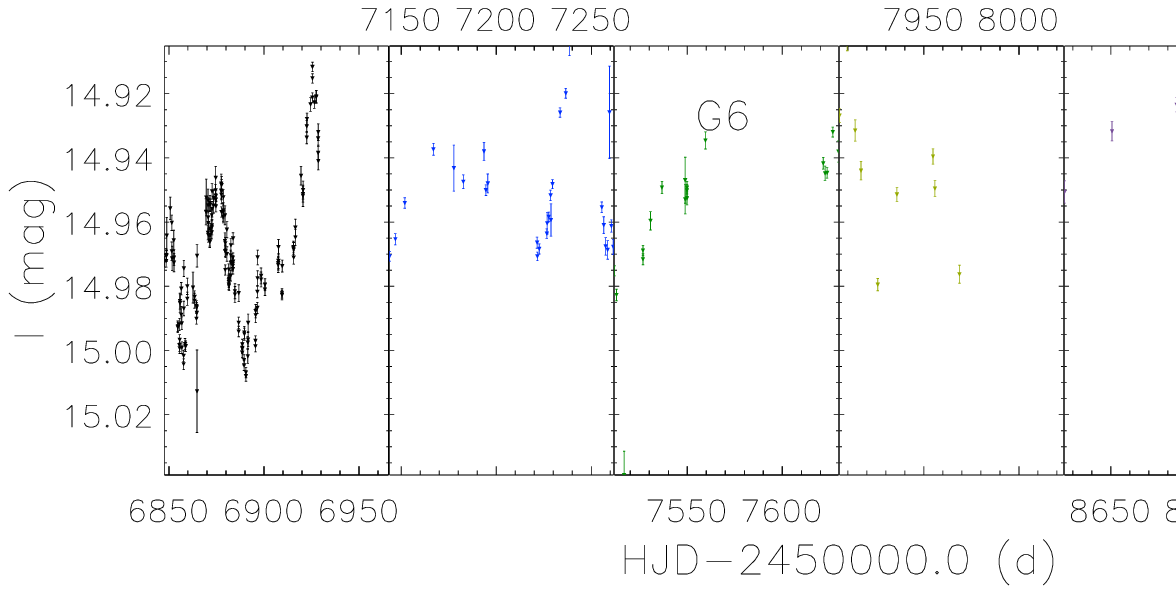} \\
    \includegraphics[width=0.474\hsize]{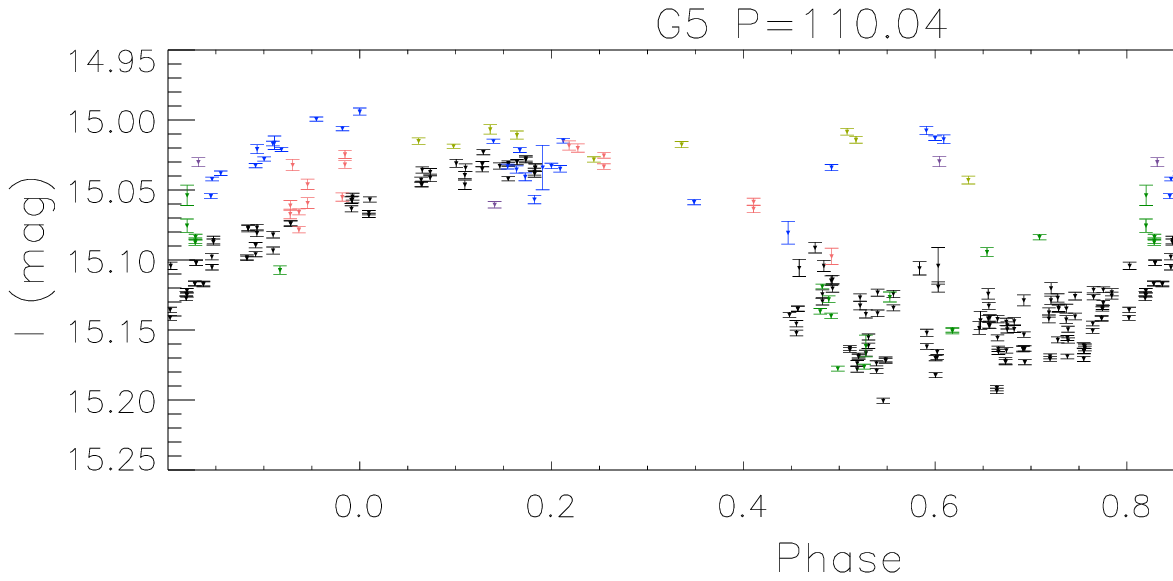} &
    \includegraphics[width=0.474\hsize]{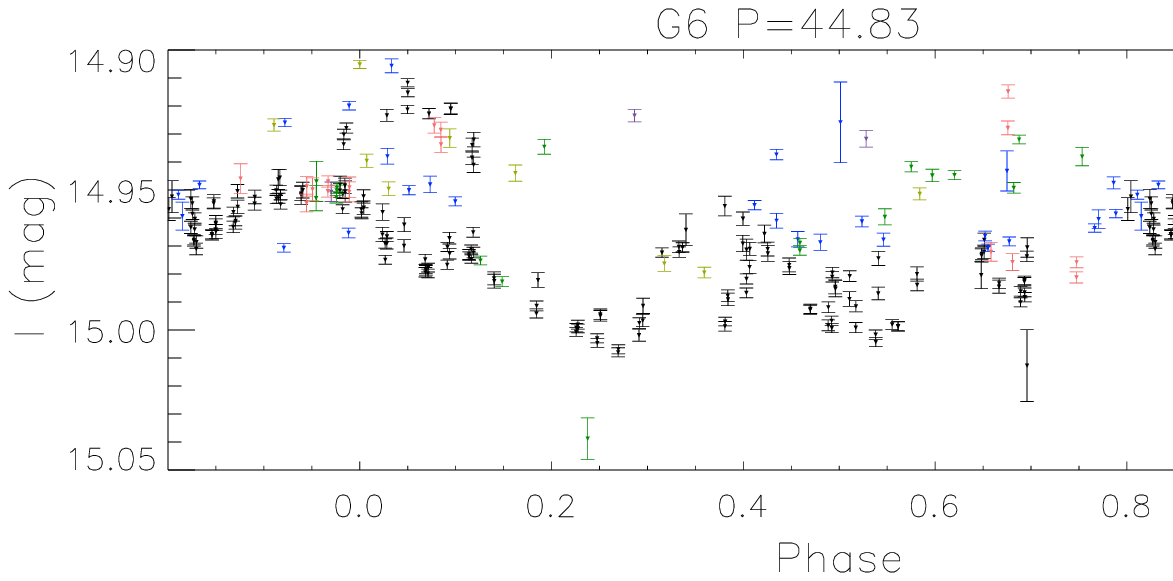} \\
    & \\
    & \\
    \includegraphics[width=0.474\hsize]{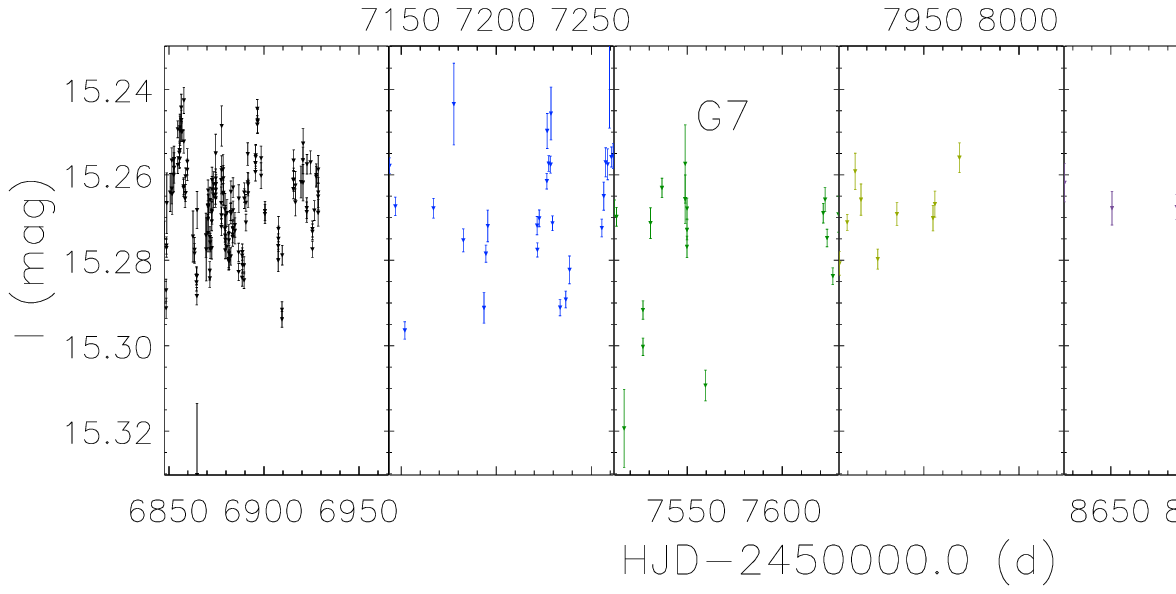} &
    \includegraphics[width=0.474\hsize]{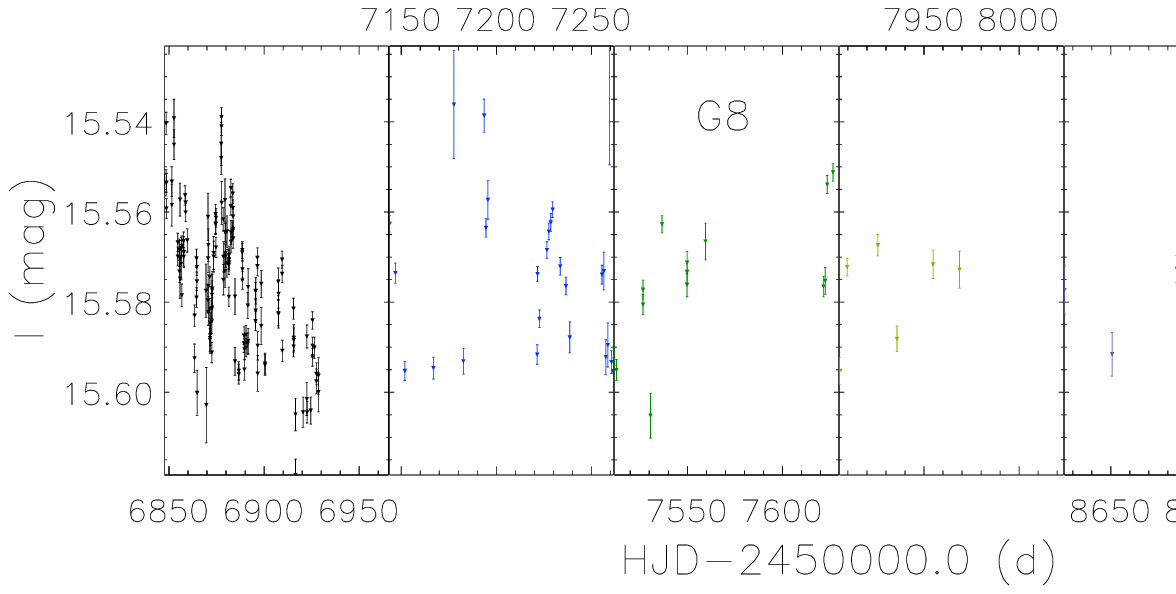} \\
    \includegraphics[width=0.474\hsize]{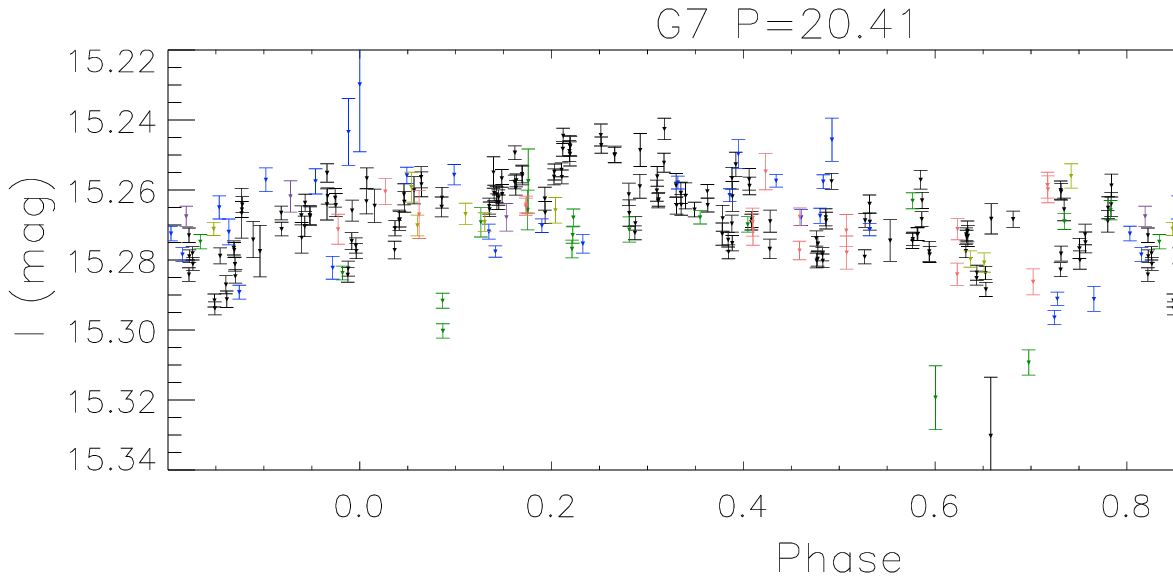} &
    \includegraphics[width=0.474\hsize]{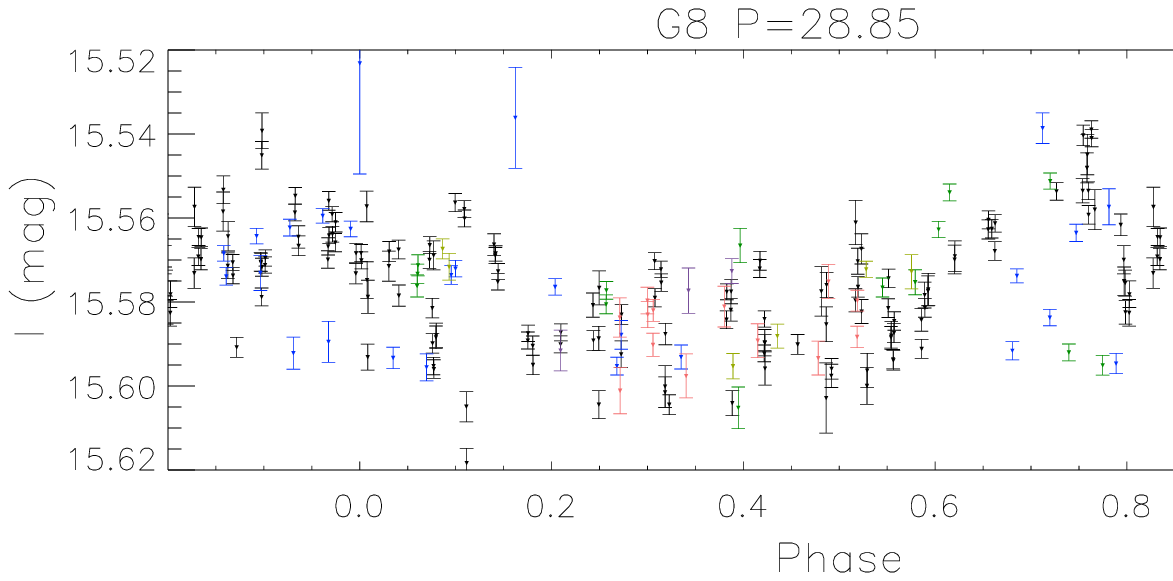} \\
  \end{tabular}
}
\caption{continued.}
\end{figure*}

\begin{figure*}
\ContinuedFloat
\centering
\subfloat[][]{
  \begin{tabular}{c}
    \includegraphics[width=0.474\hsize]{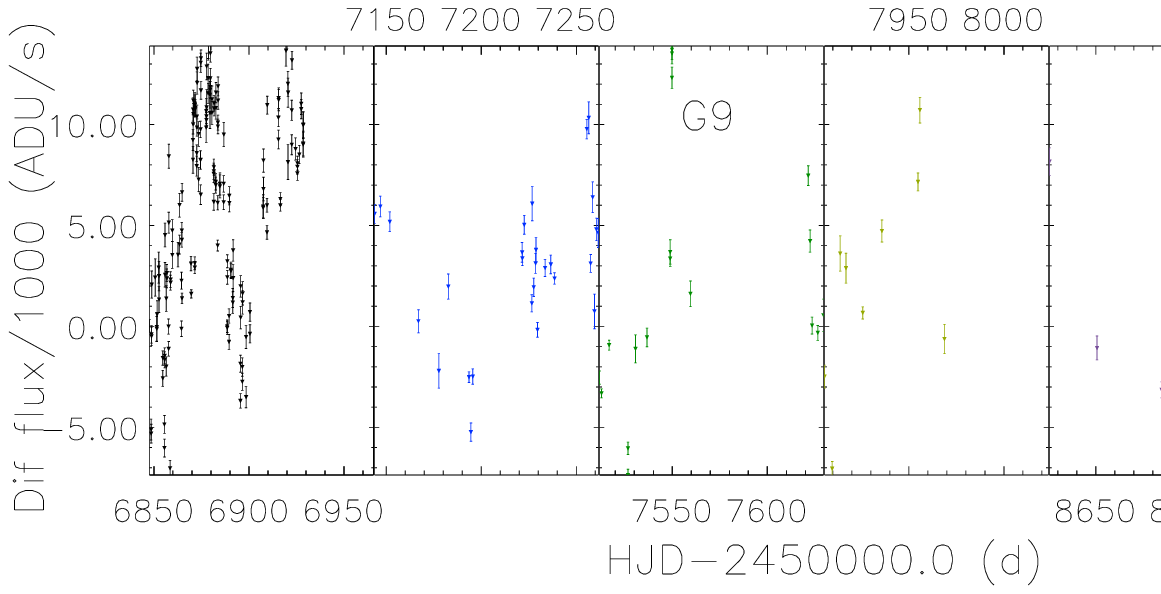} \\
    \includegraphics[width=0.474\hsize]{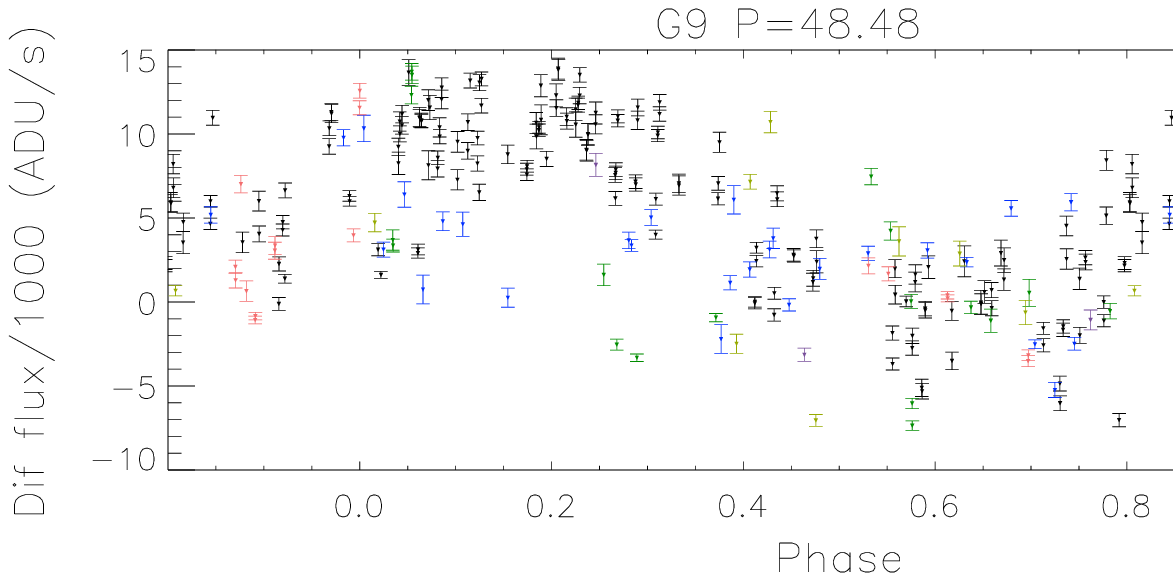} \\
  \end{tabular}
}
\caption{continued.}
\end{figure*}

\subsection{Other candidate variables}

\begin{figure*}[h]
  \centering
  \begin{tabular}{cc}
    \includegraphics[width=0.474\hsize]{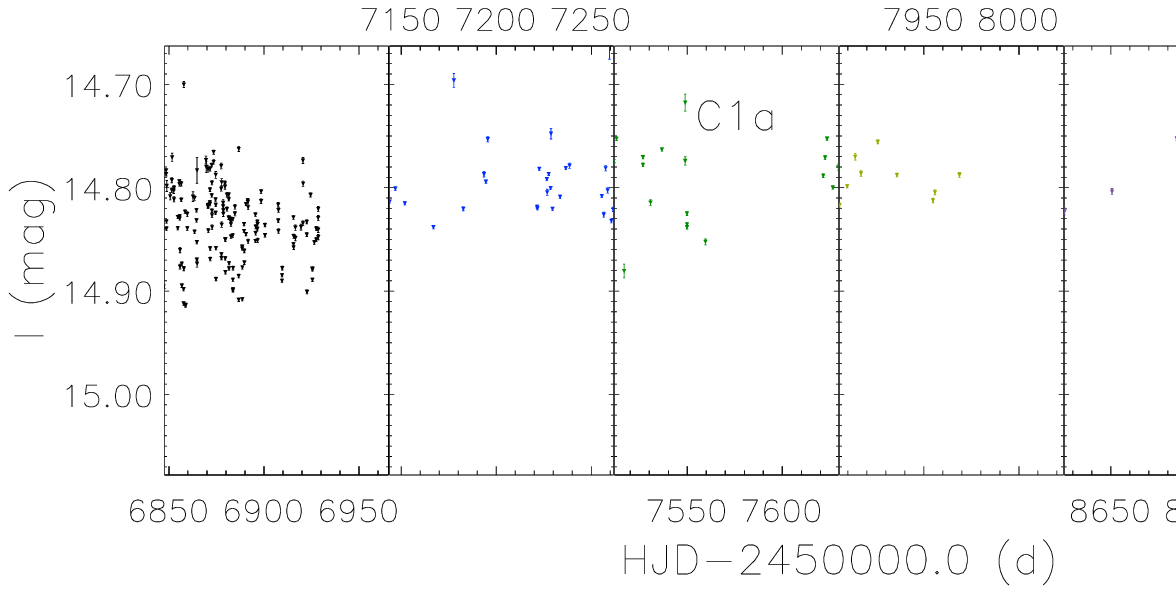} &
    \includegraphics[width=0.474\hsize]{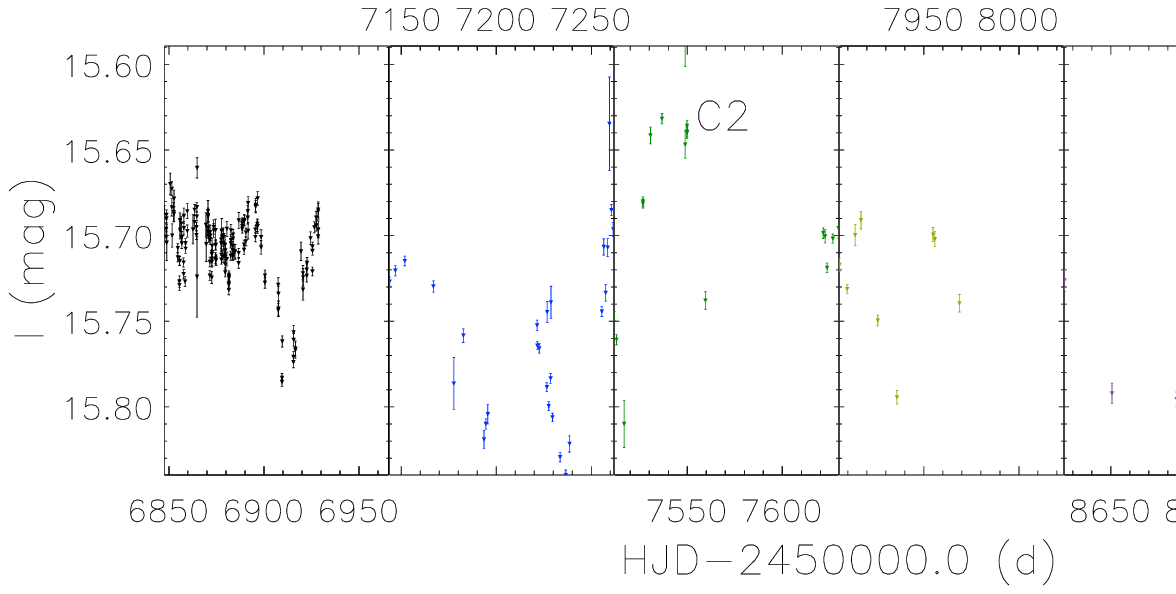} \\
    \includegraphics[width=0.474\hsize]{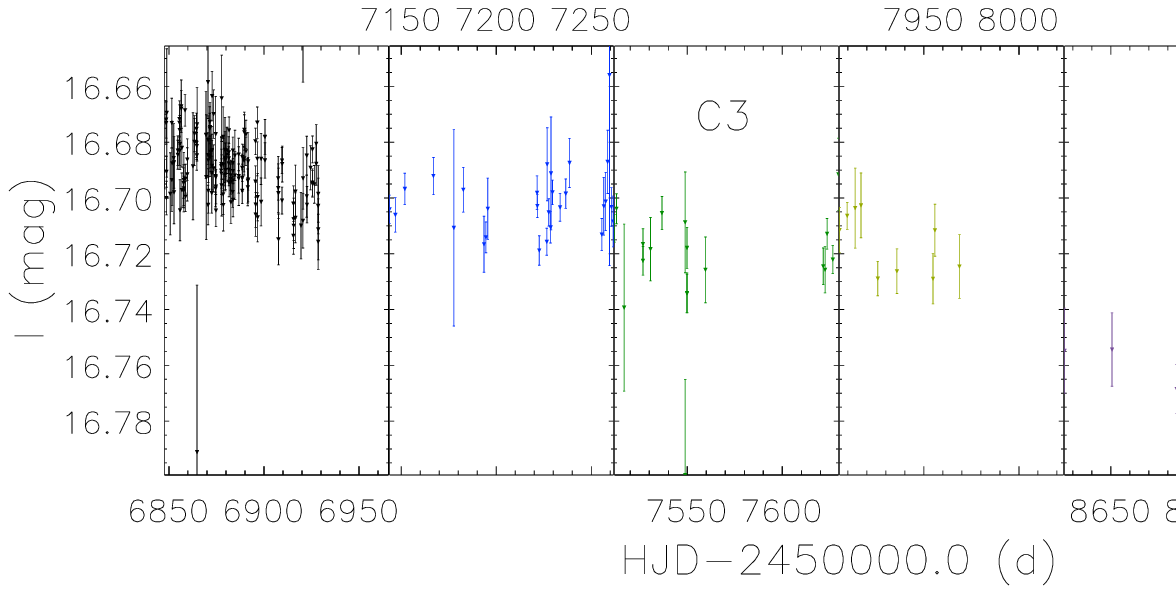} &
    \includegraphics[width=0.474\hsize]{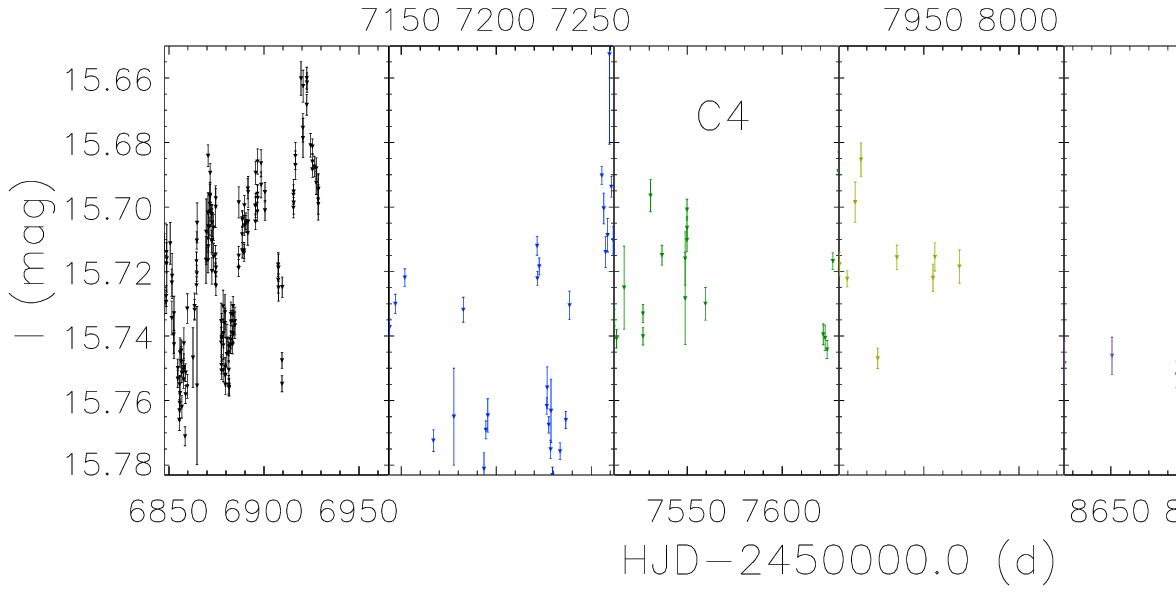} \\
  \end{tabular}
  \caption{Light curves in $I$ magnitude as a function of HJD for candidate variables C1a-C4 . The colours of the plots match those in Fig.~\ref{fig:histo_obs}.}
         \label{fig:candidates}
\end{figure*}

(C1a-C4): Although some stars appeared to exhibit actual variation in their light curve shapes, they were found to be very close to (either above or below) the RMS or $\cal S_B$ index thresholds defined in Sect.~\ref{sec:var_det}. Stars whose photometry was affected by any systematic issue in the difference images or other errors were classified as candidate variables to aid in future research and analysis. These candidate variables are highlighted by orange star-shaped points in Figs.~\ref{fig:rms_sb_terzan5} and \ref{fig:cmd_terzan5}, and are the black square labels in the finding chart presented in Fig.~\ref{fig:finding_chart_terzan5}. Their light curves are plotted in Fig.~\ref{fig:candidates}, and their ephemerides can be found in Table~\ref{table:ephemerides_var}.

Of note among these candidates, C4 corresponds to star 1 listed in \cite{origlia11+08} as a red giant star in Terzan 5. Due to lack of availability of V and I data, it was not possible to plot C1a on the cluster CMD presented in Fig.~\ref{fig:cmd_terzan5}. Additionally, C1a is one of the stars discussed in more detail in Sect.~\ref{sec:motion}.

\subsection{The C1a and C1b case}\label{sec:motion}

By comparing the position of the stars in our reference frame (EMCCD) with their positions in the images used in Sects.~\ref{subsec:astrometry} ({\em HST}) and \ref{sec:cmd} (NTT), we noticed that there is a very strong displacement at the positions of two sources, which we labelled as C1a and C1b in the finding chart given in Fig.~\ref{fig:finding_chart_terzan5}. 

\begin{figure*}[h]
  \centering
    \includegraphics[width=\hsize]{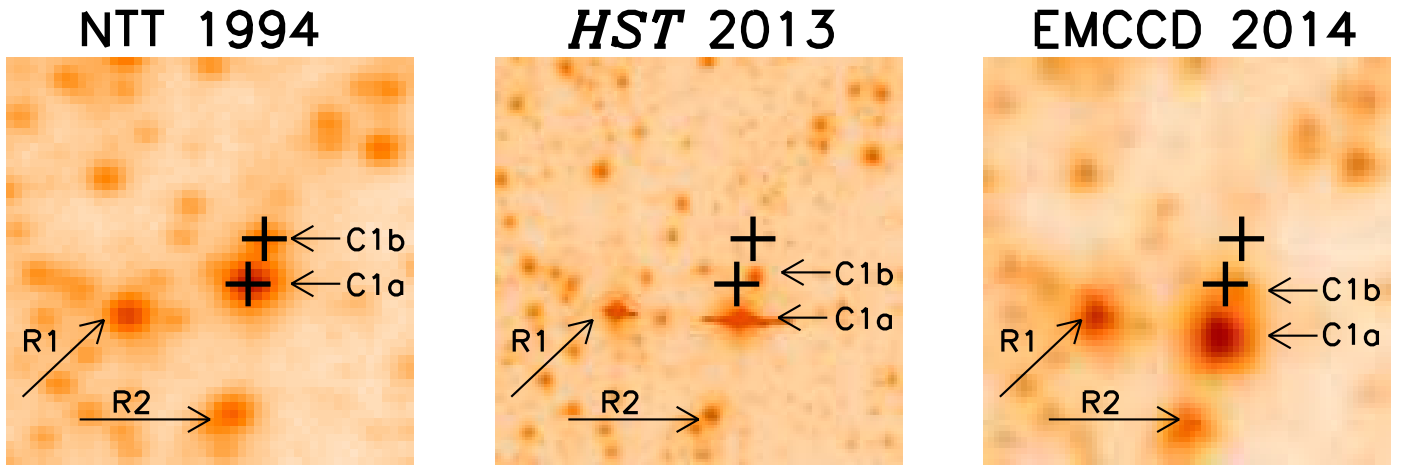}
  \caption{Comparison of the positions of C1a and C1b in a section of the NTT image from 1994 (on the left) with the positions in the {\em HST} image taken in 2013 (in the middle), and the EMCCD image taken in 2014 (on the right). The crosses represent the star's positions measured in the NTT image. The stars labeled R1 and R2 are for comparison. The size of the images is about 7${\arcsec}$. The colour scale is not exactly the same in the three plots. North is up and east is left.}
  \label{fig:shifted_stars}
\end{figure*}

In Fig.~\ref{fig:shifted_stars}, it is possible to visually compare how they changed their positions from the image (taken in 1994) used in \cite{ortolani96+02}, the {\em HST} image in 2013, and the date of the first image used in our reference frame (taken in 2014). From the visual inspection, it is possible to see that both sources seem to be moving in the same direction and with approximately the same displacement, giving the impression that they might be physically associated.

In Table~\ref{tab:shifted_stars}, we register their positions in the three images in which we noticed this displacement: The EMCCD image (2014), the {\em HST} (2013), and the NTT (1994). Considering the position of the centre of Terzan 5 (RA(J200) $=$ 17:48:04.80 and Dec(J2000) $=$ $-$24:46:45) taken from \citet[][2010 version]{harriscatalog} we calculated the angular distance of C1a and C1b from the cluster centre, noticing that in 1994 they were about 18$\farcs$33 and 18$\farcs$23 away respectively, while in 2014 they were about 18$\farcs$45 and 18$\farcs$46 away, respectively from the cluster centre. The displacements registered in the positions of C1a and C1b in the EMCCD image with respect to their position in the NTT image are $\sim0\farcs62$ and $0\farcs59$, respectively.

Finally, the total proper motion calculated for C1a between NTT and {\em HST} images was about 22.9 mas$/$yr, and 28.4 mas$/$yr between NTT and EMCCD images. Similarly, in the case of C1b, the total proper motion was about 23.4 mas$/$yr using the NTT and {\em HST} images, while it was 27.5 mas$/$yr using the NTT and the EMCCD images. This gives us mean values of 25.7 mas$/$yr and  25.5 mas$/$yr for C1a and C1b, respectively, strongly indicating that these two stars are not cluster members.

\begin{table*}
\caption{Celestial coordinates (J2000) and angular distances from the centre of Terzan 5 measured for C1a and C1b discussed in Sect.~\ref{sec:motion} in each of the images taken with the EMCCD, {\em HST}, and NTT at the given dates.}
\label{tab:shifted_stars}     
\centering
\begin{tabular}{c c c c c c}
\hline
\hline
 & \multicolumn{2}{c}{C1a}               & \multicolumn{2}{c}{C1b}      &   Date     \\
\hline
      &  RA(J2000)    &  DEC(J2000)  &  RA(J2000)   & DEC(J2000)   & \\
\hline
EMCCD &  17:48:03.548 & -24:46:52.05 & 17:48:03.527 & -24:46:51.35 & 2014-07-09 \\
HST   &  17:48:03.547 & -24:46:51.91 & 17:48:03.529 & -24:46:51.24 & 2013-08-18 \\
NTT   &  17:48:03.54  & -24:46:51.44 & 17:48:03.53  & -24:46:50.76 & 1994-05-16 \\
\hline
\multicolumn{6}{c}{Angular distance from cluster center} \\
\hline
      &  \multicolumn{2}{c}{Arcsec}          & \multicolumn{2}{c}{Arcsec}          & Date \\
\hline
EMCCD &  \multicolumn{2}{c}{18.45} & \multicolumn{2}{c}{18.46} & 2014-07-09 \\
HST   &  \multicolumn{2}{c}{18.41} & \multicolumn{2}{c}{18.40} & 2013-08-18 \\
NTT   &  \multicolumn{2}{c}{18.33} & \multicolumn{2}{c}{18.23} & 1994-05-16 \\
\hline
    
\end{tabular}
\end{table*}

\begin{table*}

\caption{Ephemerides and main properties of variable stars studied in this work. Column 1 is the ID assigned to each variable. Columns 2-3 are the right ascension and declination. The epoch used in the phased light curves is given in Column 4. The period estimation obtained in this work is given in Column 5 while the numbers in parentheses correspond to the uncertainties in the last decimal place in the reported periods. Column 6 is the median I magnitude. Column 7 corresponds to the peak-to-peak amplitude. The number of epochs is given in Column 8. Column 9 shows the classification of the variable.}            
\label{table:ephemerides_var}     
\centering

{
\begin{tabular}{c c c c c c c c c}
\hline
\hline
Var &   RA  &  Dec  &Epoch & P & $I_{median}$ & $A_{\mathrm{i}^{\prime}+\mathrm{z}^{\prime}}$  & $N$  & Type \\
ID  & J2000 & J2000 & HJD  &  d  &      mag    &  mag  &  &   \\  
(1) &   (2)  &  (3)  & (4) & (5) & (6) & (7) & (8) & (9) \\
\hline
V3 & 17:48:05.127 & -24:46:38.58 & 2456883.6253 & 0.590887(13)$^{a, c}$ & 18.367 & 0.57 & 242 & RR0 \\
N1 & 17:48:04.914 & -24:46:53.15 & --- & --- & --- & --- & 242 & Psr \\
CX3 & 17:48:05.213 & -24:46:47.57 & --- & --- & --- & --- & 242 & LMXB \\
\hline
V14 & 17:48:04.386 & -24:46:41.69 & 2457221.6313 & 120.37(55)$^{c}$ & 14.656 & 0.28 & 242 & SR \\
V15 & 17:48:04.611 & -24:46:42.88 & 2459437.7719 & 77.70(23)$^{b}$ & 14.879 & 0.26 & 242 & SR \\
V16 & 17:48:03.851 & -24:46:39.31 & 2459464.4982 & 41.9(1)$^{c}$ & 16.173 & 0.18 & 242 & SR \\
V17 & 17:48:04.683 & -24:46:42.73 & 2459374.6236 & 55.16(12)$^{c}$ & 15.275 & 0.14 & 242 & SR \\
\hline
G1 & 17:48:05.382 & -24:46:58.43 & 2456848.5165 & 110.51(47)$^{c}$ & 14.735 & 0.46 & 242 & SR \\
G2 & 17:48:04.513 & -24:46:44.28 & 2459374.6236 & 154.08(91)$^{b}$ & 14.759 & 0.36 & 242 & SR \\
G3 & 17:48:05.063 & -24:46:59.96 & 2459437.7804 & 48.64(9)$^{b}$ & 15.270 & 0.25 & 242 & SR \\
G4 & 17:48:05.314 & -24:46:50.98 & 2457536.6769 & 265.96(2.70)$^{b}$ & 15.452 & 0.24 & 242 & SR \\
G5 & 17:48:05.342 & -24:46:46.30 & 2457238.5792 & 110.04(46)$^{c}$ & 15.114 & 0.21 & 242 & SR \\
G6 & 17:48:05.062 & -24:46:48.40 & 2457909.6093 & 44.83(8)$^{c}$ & 14.966 & 0.13 & 242 & SR \\
G7 & 17:48:04.377 & -24:46:46.08 & 2457259.5671 & 20.40(2)$^{c}$ & 15.268 & 0.10 & 242 & SR \\
G8 & 17:48:05.427 & -24:47:02.79 & 2457259.5671 & 28.85(3)$^{c}$ & 15.575 & 0.10 & 212 & SR \\
G9 & 17:48:04.793 & -24:46:41.27 & 2459437.7804 & 48.48(9)$^{c}$ & --- & --- & 242 & SR \\
G10 & 17:48:03.500 & -24:47:03.48 & 2456925.4895 & 0.301770(4)$^{a}$ & 15.618 & 0.11 & 038 & EW? \\
G11 & 17:48:03.576 & -24:46:59.31 & --- & --- & 17.220 & 0.14 & 242 & LPV \\
G12 & 17:48:04.511 & -24:46:55.45 & --- & --- & 16.448 & 0.16 & 242 & LPV \\
G13 & 17:48:05.787 & -24:46:40.65 & --- & --- & 16.931 & 0.07 & 242 & LPV \\
G14 & 17:48:04.951 & -24:46:36.99 & --- & --- & 15.717 & 0.08 & 242 & LPV \\
G15 & 17:48:03.946 & -24:46:39.33 & --- & --- & 15.113 & 0.17 & 242 & LPV \\
G16 & 17:48:04.098 & -24:46:42.69 & --- & --- & 15.250 & 0.10 & 242 & LPV \\
G17 & 17:48:03.846 & -24:46:42.35 & --- & --- & 15.424 & 0.08 & 242 & LPV \\
G18 & 17:48:06.105 & -24:46:49.59 & --- & --- & 15.480 & 0.11 & 242 & LPV \\
G19 & 17:48:04.053 & -24:47:02.63 & --- & --- & 15.492 & 0.16 & 188 & LPV \\
G20 & 17:48:05.299 & -24:46:42.07 & --- & --- & 15.637 & 0.09 & 242 & LPV \\
G21 & 17:48:05.849 & -24:46:45.79 & --- & --- & 15.658 & 0.08 & 242 & LPV \\
G22 & 17:48:05.087 & -24:46:43.61 & --- & --- & 15.683 & 0.10 & 242 & LPV \\
G23 & 17:48:05.513 & -24:46:53.70 & --- & --- & 15.734 & 0.16 & 242 & LPV \\
G24 & 17:48:06.104 & -24:46:54.51 & --- & --- & 15.742 & 0.20 & 242 & LPV \\
G25 & 17:48:03.718 & -24:46:47.03 & --- & --- & 15.779 & 0.16 & 242 & LPV \\
G26 & 17:48:04.341 & -24:46:26.90 & --- & --- & 15.804 & 0.22 & 242 & LPV \\
G27 & 17:48:04.323 & -24:46:37.25 & --- & --- & 15.808 & 0.08 & 242 & LPV \\
G28 & 17:48:04.230 & -24:46:57.25 & --- & --- & 15.918 & 0.20 & 242 & LPV \\
G29 & 17:48:04.752 & -24:46:56.65 & --- & --- & 16.574 & 0.24 & 242 & LPV \\
G30 & 17:48:05.701 & -24:46:28.97 & --- & --- & 16.622 & 0.13 & 242 & LPV \\
G31 & 17:48:03.999 & -24:46:53.02 & --- & --- & 16.840 & 0.17 & 242 & LPV \\
G32 & 17:48:06.065 & -24:46:37.30 & --- & --- & 16.913 & 0.22 & 242 & LPV \\
G33 & 17:48:06.033 & -24:46:30.10 & --- & --- & 17.271 & 0.43 & 242 & LPV \\
G34 & 17:48:03.711 & -24:46:27.44 & --- & --- & 18.195 & 0.60 & 242 & LPV \\
\hline
C1a & 17:48:03.546 & -24:46:52.02 & --- & --- & 14.825 & 0.41 & 242 & C \\
C2 & 17:48:04.870 & -24:46:50.75 & --- & --- & 15.706 & 0.25 & 242 & C \\
C3 & 17:48:04.588 & -24:46:39.36 & --- & --- & 16.694 & 0.15 & 242 & C \\
C4 & 17:48:04.372 & -24:46:43.42 & --- & --- & 15.720 & 0.13 & 242 & C \\
\hline\\
\multicolumn{9}{c}{Periods found using techniques: $^{a}$string length, $^{b}$ fast $\chi^2$, $^{c}$least squares}\\
\hline
\end{tabular}}
\end{table*}

\section{{\em Gaia} variables}\label{sec:gaia}

An exploration of the {\em Gaia} DR3 survey \citep{Eyer23+GDR3} pointed out that some stars were flagged as variables. Matching the position of these stars in our reference frame helped us to identify 34 stars in the field covered by our observations. These stars are labelled as G1-G34 in Fig.~\ref{fig:finding_chart_terzan5} and their ephemerides are also given in Table~\ref{table:ephemerides_var}. A comparison between the ID used in this work and the unique ID assigned in {\em Gaia} is shown in Table~\ref{table:gaia_id}. All these stars appear classified in {\em Gaia} as Long Period Variable (LPV) stars of types: omicron Ceti (Mira), OGLE Small Amplitude Red Giants, and semiregular. Their subtypes, amplitudes, and periods are not reported.

\begin{table}

\caption{Stars flagged as variables in {\em Gaia} DR3. Column 1 corresponds to the labels used in this work while column 2 is the unique source identification assigned in {\em Gaia}.}            
\label{table:gaia_id}     
\centering

{
\begin{tabular}{c c}
\hline
\hline
Var ID & Source ID \\   
This work & {\em Gaia} \\
\hline
G1 & 4067310462871663488 \\
G2 & 4068061085762352640 \\
G3 & 4067310462871669632 \\
G4 & 4067310462871654784 \\
G5 & 4067310462871687808 \\
G6 & 4067310462871656576 \\
G7 & 4068061085762342528 \\
G8 & 4067310462871677440 \\
G9 & 4068061085762351104 \\
G10 & 4068061085694253312 \\
G11 & 4068061085694255744 \\
G12 & 4067310462871687424 \\
G13 & 4067310497231423360 \\
G14 & 4068061085762323328 \\
G15 & 4068061085762325504 \\
G16 & 4068061085762348288 \\
G17 & 4068061085762316288 \\
G18 & 4067310497214309120 \\
G19 & 4067310462871652736 \\
G20 & 4068061085762349184 \\
G21 & 4067310497231392512 \\
G22 & 4068061085762361344 \\
G23 & 4067310462871671424 \\
G24 & 4067310462871658752 \\
G25 & 4068061085762340224 \\
G26 & 4068061081387257344 \\
G27 & 4068061085762372736 \\
G28 & 4068061085762340352 \\
G29 & 4067310462871659904 \\
G30 & 4068061120122086144 \\
G31 & 4068061085762340608 \\
G32 & 4067310497231386240 \\
G33 & 4068061120122317952 \\
G34 & 4068061085808683520 \\
\hline
\end{tabular}}
\end{table}

(G1-G9): Several of these variables were also detected in our analysis for variable star detection and was possible to classify them as semiregular variables, similar to what was done in Sect.~\ref{subsec:SR_var}. Amplitudes range from about 0.1 to 0.46 mag. and periods from about 20 to 266 d. G9 was not automatically detected by the pipeline in the reference frame, so their variation was detected directly in the difference images. Its peak-to-peak amplitude is about 21250 ADU/s (analogue to digital units per second) and the period found is about 48 d. 

\begin{table}
\caption{Frequencies found in the analysis of the stars classified as semiregular variables. Column 1 represents the star ID. Column 2 shows the frequency obtained using the Fast $\chi^2$ method while Column 3 displays the frequency derived from the least squares technique. Column 4 indicates the signal-to-noise ratio measured for the frequencies listed in Column 3.}
\label{tab:sr_freq}
\centering
\begin{tabular}{cccc}
\hline\hline
ID & Fast $\chi^2$ & Least squares & SNR \\
\hline
V14 & 0.00517 & 0.00831 & 6 \\
V15 & 0.01283 & 0.01128 & 6 \\
V16 & 0.02273 & 0.02389 & 5 \\
V17 & 0.01011 & 0.01813 & 5 \\
G1  & 0.00595 & 0.00905 & 6 \\
G2  & 0.00649 & 0.00926 & 6 \\
G3  & 0.02056 & 0.00145 & 7 \\
G4  & 0.00376 & 0.00064 & 6 \\
G5  & 0.00405 & 0.00909 & 7  \\
G6  & 0.00613 & 0.02230 & 4 \\
G7  & 0.04547 & 0.04899 & 5 \\
G8  & 0.00524 & 0.03466 & 4 \\
G9  & 0.01786 & 0.02063 & 6 \\
\hline

\end{tabular}
\end{table}

\begin{figure}
  \centering
  \subfloat[][]{
  \begin{tabular}{c}
    \includegraphics[width=0.96\hsize]{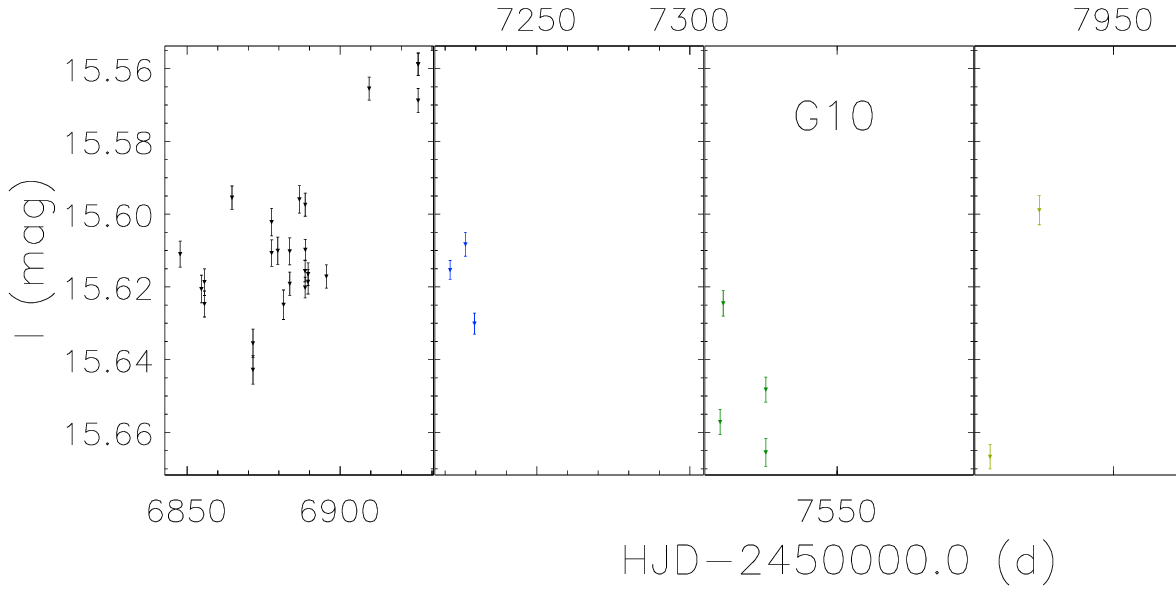} \\
    \includegraphics[width=0.96\hsize]{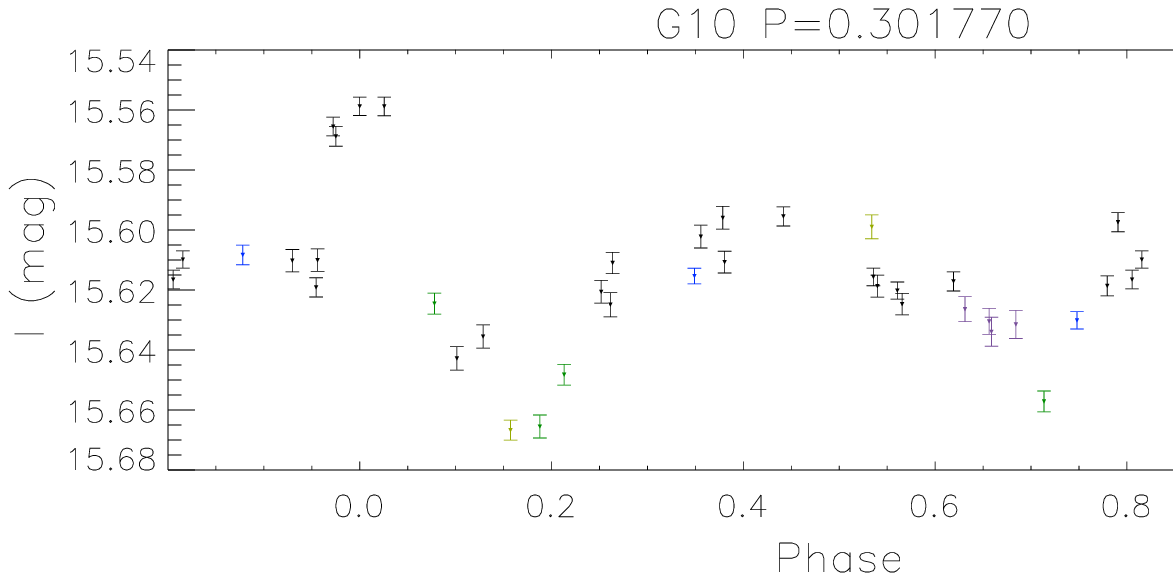} \\
  \end{tabular}}
  \caption{Light curve of variable star G10. The top panel shows the light curve in $I$ magnitude versus HJD while the bottom panel shows the phased light curve with the periods and epochs given in Table~\ref{table:ephemerides_var}. Colours are the same as in Fig.~\ref{fig:histo_obs}.}
  \label{fig:phased_G10}
\end{figure}

(G10): This star, rather than showing a long period, seems to exhibit a variation similar to those found in short-period variables. The least squares method found a period of 0.497880 days (close to half a sidereal day) with a signal-to-noise ratio of about 3.2. However, the light curve appears better phased with a period of 0.301770 days, as found using the string length method. This period leads to a better defined light curve shape, resembling that found in eclipsing binaries of the W Ursae Majoris type. The amplitude is approximately 0.11 mag. The period found is tentative, and a proper classification of this star is unclear. Further analysis is encouraged. Phased light curve for this star is shown in Fig.~\ref{fig:phased_G10}.

Finally, we did not find coherent periods for the remaining stars flagged as variables in {\em Gaia} survey (G11-G34) and were left classified as LPV. Light curves for these stars are shown in Fig.~\ref{fig:LVP}.

\section{Semiregular variables period-luminosity (PL) relation}\label{sec:PL}

\begin{figure}[h]
  \centering
  \includegraphics[width=\hsize]{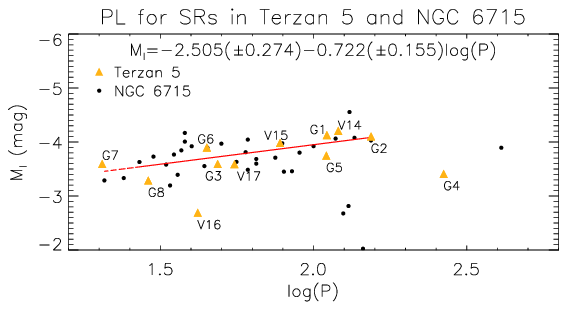}
  \caption{Period-luminosity diagram for semiregular variables studied in this work (dark-yellow triangles), while the black points correspond to the semiregular variables listed in the \citet[][August 2019 update]{Clement01+09} catalog for the globular cluster NGC~6715. The red line represents the best-fit line for both datasets.}
         \label{fig:pld_SR}
\end{figure}

All stars classified as semiregular variables in Table~\ref{table:ephemerides_var} were plotted on a period-luminosity diagram. For comparison, we included all semiregular variables \citep[discovered by][]{rosino1952, rosino1958+01, layden00+01, sollima10+03, figuera16+mindstepb, hamanowicz16+ogle} available in the catalogue of \citet[][August 2019 update]{Clement01+09} for NGC~6715 (M54). A clear linear relation between both groups of semiregulars was observed. To adjust the zero points, we transformed the apparent magnitudes to absolute magnitudes using the distances to the globular clusters of 26280 $\pm$ 330 pc and 6620 $\pm$ 150 pc for NGC 6715 and Terzan 5, respectively \citep{baumgardt21+01}. Similarly, an extinction of $A_I=0.23$ \citep{schlafly11+01} mag was adopted for NGC~6715. However, as previously mentioned, Terzan 5 is a cluster that is affected by a large but uncertain amount of foreground extinction. Using, for instance, the reddening reported in the \citet{harriscatalog}, and assuming a standard extinction law, one finds $A_I = 3.42$~mag. In like vein, the reddening map provided by \citet{massari12+09} gives an average extinction towards the cluster of $A_I \approx 3.78$~mag. The \citep{schlafly11+01} maps, on the other hand, report $A_I=5.91$~mag.  
None of these extinction values lead to an alignment of the NGC~6715 and Terzan~5 relations. Accordingly, in our analysis, we assumed that both clusters share the same PL relation, and let the fit determine the best extinction value for Terzan 5 while fitting both datasets simultaneously. This led to an extinction value for Terzan 5 of $A_I=4.76 \pm 0.09$ mag, which is intermediate between the values quoted above. The final PL relation, as shown in Fig.~\ref{fig:pld_SR}, is
\begin{equation}\label{eq:plr_SR_ma}
    M_\mathrm{I}=-2.505(\pm 0.274)-0.722(\pm0.155)\log(P),
\end{equation}
\noindent where $P$ is the period of the star and $M_\mathrm{I}$ is the absolute magnitude. The fit did not include the outlier stars (e.g., V16 and G4). These outliers might follow a different PL relation, as known to happen in these types of stars \citep{Soszynski13+01, trabucchi21+02}.

\begin{figure*}
\centering
\subfloat[][]{
  \begin{tabular}{cc}
    \includegraphics[width=0.474\hsize]{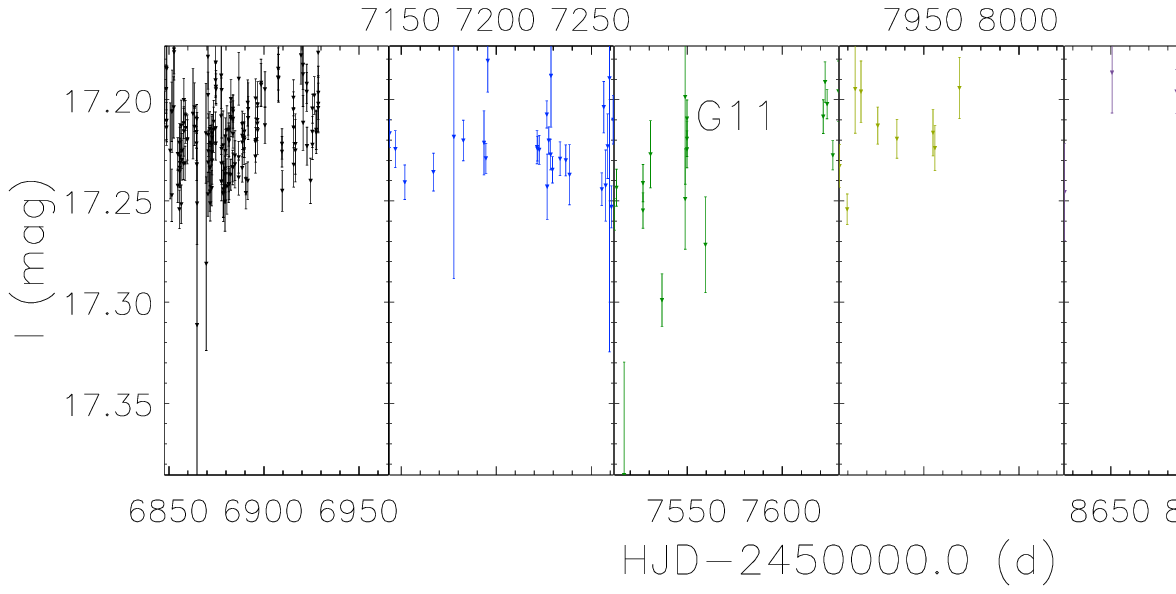} &
    \includegraphics[width=0.474\hsize]{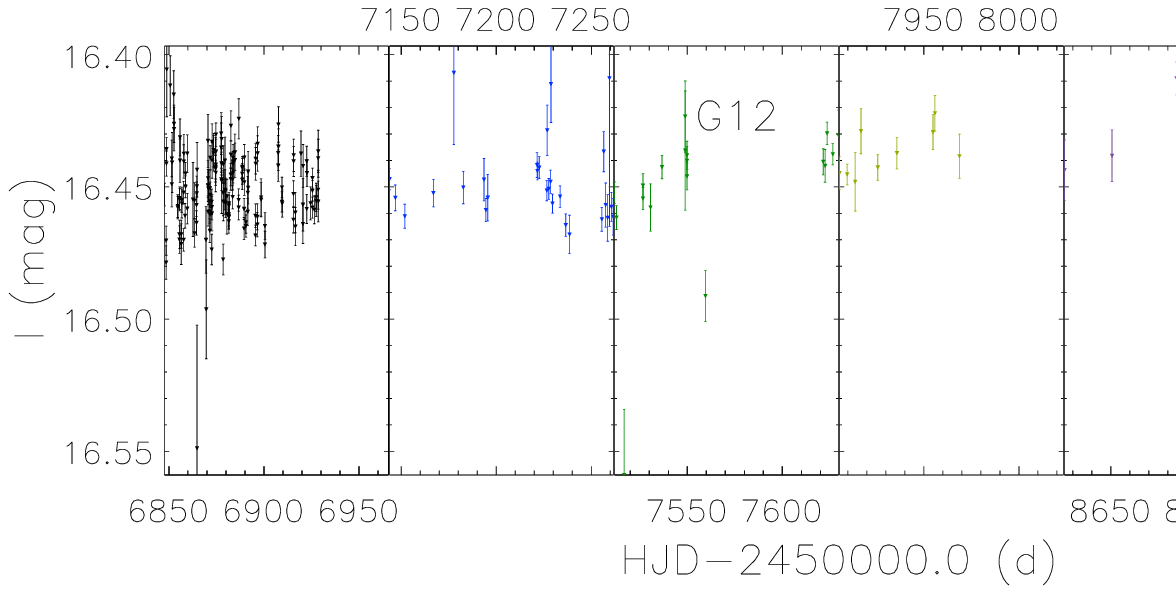} \\
    \includegraphics[width=0.474\hsize]{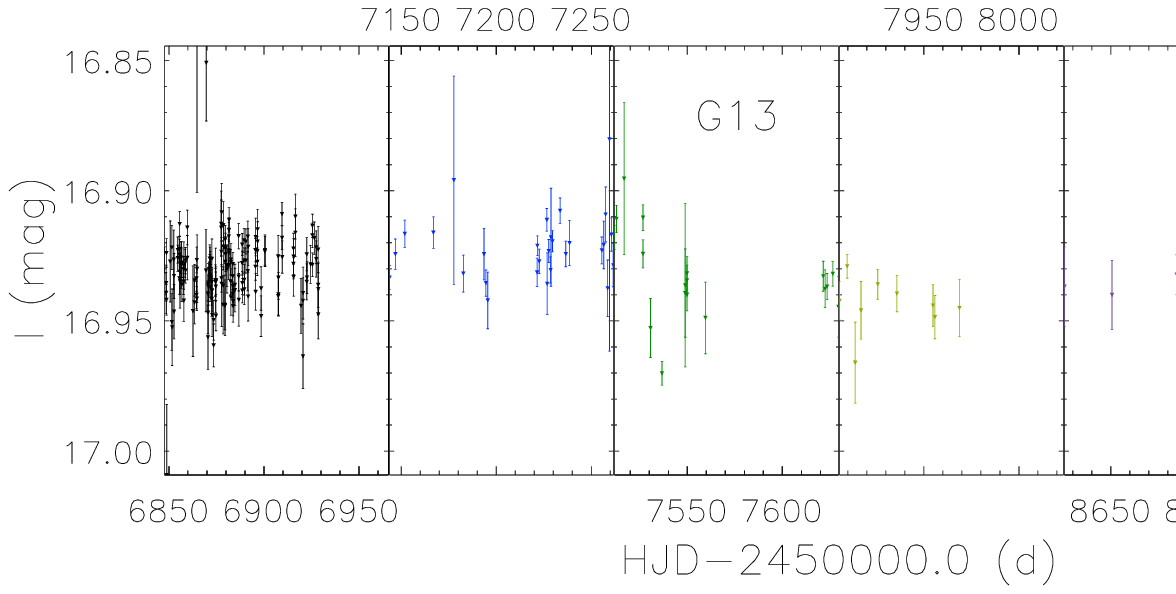} &
    \includegraphics[width=0.474\hsize]{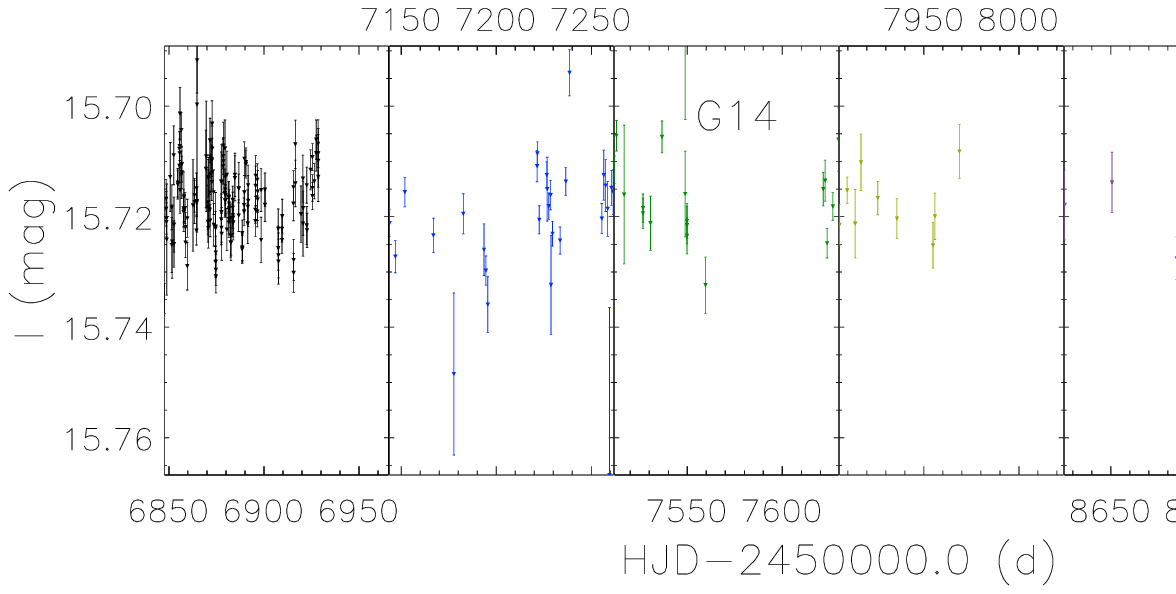} \\
    \includegraphics[width=0.474\hsize]{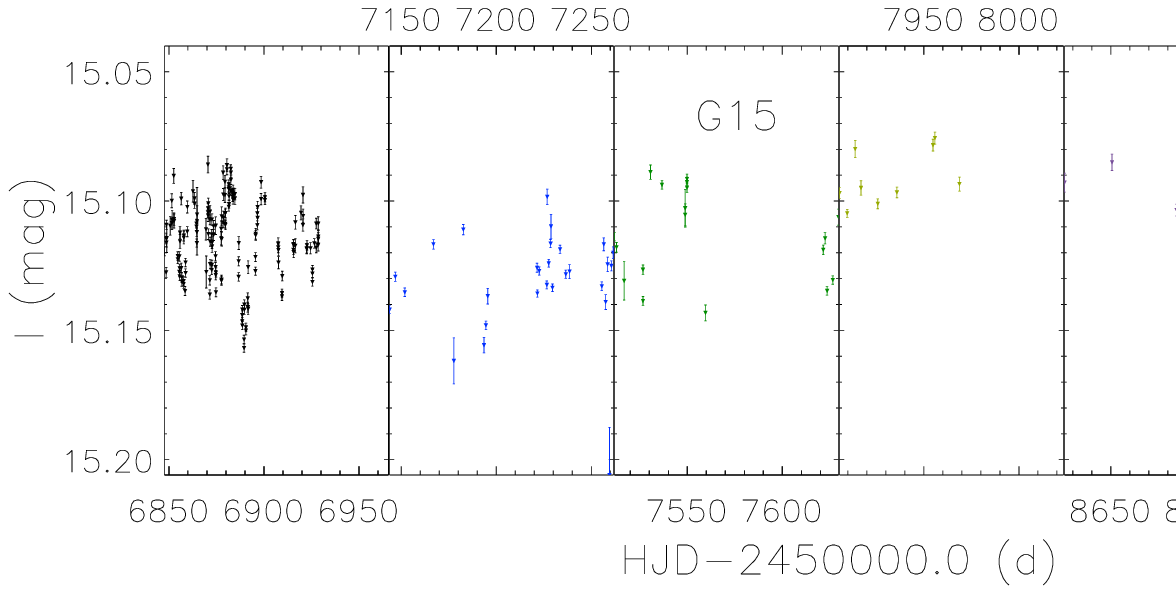} &
    \includegraphics[width=0.474\hsize]{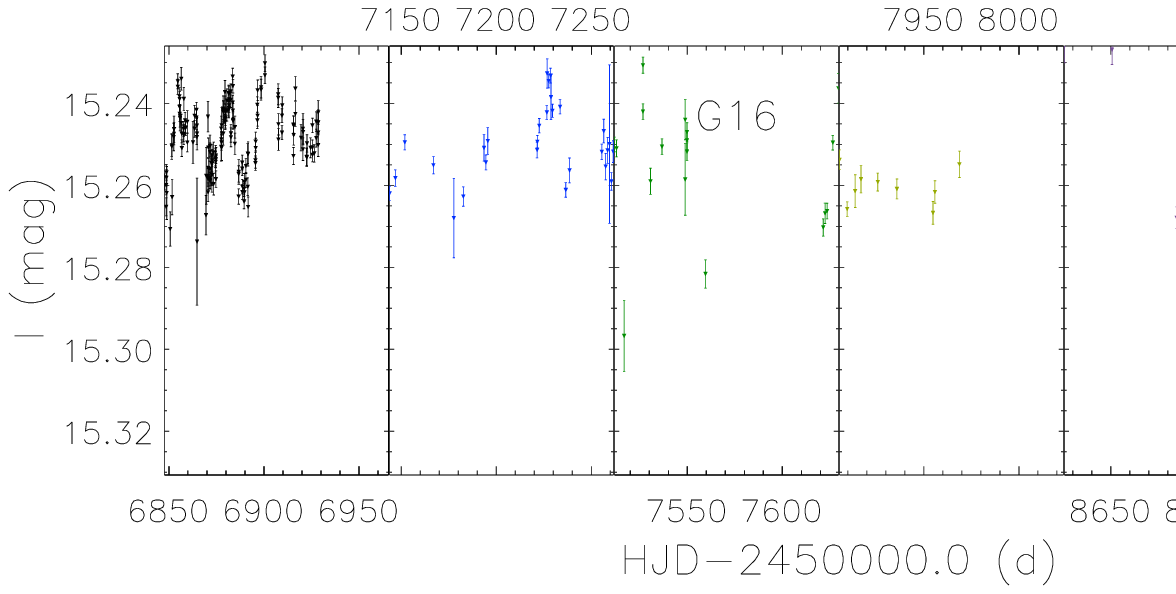} \\
    \includegraphics[width=0.474\hsize]{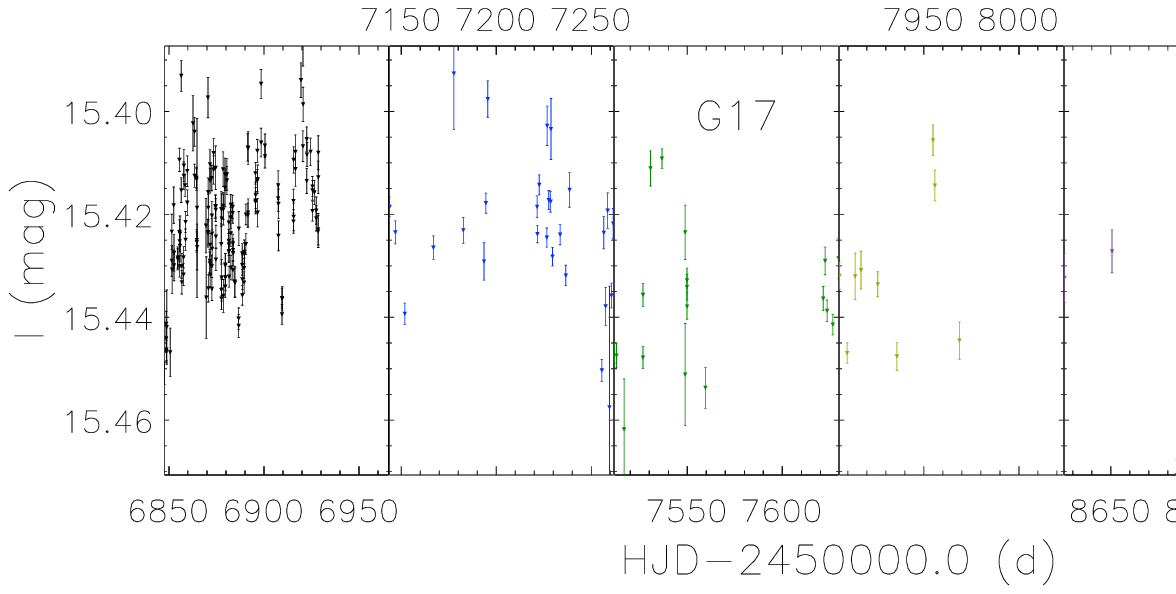} &
    \includegraphics[width=0.474\hsize]{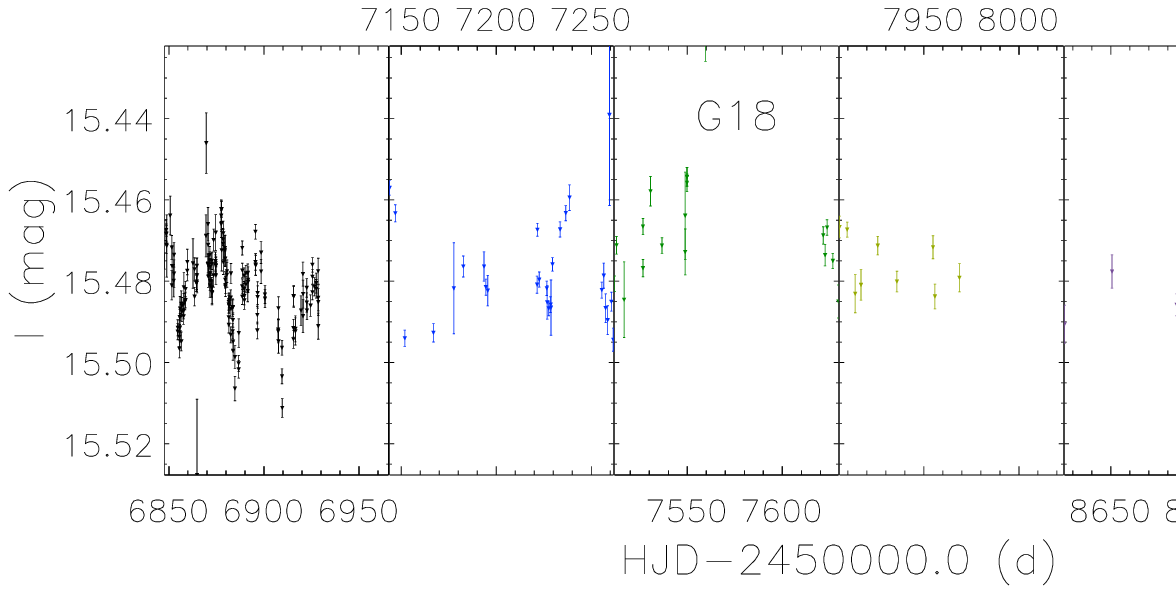} \\
    \includegraphics[width=0.474\hsize]{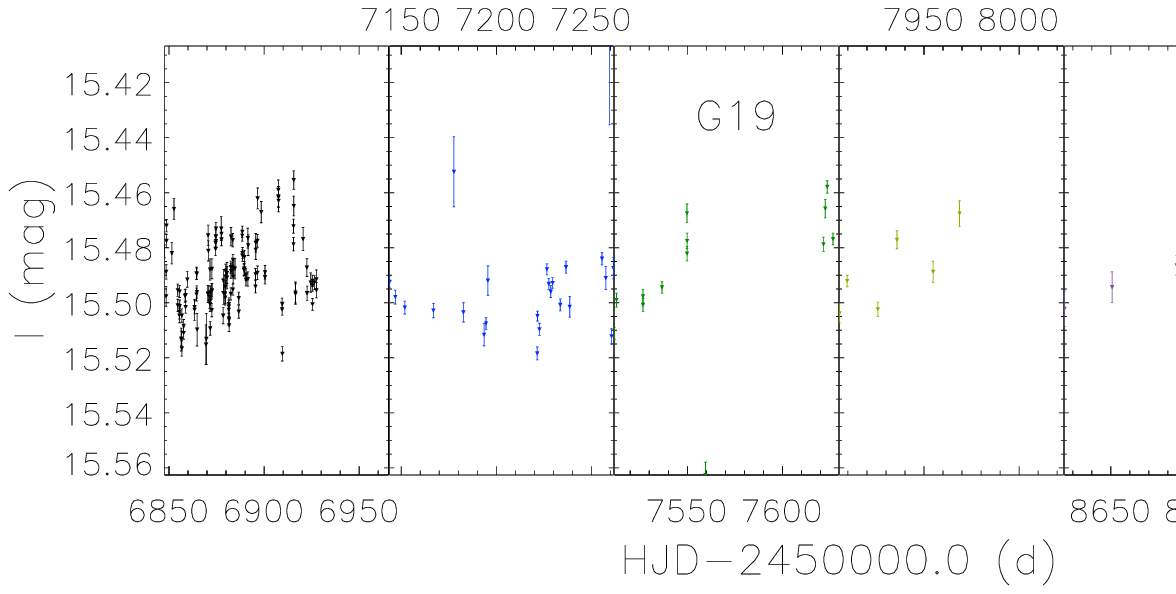} &
    \includegraphics[width=0.474\hsize]{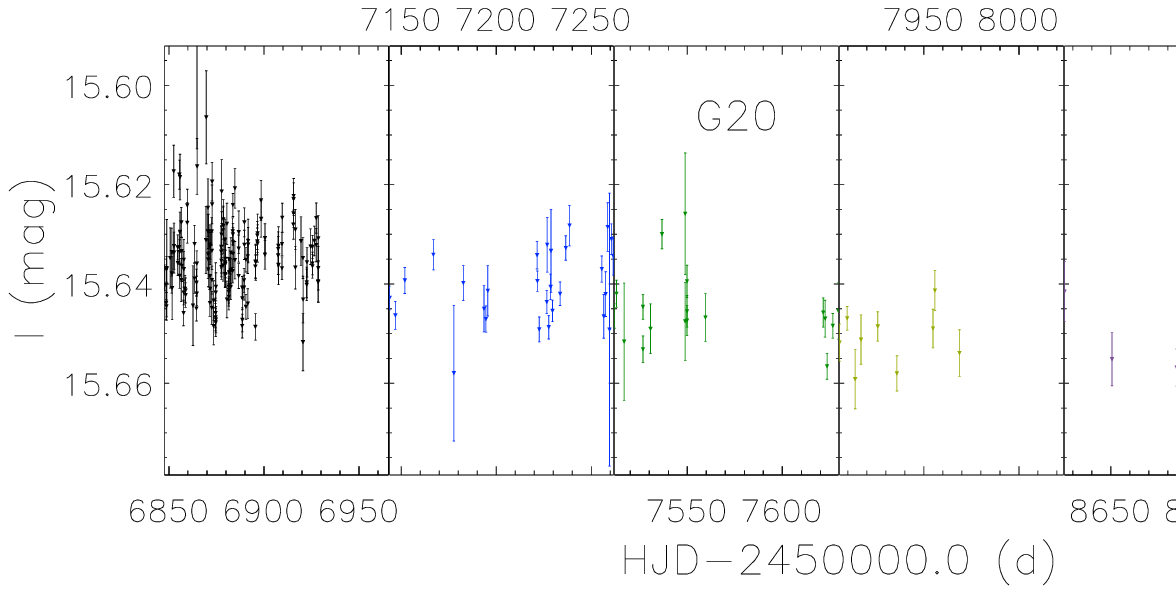} \\
    \includegraphics[width=0.474\hsize]{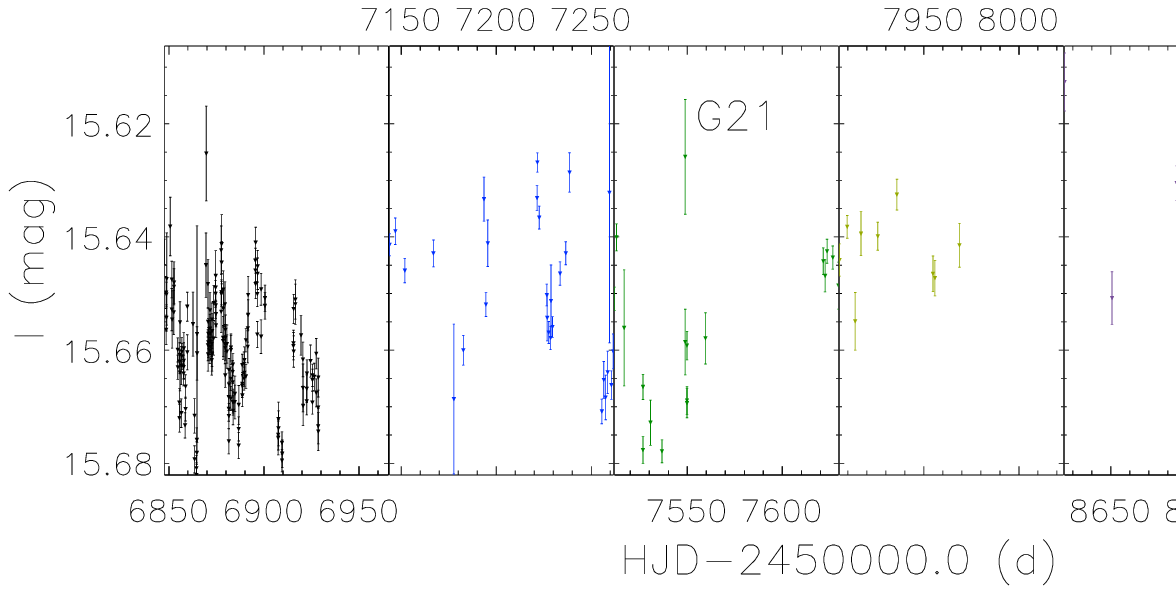} &
    \includegraphics[width=0.474\hsize]{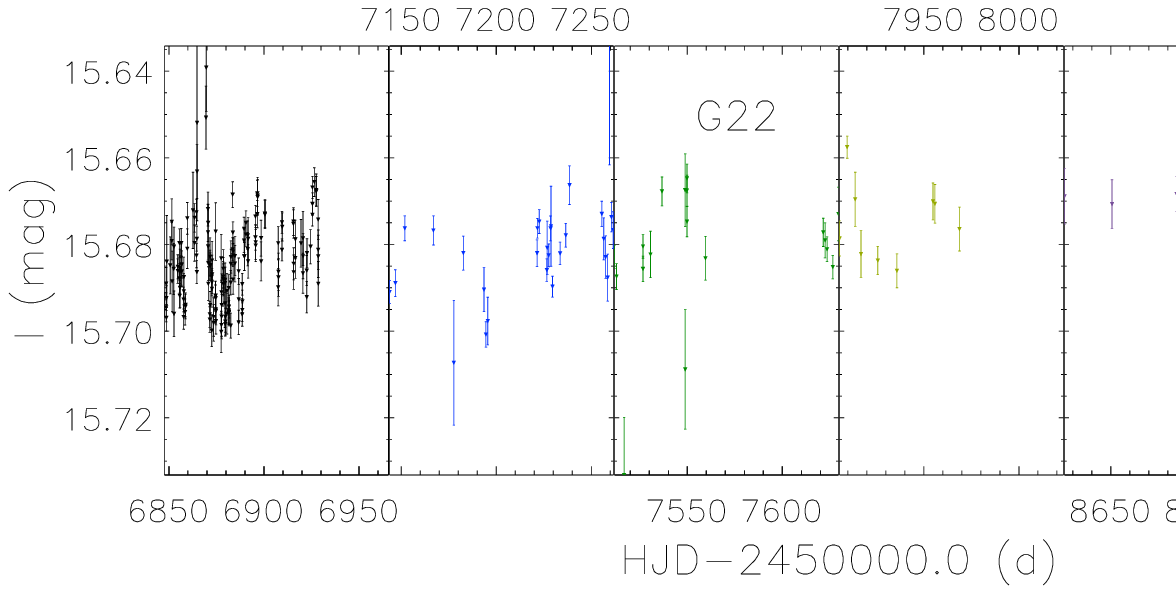} \\
  \end{tabular}
}
\caption{Light curves in $I$ magnitude as a function of HJD for stars flagged as LPV variables in {\em Gaia}. The colours of the plots match those in Fig.~\ref{fig:histo_obs}.}
\label{fig:LVP}
\end{figure*}

\begin{figure*}
\ContinuedFloat
\centering
\subfloat[][]{
  \begin{tabular}{cc}
    \includegraphics[width=0.474\hsize]{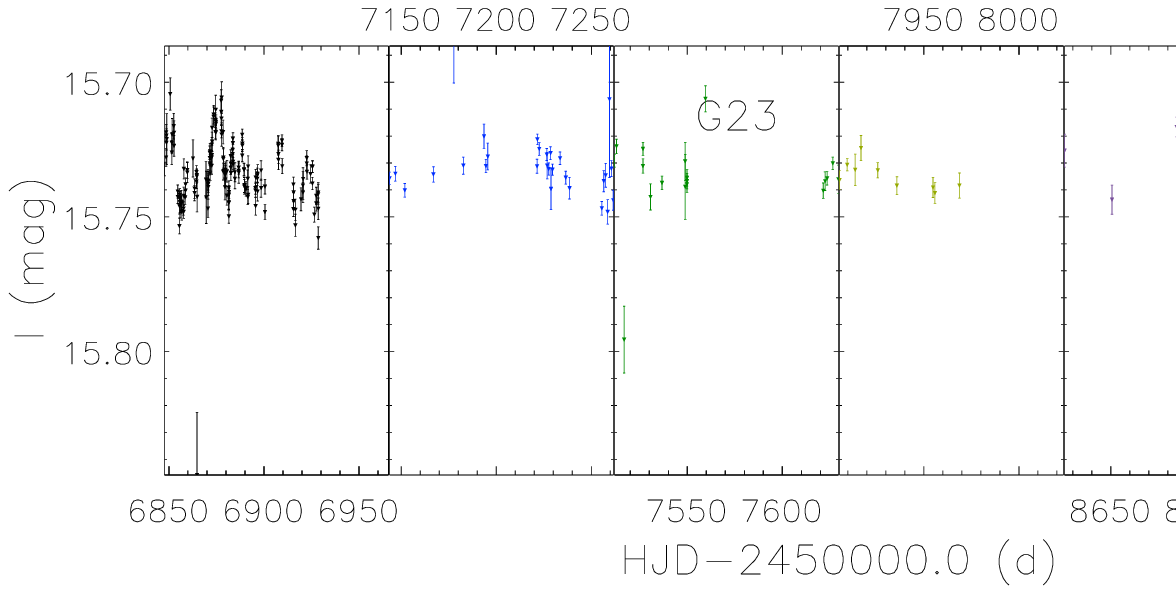} &
    \includegraphics[width=0.474\hsize]{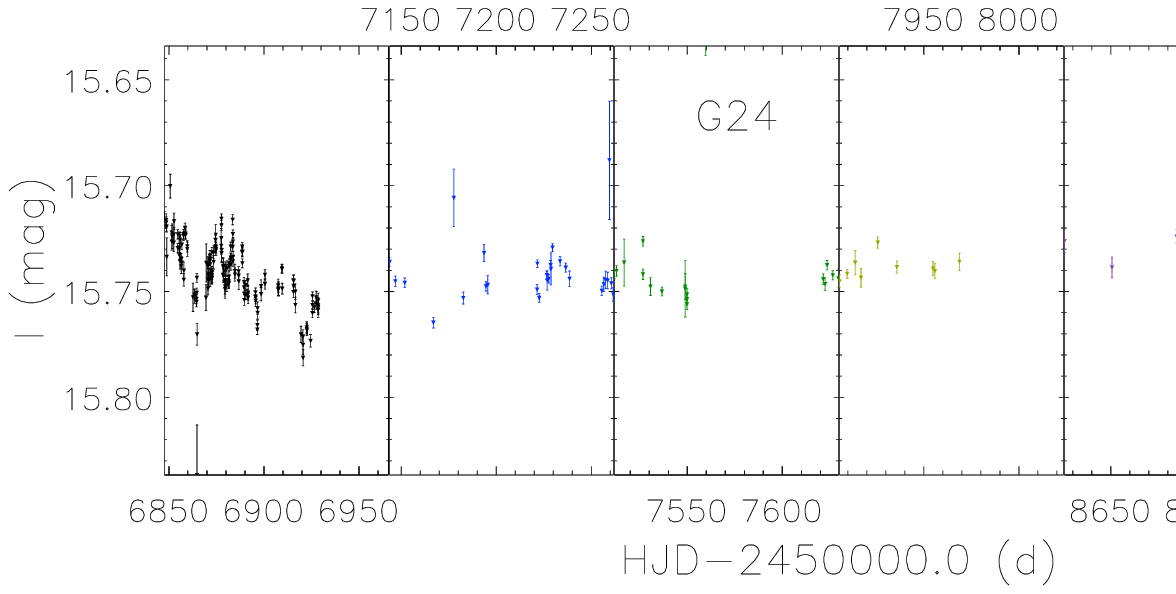} \\
    \includegraphics[width=0.474\hsize]{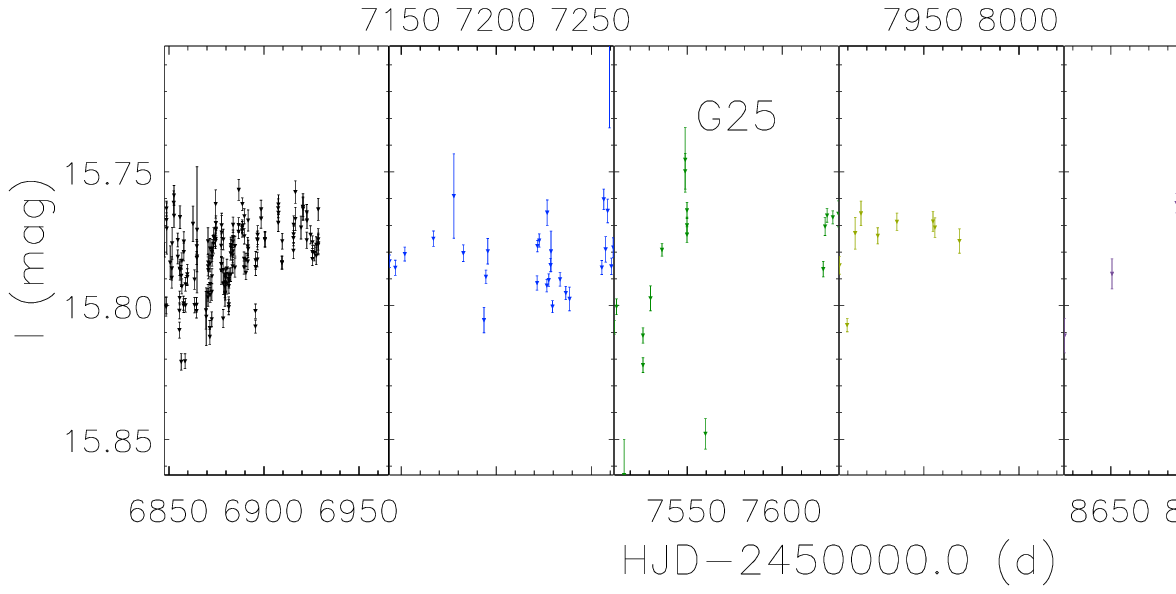} &
    \includegraphics[width=0.474\hsize]{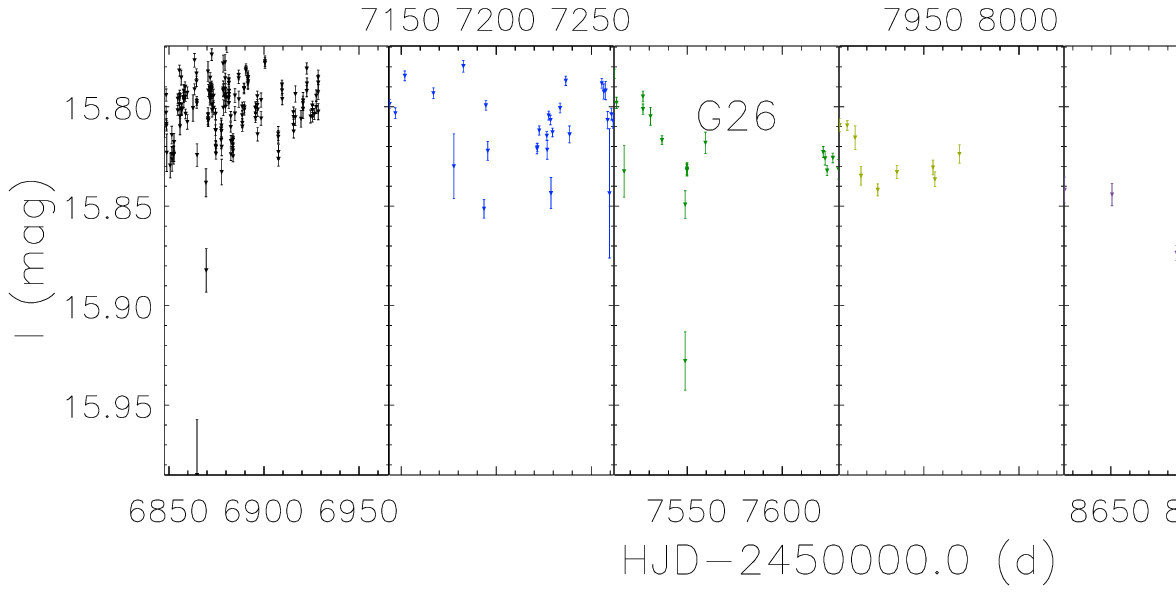} \\
    \includegraphics[width=0.474\hsize]{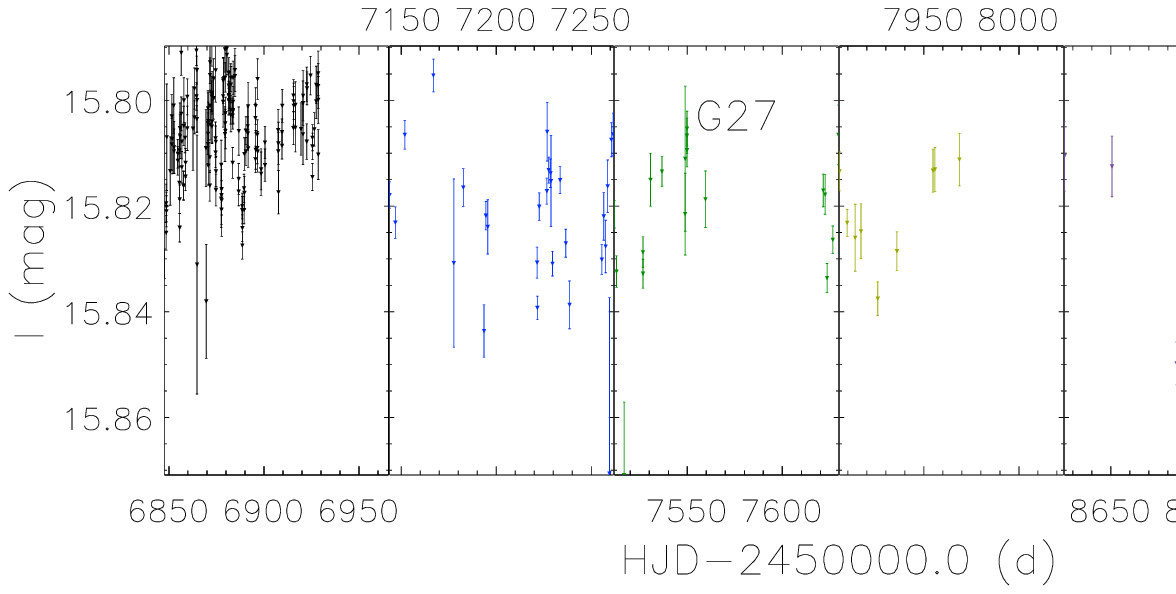} &
    \includegraphics[width=0.474\hsize]{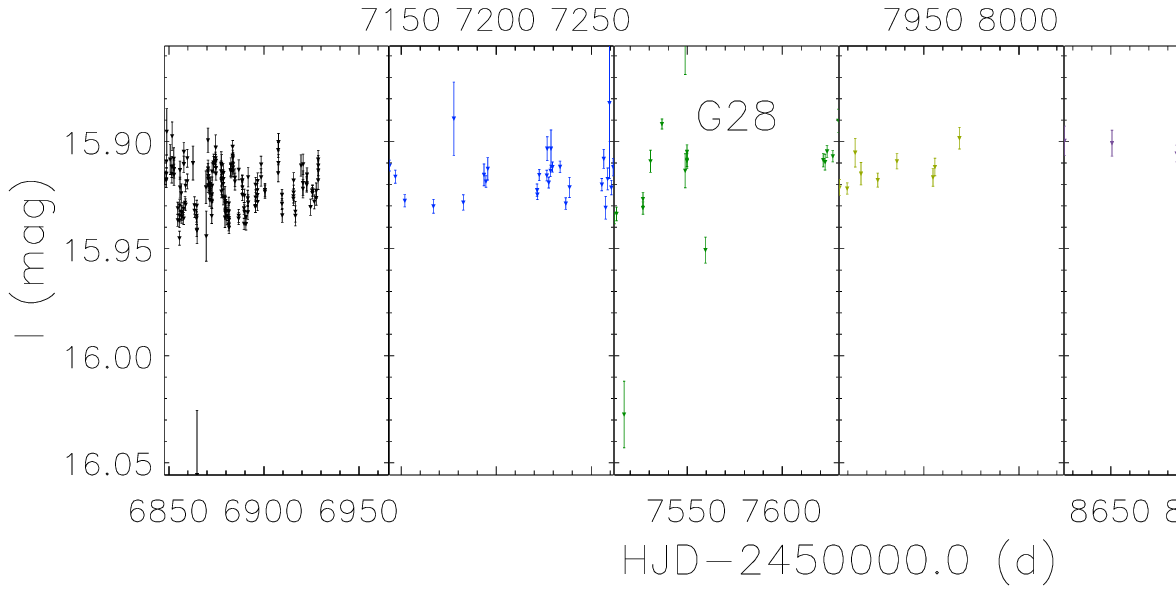} \\
    \includegraphics[width=0.474\hsize]{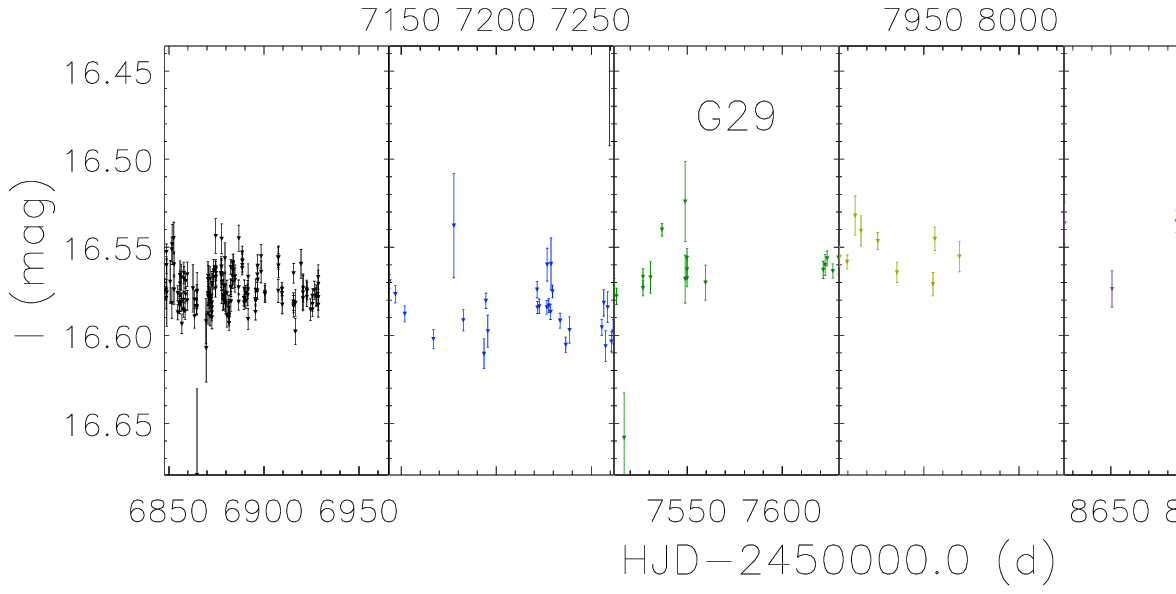} &
    \includegraphics[width=0.474\hsize]{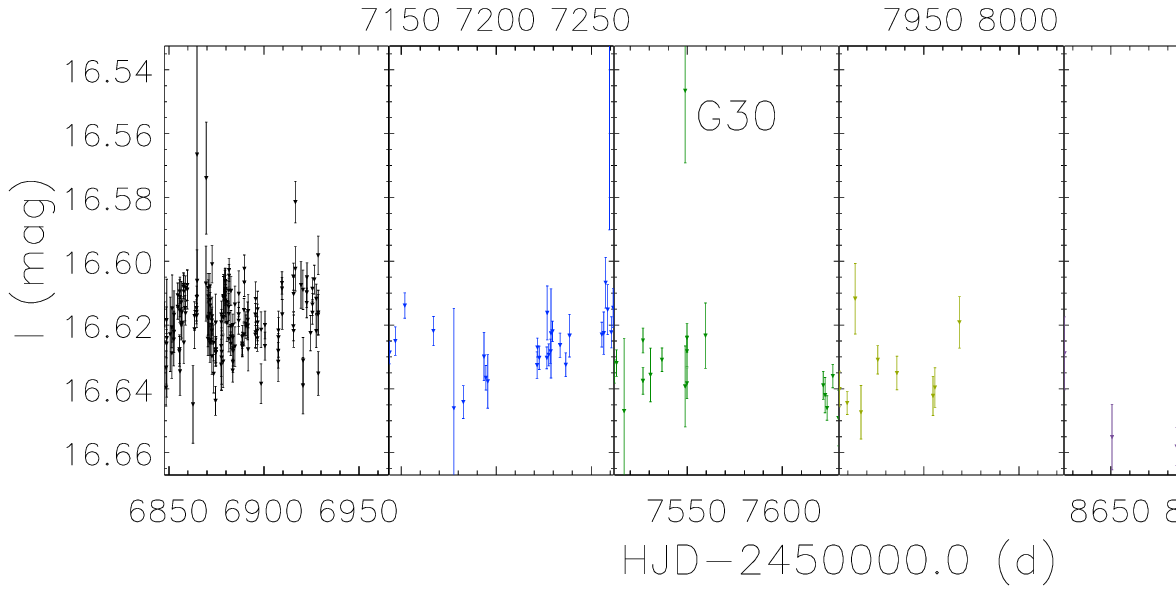} \\
    \includegraphics[width=0.474\hsize]{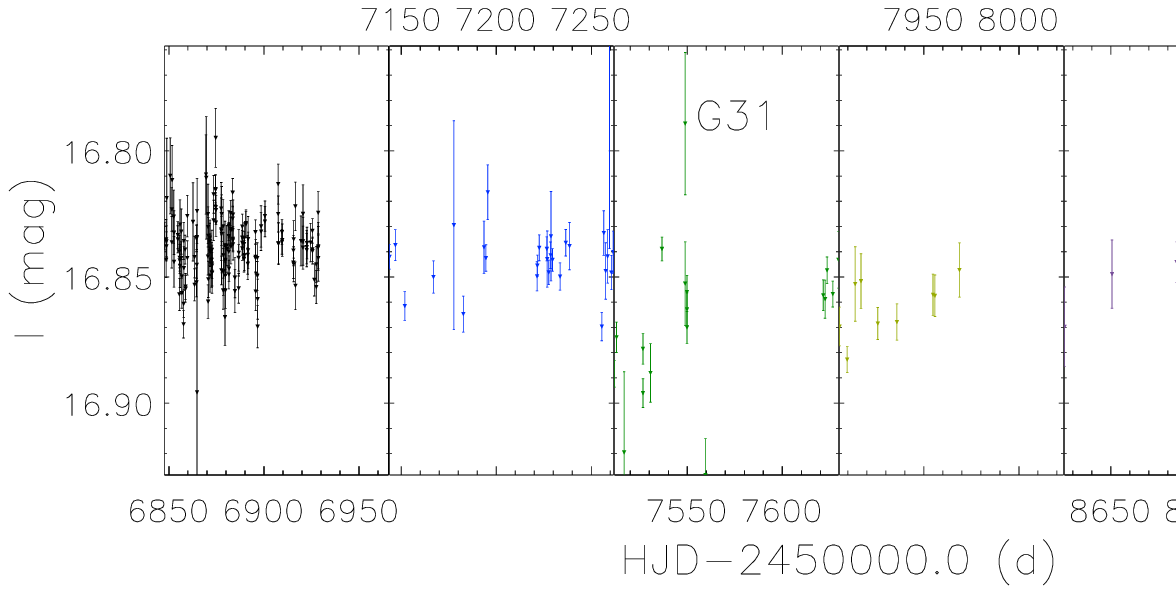} &
    \includegraphics[width=0.474\hsize]{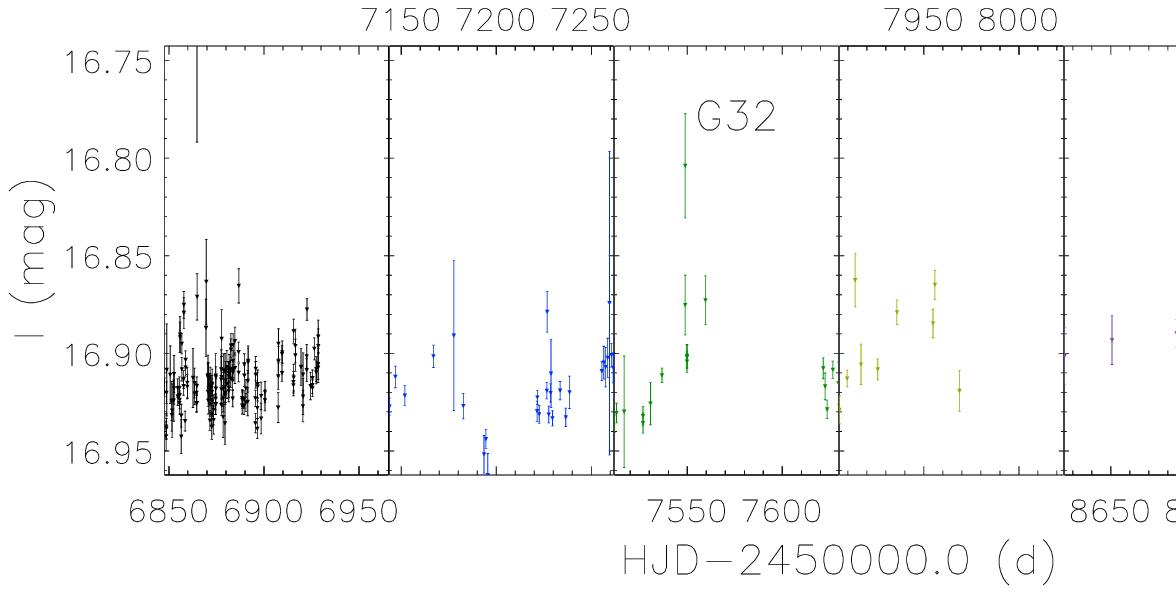} \\
    \includegraphics[width=0.474\hsize]{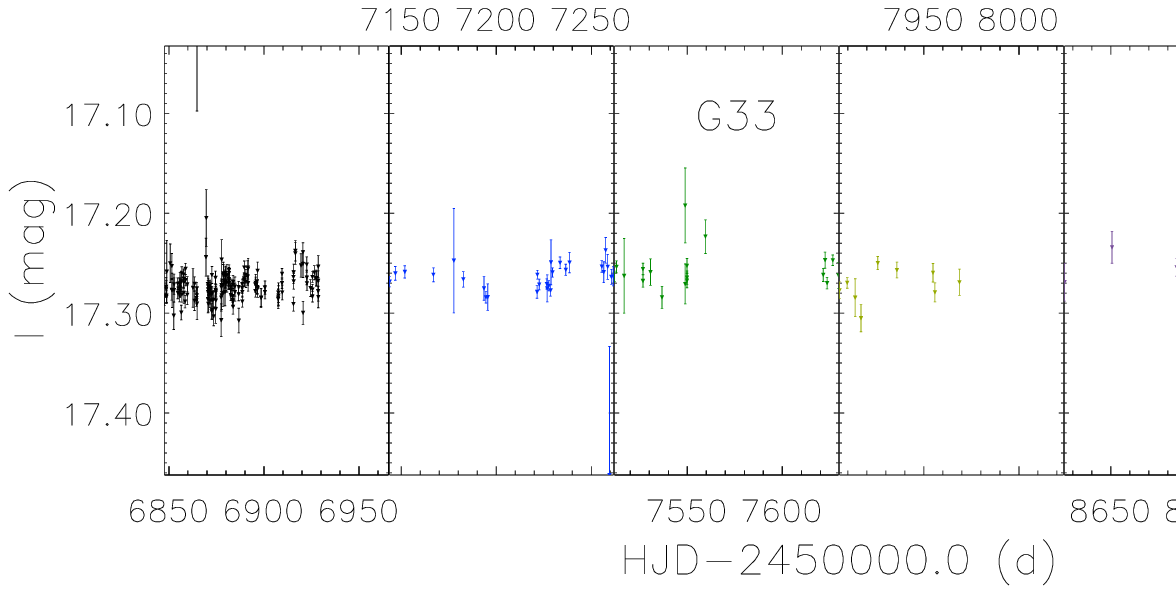} &
    \includegraphics[width=0.474\hsize]{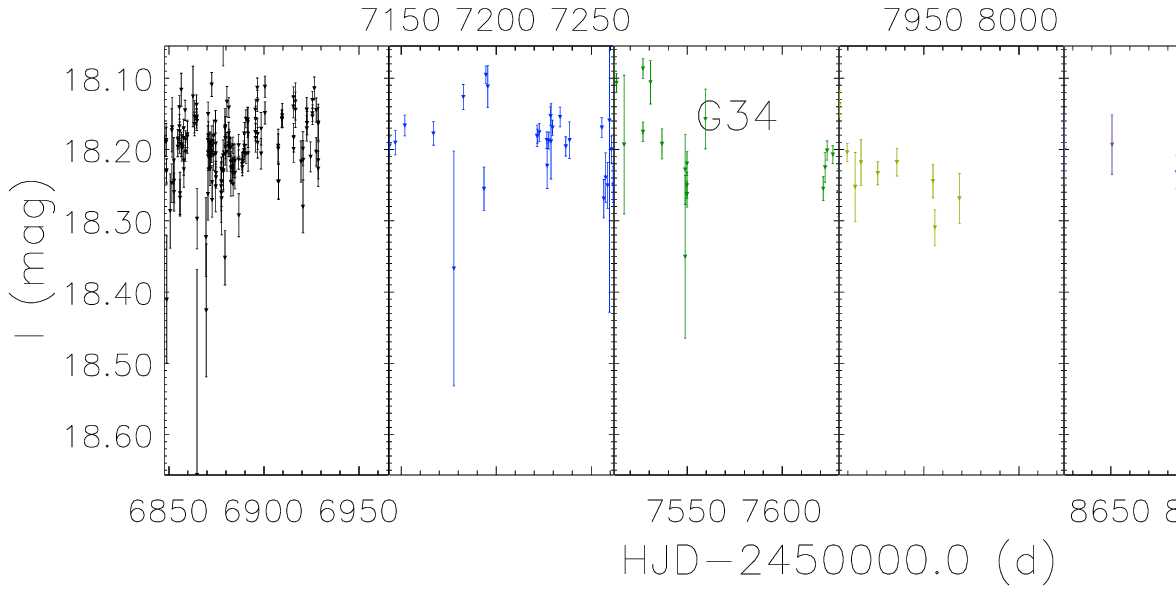} \\
  \end{tabular}
}
\caption{continued.}
\end{figure*}

\begin{figure*}
\centering
\subfloat[][]{
  \begin{tabular}{cc}
    \includegraphics[width=0.474\hsize]{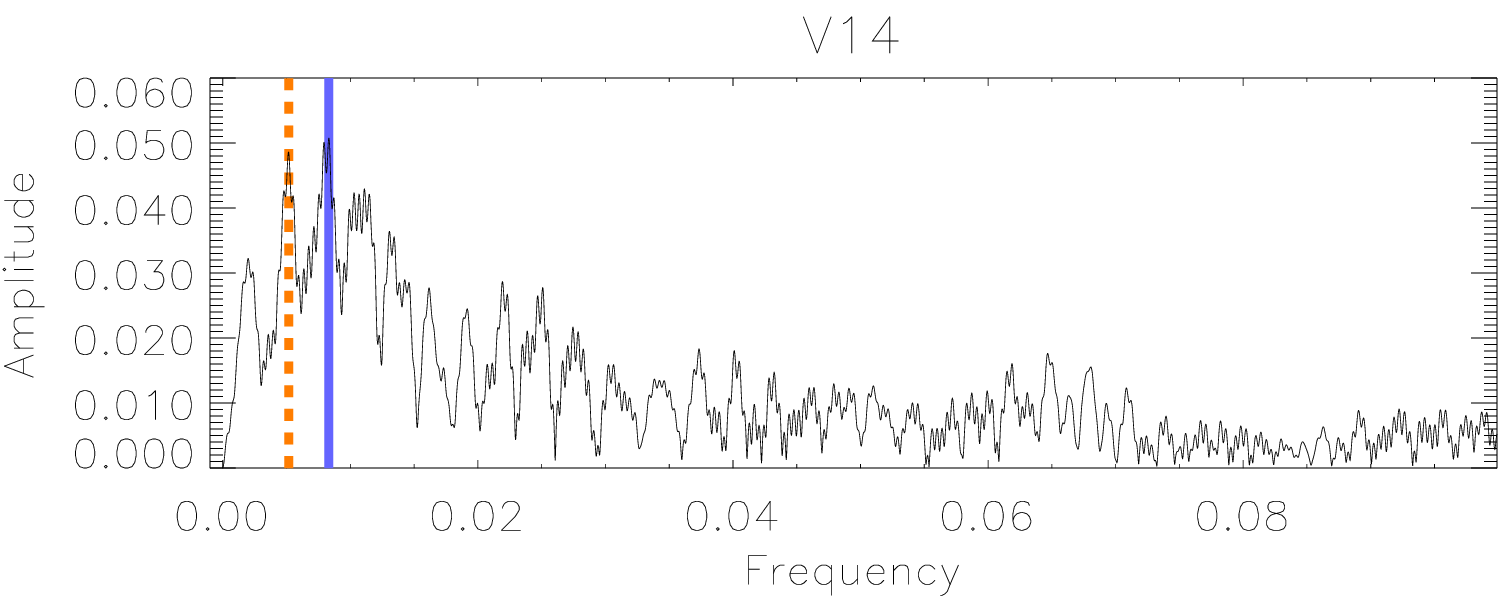} &
    \includegraphics[width=0.474\hsize]{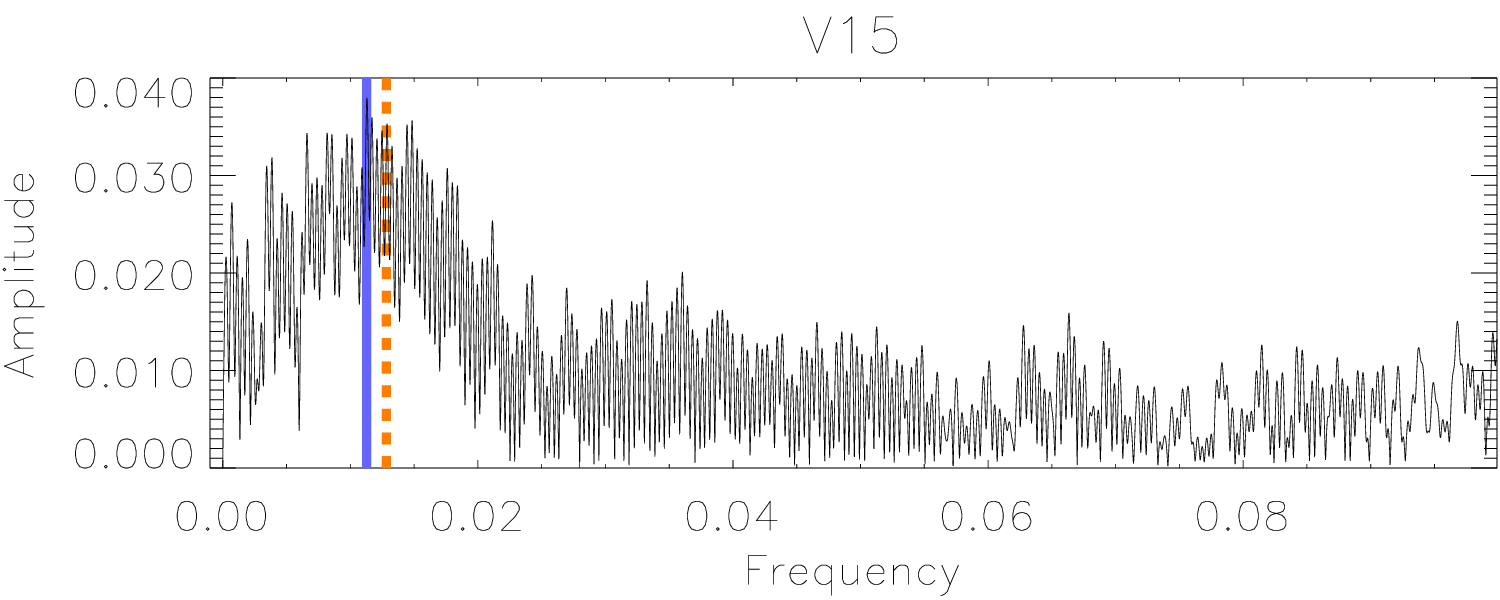} \\
    \includegraphics[width=0.474\hsize]{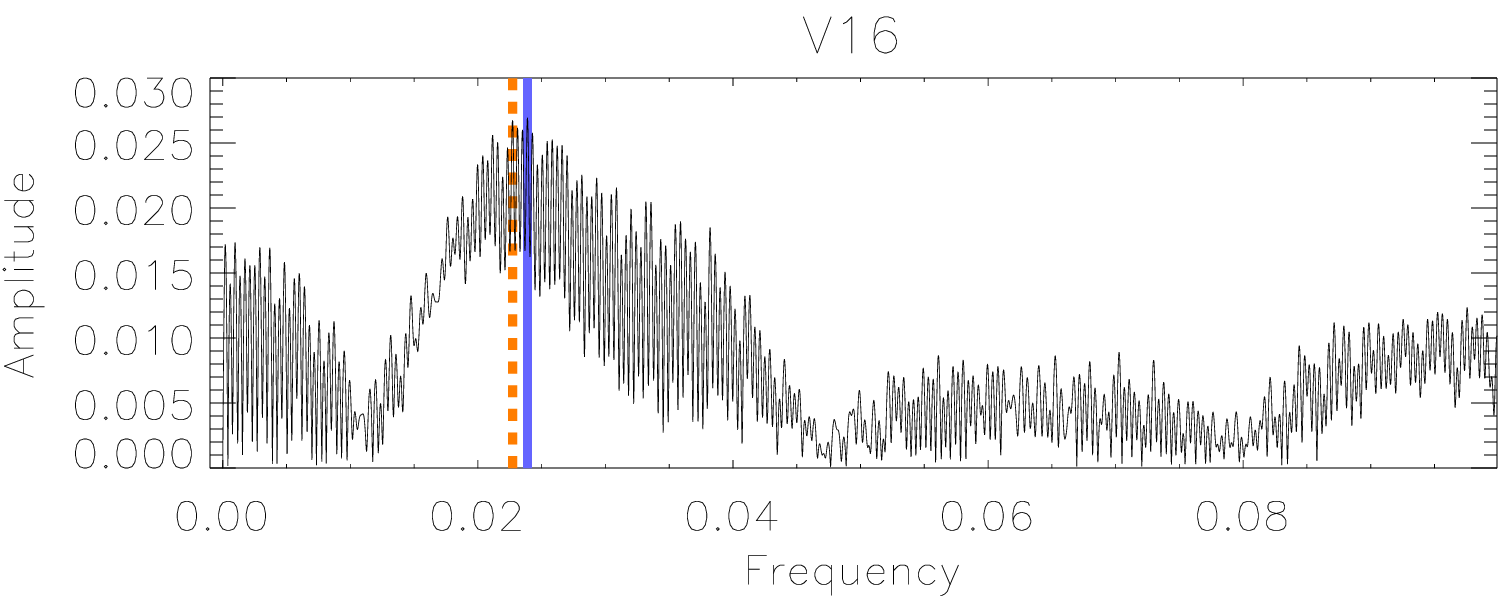} &
    \includegraphics[width=0.474\hsize]{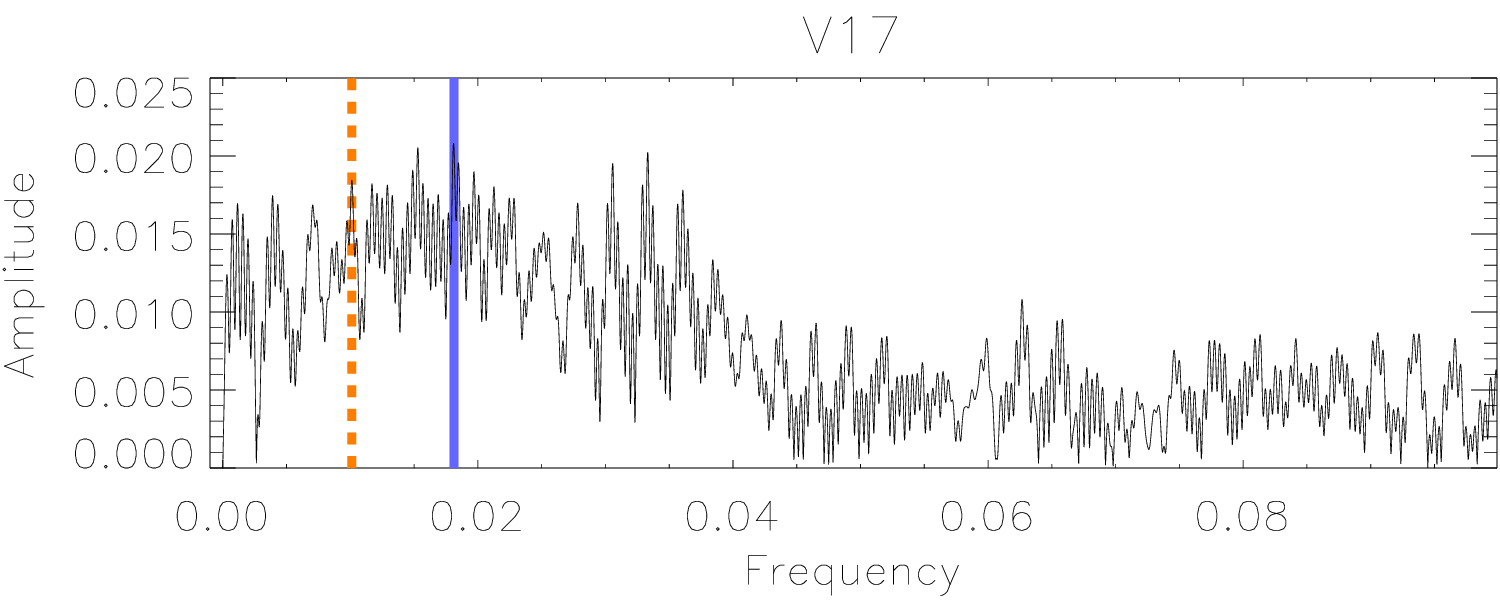} \\
    \includegraphics[width=0.474\hsize]{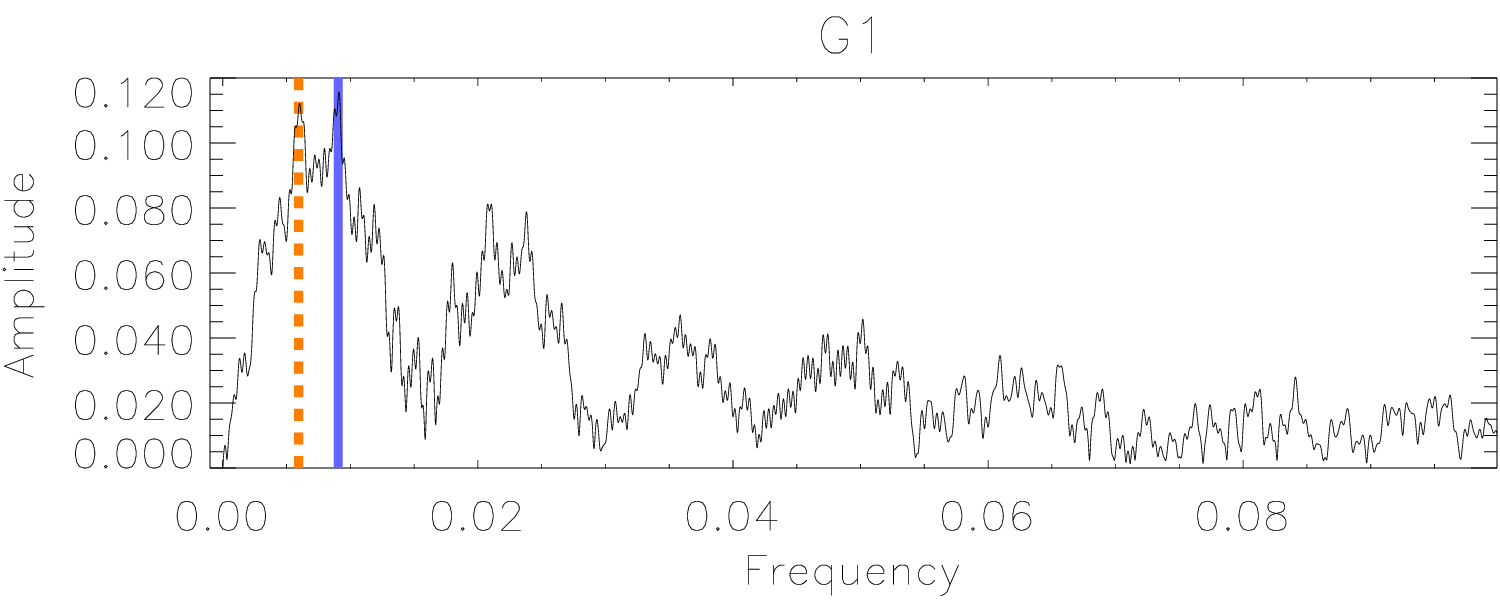} &
    \includegraphics[width=0.474\hsize]{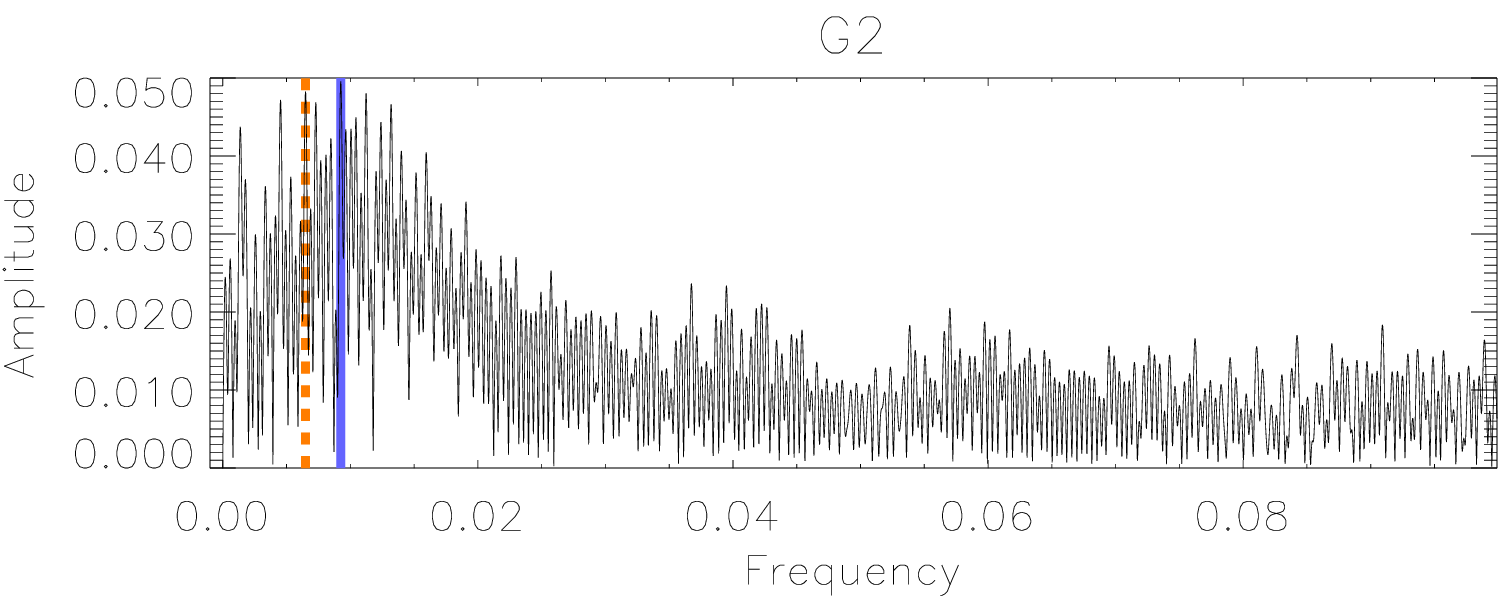} \\
    \includegraphics[width=0.474\hsize]{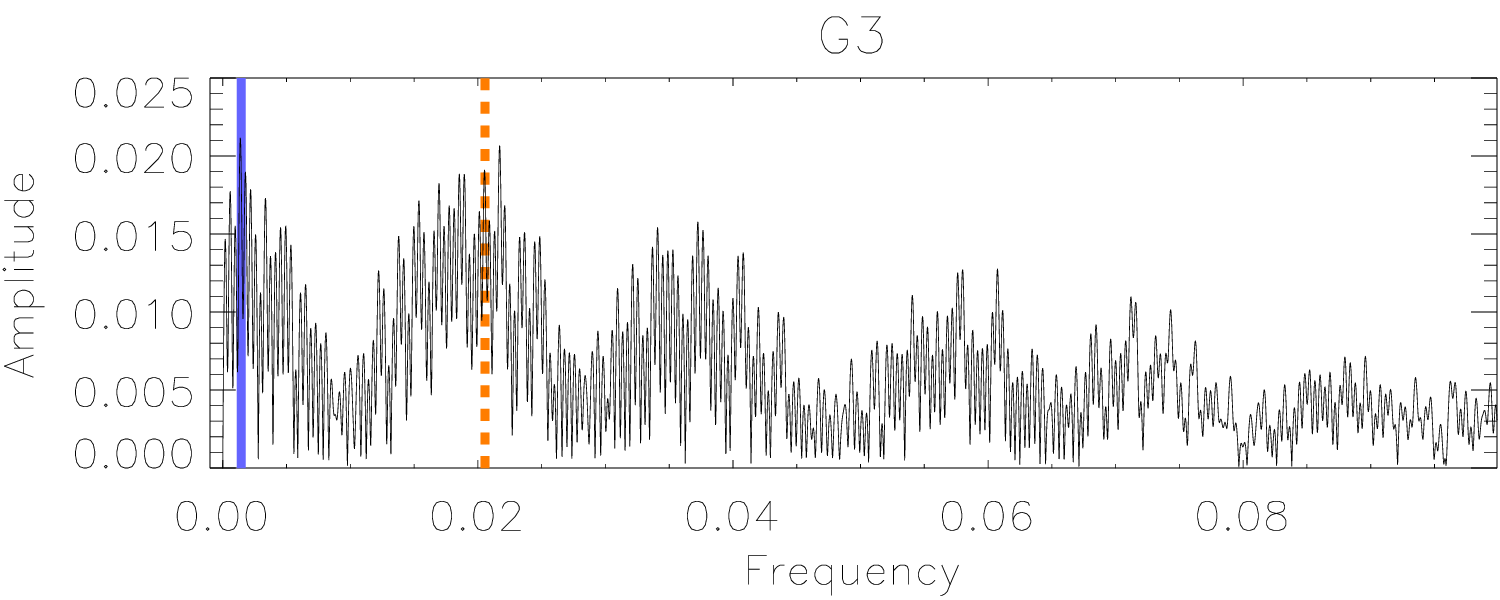} &
    \includegraphics[width=0.474\hsize]{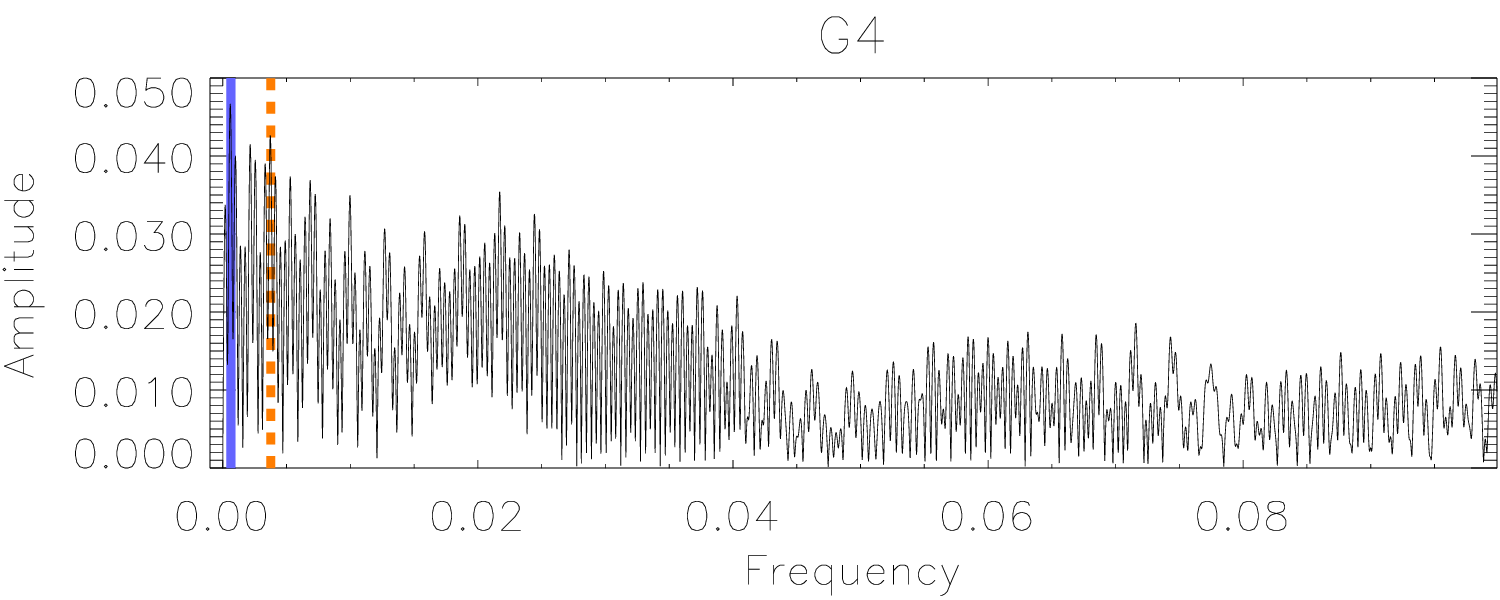} \\
    \includegraphics[width=0.474\hsize]{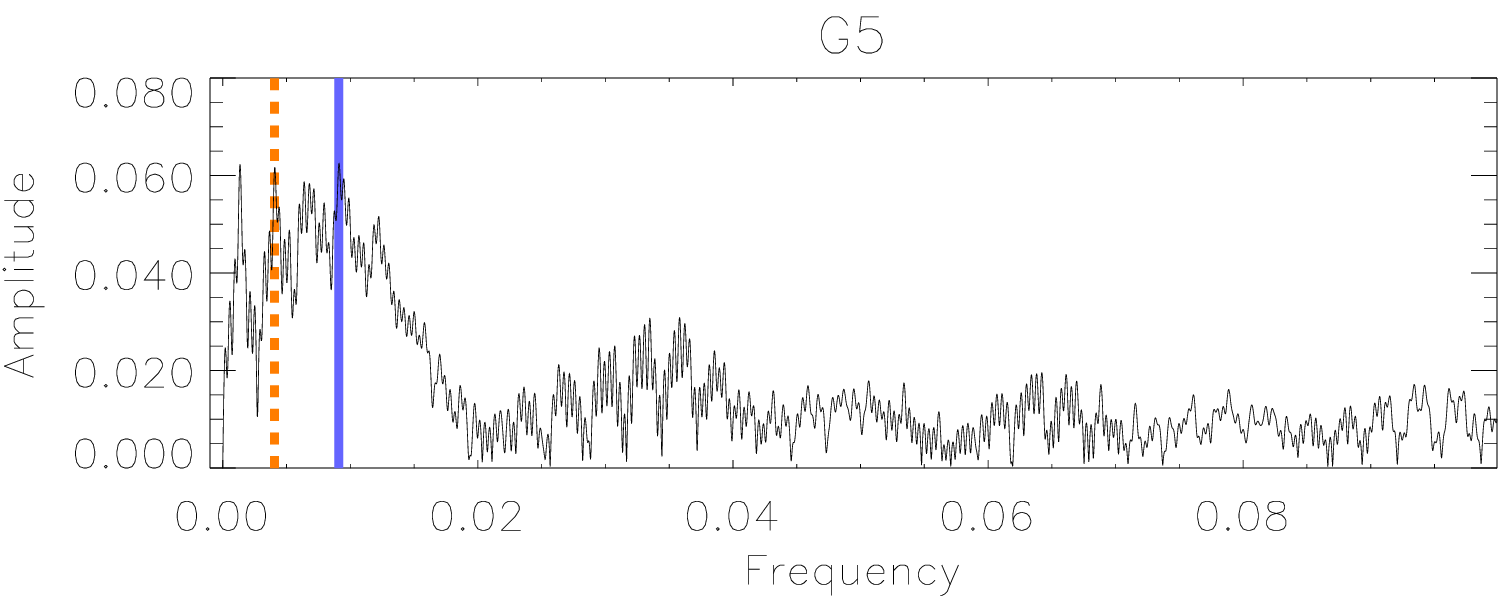} &
    \includegraphics[width=0.474\hsize]{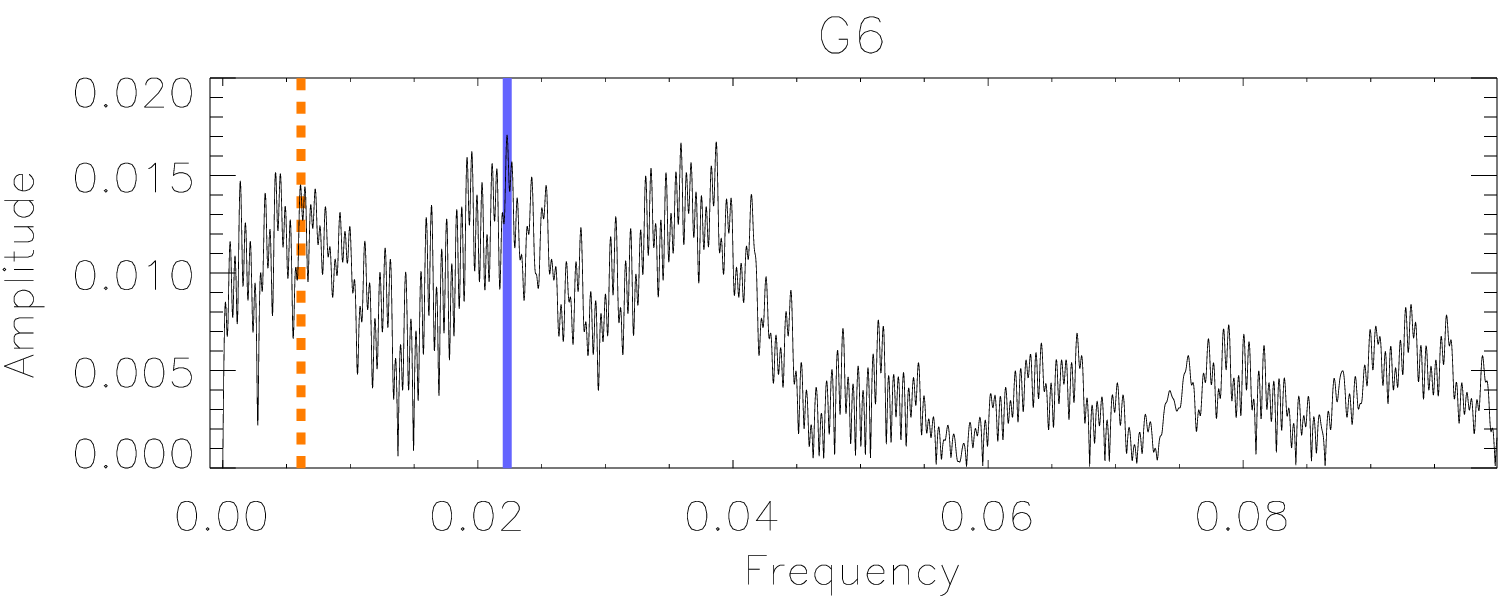} \\
    \includegraphics[width=0.474\hsize]{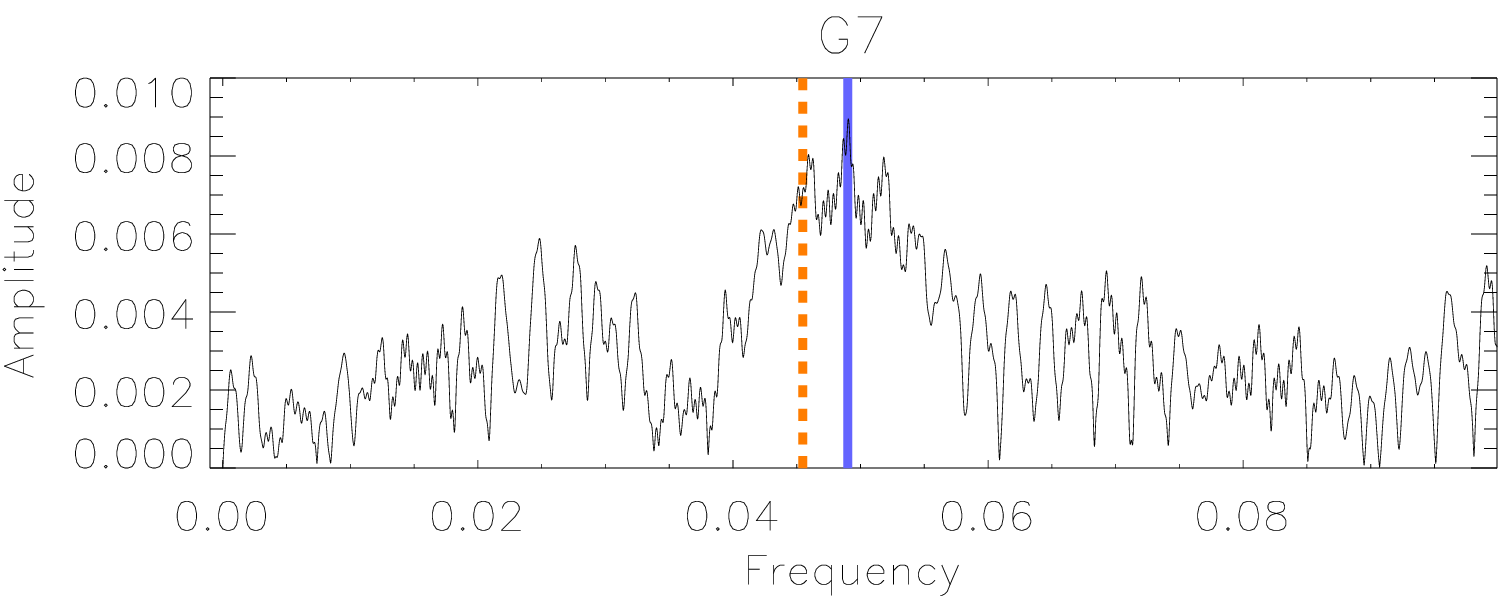} &
    \includegraphics[width=0.474\hsize]{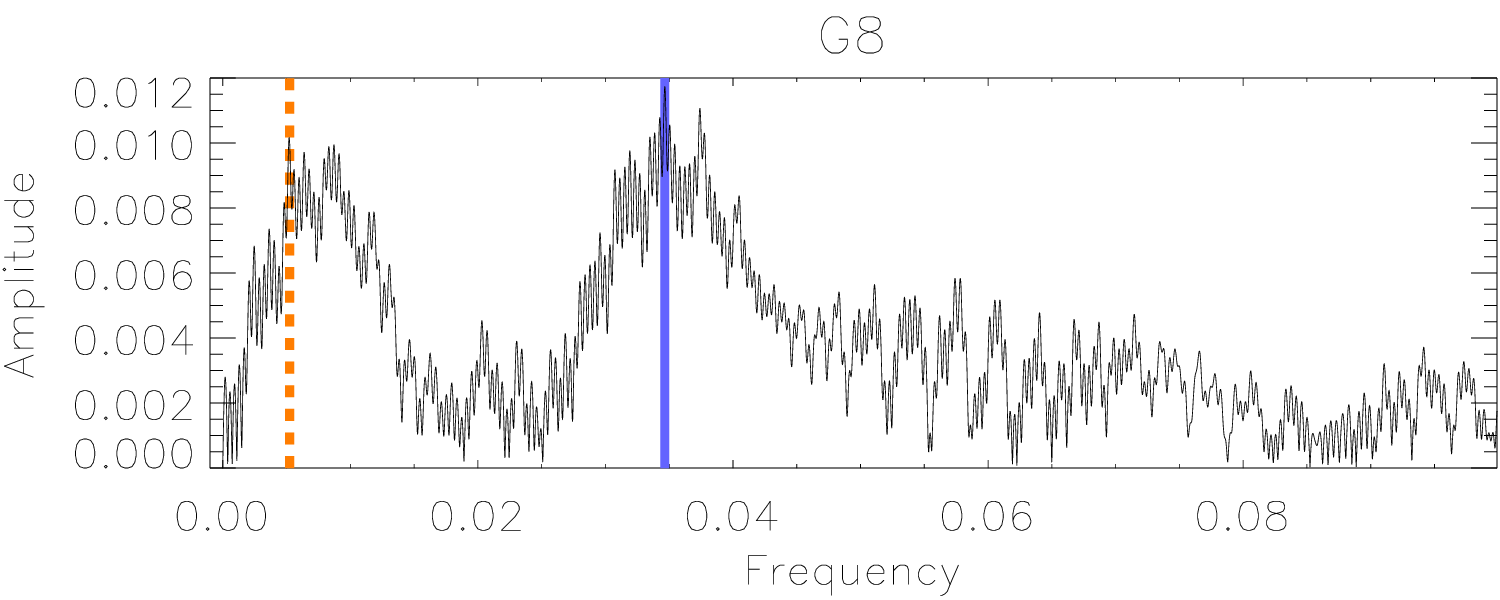} \\

  \end{tabular}
}
\caption{Periodograms for all stars classified as semi-regular in this study, including the new discovered variables (V14-V17) and earlier stars flagged as variables in {\em Gaia} (G1-G9). The vertical solid purple line represents the strongest frequency found in the periodogram using the least square technique, while the dashed orange line represents the frequency found using the fast $\chi^2$ method.}
\label{fig:periodogram}
\end{figure*}

\begin{figure*}
\ContinuedFloat
\centering
\subfloat[][]{
  \begin{tabular}{c}
\includegraphics[width=0.474\hsize]
    {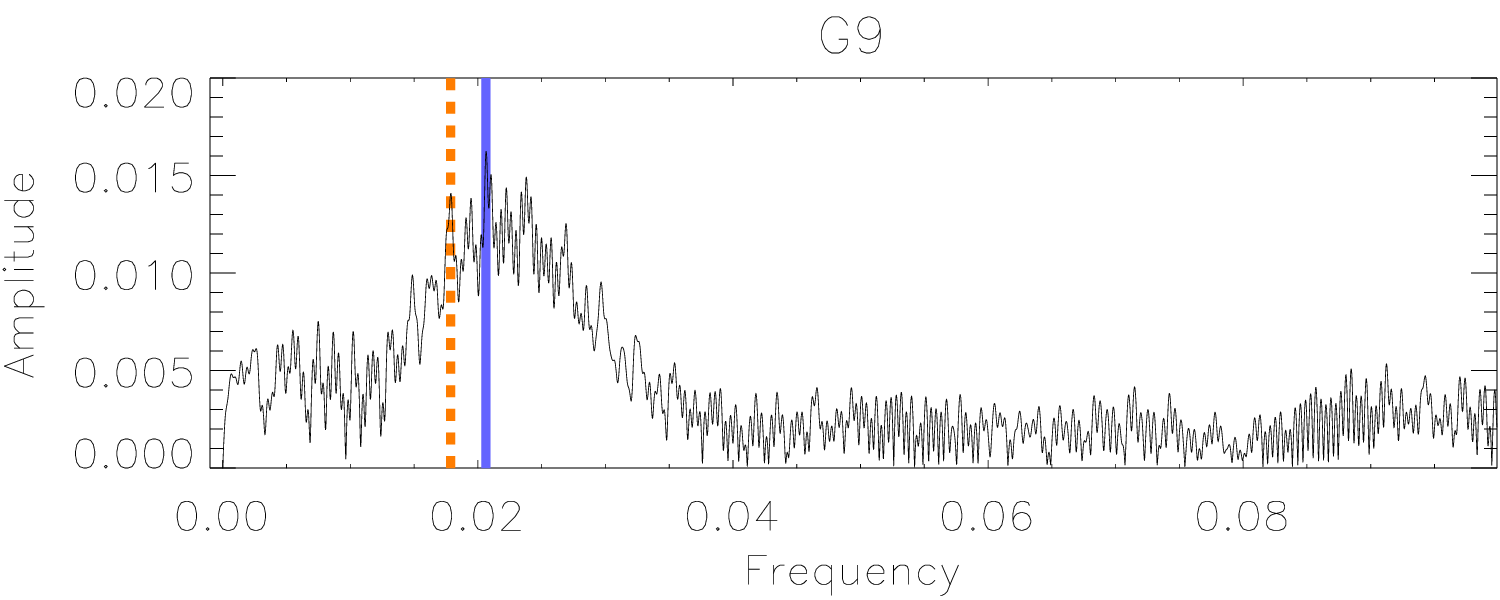} \\
  \end{tabular}
}
\caption{continued.}
\end{figure*}
\section{Conclusions}\label{sec:concl}

    \begin{enumerate}

       \item The analysis of the very crowded central region in Terzan 5 with the very high angular resolution obtained in this work drove us to several interesting discoveries.       
       \item The use of EMCCD and proper manipulation of the obtained images has shown to be a very efficient alternative to mitigate the effect of atmospheric turbulence.
       \item We were able to produce high-angular resolution images with ground-based telescopes.
       \item  It was possible to avoid saturation of the brightest stars presented in the field observed.       
       \item The employment of very sophisticated pipelines to process the images and produce the photometry, such as difference image analysis, and the implementation of our techniques for variable star detection helped us automatically or semi-automatically analyze around 1670 stars in the very crowded central region of Terzan 5. 
       \item As a result of the analysis employed in this work and the 242 observations obtained after the stacking, it was possible to produce a very good coverage of the light curve variation of the only RR Lyrae previously known in the field covered by our reference frame in Terzan 5 which clearly proved to be of the RR0 type. It was also possible to calculate an accurate period for this star.
       \item The discovery of four semiregular variable stars, and the classification of stars flagged as variables in the {\em Gaia} survey, are presented.
       \item We observed an outburst in the visual, and its position appears to coincide with MSP J1748-2446N. We also obtained a visual light curve of the detected 2015 outburst in the LMXB, known as CX 3.

       \end{enumerate}

\begin{acknowledgements}

Support for this project is provided by ANID's Millennium Science Initiative through grant ICN12\textunderscore 009, awarded to the Millennium Institute of Astrophysics (MAS), and by ANID's Basal project FB210003. M.C. acknowledges additional support from FONDECYT Regular grant \#1171273. This research has received funding from the Europlanet 2024 Research Infrastructure (RI) programme. The Europlanet 2024 RI provides free access to the world’s largest collection of planetary simulation and analysis facilities, data services and tools, a ground-based observational network and programme of community support activities. Europlanet 2024 RI has received funding from the European Union’s Horizon 2020 research and innovation programme under grant agreement No. 871149. N.P. acknowledge financial support by FCT--Fundação para a Ciência e a Tecnologia through Portuguese national funds and by FEDER through COMPETE2020-Programa Operacional Competitividade e Internacionalização by the grants UIDB/04434/2020 and UIDP/04434/2020. ChatGPT (powered by OpenAI’s language model, GPT-4; \url{http://openai.com} was used to generate code to produce the fit described in Sect.~\ref{sec:PL}. 

\end{acknowledgements}

%\bibliography{references}

\end{document}